\documentclass[twocolumn,prd,noshowpacs,nofootinbib,amsmath,amssymb]{revtex4}
\usepackage{graphicx,epsfig,psfrag,axodraw,bm,amssymb}
\usepackage{dcolumn}
\usepackage{bm}
\usepackage{color}
\usepackage{mathrsfs,amsfonts,hepunits, color}
\usepackage{mciteplus}
\usepackage{tikz}
\usetikzlibrary{arrows,shapes}
\usetikzlibrary{trees}
\usetikzlibrary{matrix,arrows} 				
\usetikzlibrary{positioning}				
\usetikzlibrary{calc,through}				
\usetikzlibrary{decorations.pathreplacing}  
\usepackage{pgffor}							

\usetikzlibrary{decorations.pathmorphing}	
\usetikzlibrary{decorations.markings}
\usetikzlibrary{snakes}

\tikzset{
    vector/.style={decorate, decoration={snake}, draw},
	provector/.style={decorate, decoration={snake,amplitude=2.5pt}, draw},
	antivector/.style={decorate, decoration={snake,amplitude=-2.5pt}, draw},
    fermion/.style={draw, postaction={decorate},
        decoration={markings,mark=at position .55 with {\arrow[draw]{>}}}},
    fermionbar/.style={draw, postaction={decorate},
        decoration={markings,mark=at position .55 with {\arrow[draw=black]{<}}}},
    fermionnoarrow/.style={draw},
    gluon/.style={decorate, draw,decoration={coil,amplitude=4pt, segment length=6pt}, line width=1},
    scalar/.style={dashed,draw, postaction={decorate},
        decoration={markings,mark=at position .55 with {\arrow[draw]{>}}}},
    scalarbar/.style={dashed,draw, postaction={decorate},
        decoration={markings,mark=at position .55 with {\arrow[draw]{<}}}},
    scalarnoarrow/.style={dash pattern = on 6 pt off 3 pt,draw},
    electron/.style={draw, postaction={decorate},
        decoration={markings,mark=at position .55 with {\arrow[draw]{>}}}},
	bigvector/.style={decorate, decoration={snake,amplitude=4pt}, draw},
	vectorscalar/.style={loosely dotted,draw, postaction={decorate}},
}
\RequirePackage{xspace}
\usepackage{relsize}
\usepackage{slashed}
\newcommand{\fb}{{\rm fb}}

\newcommand{\mub}{{\rm \mu b}}

\newcommand{\apr}{A^\prime}
\newcommand{\be}{\begin{eqnarray}}
\newcommand{\ee}{\end{eqnarray}}

\def\lsim{\mathrel{\rlap{\lower4pt\hbox{\hskip 0.5 pt$\sim$}}
    \raise1pt\hbox{$<$}}}                
\def\gsim{\mathrel{\rlap{\lower4pt\hbox{\hskip1pt$\sim$}}
    \raise1pt\hbox{$>$}}} 

\newcommand{\schi}{s_{\chi\bar\chi}}  

\newcommand{\pf}[2]{\left(\frac{#1}{#2}\right)}
\newcommand{\g}{{\rm g}}
\newcommand{\s}{{\rm s}}

\begin{document}

\title{Testing GeV-Scale Dark Matter with Fixed-Target Missing Momentum Experiments} 
 \author{Eder Izaguirre, Gordan Krnjaic, Philip Schuster, and Natalia Toro}

\vspace{2cm}
\affiliation{ \\ Perimeter Institute for Theoretical Physics, Waterloo, Ontario, Canada                  }
\vspace{5cm}
\date{\today}

\begin{abstract}
We describe an approach to detect dark matter and other invisible particles with mass below a GeV, 
exploiting missing energy-momentum measurements and other kinematic features of fixed-target production.  
In the case of an invisibly decaying MeV--GeV-scale dark photon, this approach can improve on present constraints by 2--6 orders of magnitude over the entire mass range, reaching sensitivity as low as $\epsilon^2\sim 10^{-14}$. 
Moreover, the approach can explore essentially all of the viable parameter space for thermal or asymmetric dark matter annihilating through the vector portal.  
\end{abstract}

\maketitle

%
%

\section{Introduction}
Existing techniques to search for dark matter (DM) are most effective in two regimes:  
if dark matter is heavy like a WIMP \cite{Akerib:2013tjd,Aalseth:2012if,Aprile:2013doa,Fermi-LAT:2013uma,DAMA,Chatrchyan:2012me,ATLAS:2012zim}, or if it is very 
light and coherent like an axion field \cite{Asztalos:2001tf,Bahre:2013ywa,Irastorza:2011gs,Vogel:2013bta,Irastorza:2012qf,Horns:2012jf,Stadnik:2014ala,Stadnik:2013raa,Stadnik:2014xja}. 
If dark matter is lighter than a few GeV and not coherent, then direct detection techniques are notoriously difficult.
But some of the most appealing dark matter scenarios overlap with this difficult category, 
such as the case when dark matter and baryons have a common origin with comparable number
 densities \cite{Kaplan:1991ah,Kaplan:2009ag} or DM is part of a hidden sector (see \cite{Essig:2013lka} for a review). 
This largely open field of GeV-scale dark matter possibilities offers well-motivated discovery opportunities and is ripe for experimental exploration. 

\begin{figure}[t!]
\vspace{0.2cm}
\includegraphics[width=8.4cm]{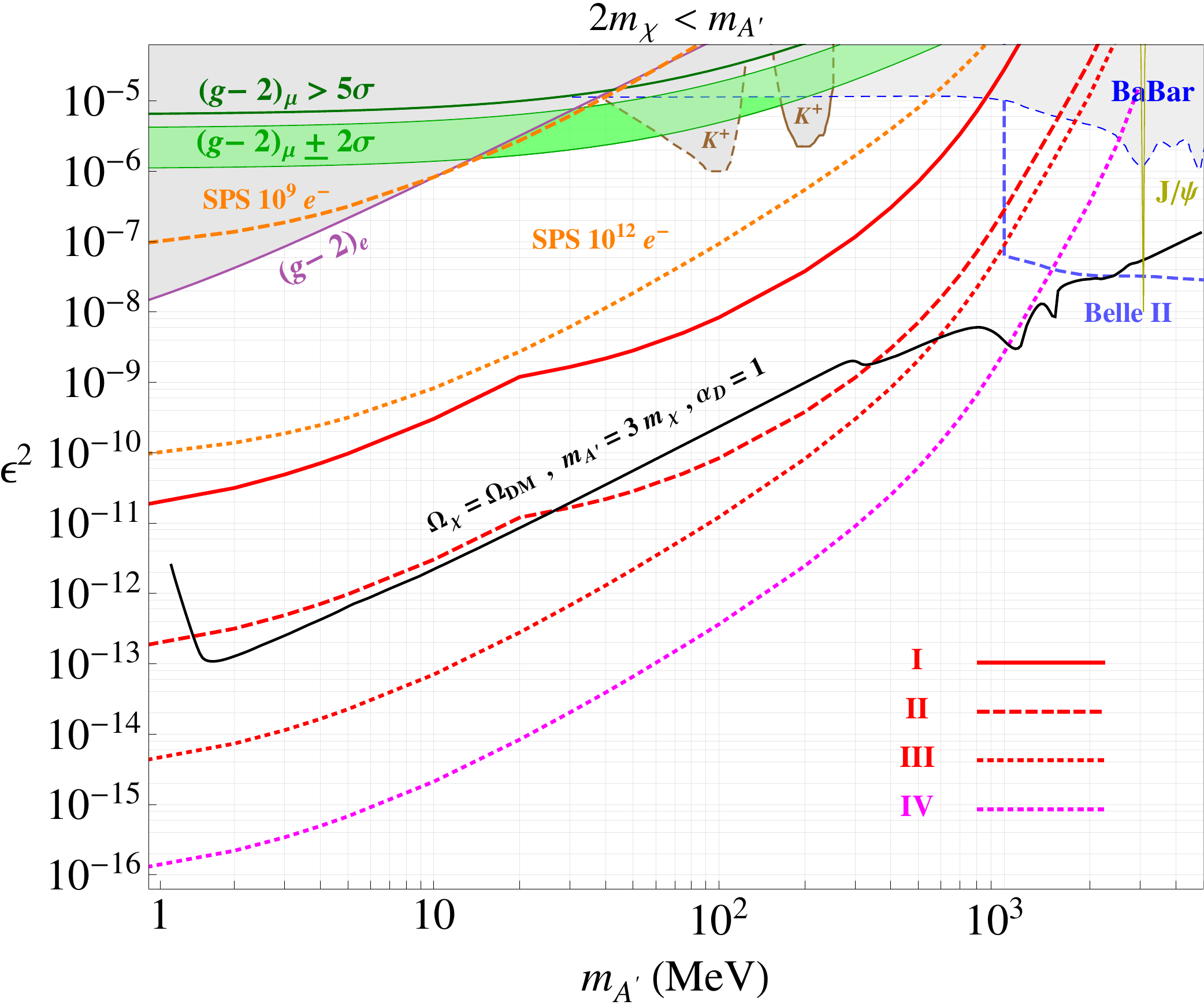}   ~ \\ 
\caption{  
Sensitivity projection for a Tungsten-based missing energy-momentum experiment in a JLab-style setup with an 11 GeV electron beam (red curves, color online) for variations of 
 Scenario B described in Sec. \ref{sec:forward} and illustrated schematically in Fig. \ref{fig:SchematicFwd}b. 
 The upper-most curve labeled I  (red, solid) represents the  90 \% confidence exclusion (2.3 event yield with zero background)
  of an experiment with target thickness of $10^{-2}X_0$ and $10^{15}$ EOT, the middle curve labeled II (red, dashed)
 represents the same exclusion for an upgraded experiment with $10^{16}$ EOT and a thicker target of $10^{-1} X_0$ with varying $P_T$ cuts on the recoiling 
 electron in different kinematic regions  (see Sec. \ref{sec:forward} for details), and the lowest curve labeled III (red, dotted) represents an ultimate target for this experimental program assuming $3 \times 10^{16}$  EOT and imposing the highest signal-acceptance $P_T$ cuts on the recoiling electron. Here $X_0$ is the radiation length of the target material. The dotted magenta
 curve labeled IV is identical to curve III, only with $10^{18}$ EOT, at which one event is expected from the irreducible neutrino trident background. 
 Also plotted are the projections for an SPS style setup \cite{Andreas:2013lya} using our Monte Carlo for $10^{9}$ and $10^{12}$ EOT. 
 The black curve is the region for which the $\chi$ has a thermal-relic annihilation cross-section for $m_{\apr} = 3 m_\chi$ assuming the aggressive value $\alpha_D = 1$;
 for smaller $\alpha_D$ and/or larger $m_{\apr}/m_{\chi}$ hierarchy the curve moves upward. Below this 
line, $\chi$ is generically overproduced in the early universe unless it avoids thermal equilibrium with the SM. The kinks in the black curves
correspond to thresholds where muonic and hadronic  annihilation channels become open; data for hadronic annihilation is taken from \cite{Agashe:2014kda}. 
 Combined with the projected sensitivity of Belle-II with a mono-photon trigger \cite{Essig:2013vha}, the missing energy-momentum
approach can decisively probe a broad class of DM models. Without making further assumptions about dark sector masses or coupling-constants, this parameter 
space is only constrained by $(g-2)_{e}$ \cite{Giudice:2012ms,Izaguirre:2013uxa}, and $(g-2)_{\mu}$ \cite{Pospelov:2008zw}. 
If  $m_\apr \gg m_\chi$, there are additional constraints from on-shell $\apr$ production in association with SM final states from 
BaBar  \cite{Izaguirre:2013uxa,Essig:2013vha},  BES  ($J/\psi$) \cite{Ablikim:2007ek}, E787 ($K^+$) \cite{Adler:2004hp}, and E949  ($K^+$) \cite{Artamonov:2009sz}. 
}\label{fig:moneyA}
\end{figure}
 
In either of the above scenarios, DM must interact with the Standard Model (SM) to avoid
overproduction in the early universe. 
Among the simplest such interactions are those mediated by a kinetically  
mixed gauge boson ($\apr$) associated with a dark sector gauge symmetry \cite{Okun:1982xi,Holdom:1985ag}. 
Light DM that primarily annihilates through an off-shell $\apr$ 
into Standard Model particles is largely unconstrained by available data \cite{Essig:2013lka}.
With light DM and mediator mass scales $m$ comparable,  
an acceptably small relic density robustly bounds the dark sector coupling $\alpha_D$ and
 kinetic mixing $\epsilon$ (see Sec. 2) by
\be\label{eq:goal}
(\alpha_D\epsilon^2)_{\mbox{relic density}} \gsim O(1)\times10^{-10} \left(\frac{m}{100 \ \MeV}\right)^2 ~. 
\ee
This is an important benchmark level of sensitivity to reach to decisively probe this broad and widely 
considered framework for light DM. 

Recently, new beam-dump experiments have aimed to produce light DM
candidates and then observe their scattering in downstream detectors 
\cite{deNiverville:2012ij,Dharmapalan:2012xp,deNiverville:2011it,Batell:2009di,Izaguirre:2013uxa,Batell:2014mga,Battaglieri:2014qoa,Batell:2014yra,Izaguirre:2014dua,Diamond:2013oda}. 
This is a compelling technique to discover light DM, 
but its reliance on a small re-scattering probability prevents this approach from reaching the 
milestone sensitivity of Eq.~\ref{eq:goal}.

Achieving the desired sensitivity requires the identification of DM production events based {\it solely} on their 
kinematics, which in fixed-target electron-nuclear collisions is quite distinctive \cite{Bjorken:2009mm}. 
Light DM candidates produced in such collisions carry most of the incident beam-energy, 
so a forward detector that can efficiently capture the energy of electron/hadron showers  
can be used to observe this signature above irreducible backgrounds (which are small) and reducible backgrounds (which require aggressive rejection).  
In fact, an effort to exploit this feature and search for light DM using a secondary beam of electrons from SPS spills at CERN was proposed in \cite{Andreas:2013lya}. 

\begin{figure}[t!]
 \vspace{0cm}
  \hspace{-0.5cm}
\includegraphics[width=8.cm]{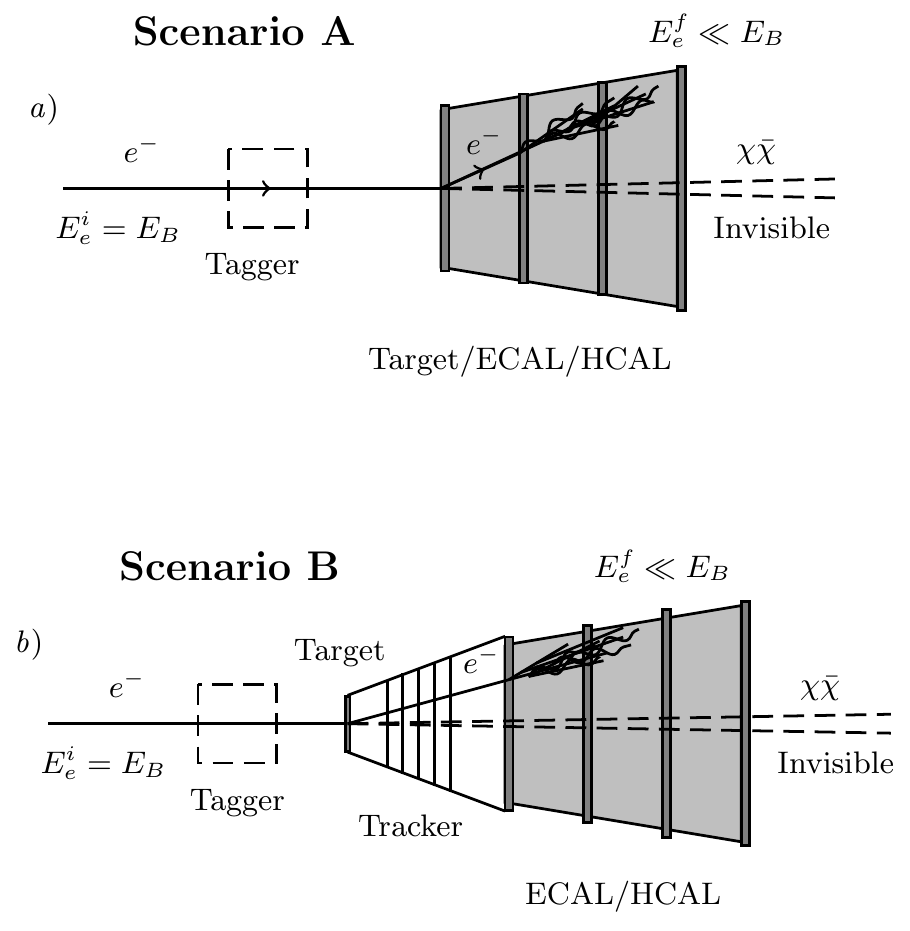}
\label{fig:Schematic}
\caption{ 
a)
Schematic diagram of Scenario A described in Sec. \ref{sec:forwardCal}. Here a  single electron  
 first passes through an upstream tagger to ensure that it carries high momentum.
 It then enters the target/calorimeter volume, and radiatively emits an $\apr$, which carries away most of the 
 beam energy and  leaves behind a feeble electron in the final state.
 b) 
 Schematic diagram of Scenario B described in Sec.~\ref{sec:forward}. In this scenario, the target is thin
 to reduce straggling and charged-current neutrino reaction backgrounds, the calorimeter is spatially
 separated from the target itself to allow clean identification of single charged particle 
 final states. Additionally, the energy-momentum measurement of the recoil electron is used for signal discrimination, 
 to reduce backgrounds associated with hard bremsstrahlung and virtual photon reactions,
 and to measure residual backgrounds \emph{in situ} with well-defined data-driven control regions.   
 For both scenarios, the production mechanism in the target is depicted in Fig. \ref{fig:ProdFig}.  
  }\label{fig:SchematicFwd}
\end{figure}

Our goals in this paper are twofold. We first evaluate the ultimate limitations for calorimeter-based fixed-target DM searches (of which \cite{Andreas:2013lya} is an example), which can use only the reduced energy of the recoiling electron and vetoes on other visible products as discriminating variables to search for light DM.  Such a setup is illustrated in Figure \ref{fig:Schematic}(a).  While neutrino production reactions set an in-principle background floor, in practice such an experiment will likely be limited by instrumental backgrounds --- specifically, detection inefficiencies that allow rare photo-production reactions to mimic the missing energy signature. The second goal of this paper is to propose and examine a modified experimental scenario that can reject such backgrounds more efficiently and robustly.  This setup, illustrated in Figure  \ref{fig:Schematic}(b), adds a tracking detector to measure the recoil electron's \emph{transverse momentum}.  We show that adding this measurement allows significantly improved kinematic background rejection and \emph{in situ} measurements of detector inefficiencies. This approach can reach the milestone sensitivity \eqref{eq:goal} to robustly test vector portal light DM over the entire mass range from $\MeV-$GeV. Moreover, a new-physics interpretation of any positive signal would be greatly bolstered by these additional kinematic handles.

Figure~\ref{fig:moneyA} summarizes the potential sensitivity for a few benchmark scenarios, including for the first time in the literature a realistic calculation of the DM signal yields. 
Belle-II could explore the remaining $m>$GeV portion of this target if mono-photon triggers are implemented \cite{Essig:2013vha}. 
Beyond dark matter physics, the approach we advocate will play an important role in improving sensitivity 
to kinetically mixed dark photons that decay invisibly, nicely complementing the ongoing program of searches for visible decays 
\cite{Bjorken:1988as,Riordan:1987aw,Bross:1989mp,Essig:2009nc,Essig:2010xa,Fayet:2007ua,Freytsis:2009bh,Batell:2009di,Essig:2010gu,Reece:2009un,Wojtsekhowski:2009vz,AmelinoCamelia:2010me,Batell:2009yf,Baumgart:2009tn,deNiverville:2011it,Merkel:2011ze,Abrahamyan:2011gv,Aubert:2009cp,Babusci:2012cr,Echenard:2012hq,Adlarson:2013eza,Hook:2010tw}.
Indeed, while the window identified six years ago for \emph{visibly} decaying dark photons to explain the muon $g-2$ anomaly has recently been closed \cite{NA48}, the corresponding parameter space for invisibly decaying dark-photons has not been fully explored. The approach outlined in this paper will cover the entire $g-2$ anomaly region for {\it invisible decays} (as does the proposal of \cite{Andreas:2013lya}) 
and has sensitivity that extends beyond any existing or planned experiment by several orders of magnitude, in a manner largely insensitive to model details.


Section \ref{sec:model} summarize our benchmark model for light dark matter interacting with the standard model through its coupling to a new gauge boson (``dark photon'') that kinetically mixes with the photon, 
and summarizes existing constraints. Section \ref{sec:signal} summarizes the essential kinematic features of dark photon and light DM production. Section \ref{sec:forwardCal} evaluates the ultimate limits of a fixed-target style missing energy-momentum approach based on calorimetry alone, and in particular identifies important physics and instrumental backgrounds.
Section \ref{sec:forward} describes our proposal for a missing energy-momentum experiment that can mitigate backgrounds using kinematic information and near-target tracking. 
Section \ref{sec:conclusion} summarizes our findings and highlights important directions for future work.

\section{Vector Portal Light Dark Matter}
\label{sec:model}
Hidden sectors with MeV--GeV light DM are a simple, natural, and widely considered extension of the Standard Model. Such sectors remain weakly constrained experimentally, though they have been studied in many contexts -- for example to address anomalies in dark matter direct 
and indirect detection \cite{Boehm:2003bt,Huh:2007zw,ArkaniHamed:2008qn,Hooper:2008im,Cirelli:2009uv}, resolve puzzles in simulations of structure formation \cite{CyrRacine:2012fz,Kaplinghat:2013xca}, modify the number of relativistic species in the early universe \cite{Kaplan:2011yj,Brust:2013ova}, explain 
the ``cosmological coincidence" between dark and visible energy-densities \cite{Kaplan:1991ah,Kaplan:2009ag}, resolve the proton charge radius and other SM anomalies
\cite{Barger:2010aj,Izaguirre:2014cza,Batell:2011qq,TuckerSmith:2010ra,Kahn:2007ru},
and explore novel hidden-sector phenomenology \cite{Galison:1983pa,Boehm:2003hm,Pospelov:2007mp,Finkbeiner:2007kk,Pospelov:2008zw,ArkaniHamed:2008qn,Alves:2009nf,Feng:2008mu,Feng:2008ya,Morrissey:2009ur,Chang:2010yk,Morris:2011dj,Falkowski:2011xh,Essig:2011nj,Kaplan:2011yj,Andreas:2011in,An:2012va,Graham:2012su,Hooper:2012cw,Cline:2012is,Foot:2014mia,Hochberg:2014dra,Shuve:2014doa,Krnjaic:2014xza,Detmold:2014qqa}.

The elaborate parameter space for this large class of theories motivates a simplified-model approach for characterizing experimental bounds 
 and projecting the sensitivities of future searches. 
To be concrete, we consider a simple dark sector consisting of a Dirac fermion DM particle $\chi$ with unit charge under a 
spontaneously broken abelian gauge group $U(1)_D$. 
The most general renormalizable Lagrangian for this scenario  contains 
\be
 \label{eq:lagrangian}
{\cal L}_{D} \supset 
 \frac{\epsilon_Y}{2} F^\prime_{\mu\nu} B_{\mu \nu} + \frac{m^2_{A^\prime}}{2} A^{\prime}_\mu A^{\prime\, \mu}+  \bar \chi ( i \displaystyle{\not}{D}- m_\chi) \chi,
 \ee
where $A^\prime$  is the $U(1)_D$ gauge boson, 
 $F^\prime_{\mu\nu} = \partial_{[\mu,} A^\prime_{\nu]}$ 
 and  $B_{\mu\nu} = \partial_{[\mu,} B_{\nu]}$ 
 are  the dark and hypercharge field strength tensors, and $m_{\chi, {A^\prime}}$ are the appropriate dark sector masses.  
 The covariant derivative $D_\mu \equiv \partial_\mu + i g_D \apr_\mu$ contains the coupling 
 constant $g_D$, and we define $\alpha_D \equiv g_D^2/4\pi$ in analogy with electromagnetism. The $\apr$-hypercharge kinetic mixing 
 parameter $\epsilon_Y$ is expected to be small ($\epsilon \ll 1$) because it most-naturally arises at loop level if {\it any} particles in nature carry charges under both $U(1)_Y$ and $U(1)_D$. 
 
After electroweak symmetry breaking, the hypercharge field is $B_{\mu} = \cos \theta_W A_{\mu} -\sin \theta_W Z_{\mu}$ in
the mass eigenbasis, so the kinetic mixing between 
dark and visible photons becomes $\frac{\epsilon}{2} F^\prime_{\mu\nu}  F_{\mu\nu}$, where $\epsilon \equiv \epsilon_Y \cos\theta_W$ and $ \theta_W$ is the weak mixing angle. 
Diagonalizing the $A, A^\prime$ field strengths, thus,  gives all charged SM particles $U(1)_D$ millicharges proportional to $\epsilon e$;
any photon in a QED Feynman diagram can be replaced with an $\apr$, with its coupling to SM states rescaled by $\epsilon$. 
This simplified model serves as a useful avatar for a generic dark sector because its parameter space
can easily be reinterpreted to constrain many other, more elaborate scenarios. 


 \begin{figure}[t!]
 \vspace{0cm}
 \hspace{1cm}

\includegraphics[width=8.cm]{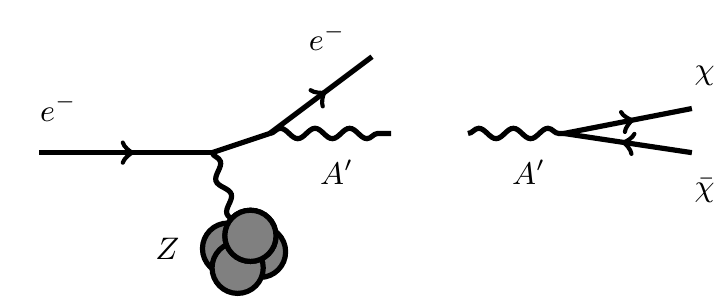}

\caption{  Radiative production of an $\apr$ in a coherent electron-nucleus collision followed by a prompt decay 
to dark sector invisible states $\apr \to  \bar\chi \chi$. Production of $\bar\chi \chi$ can also proceed through 
an off-shell $A'$ with an extra surpression of $\alpha_D/\pi$.
  }\label{fig:ProdFig}
\end{figure}


Beyond its role as a convenient parametrization for more general sectors, this scenario is also a self-contained, renormalizable theory of dark matter. 
If the DM is particle-antiparticle symmetric and $m_{\apr} > m_\chi$,  the relic density is set by $\chi \bar \chi$ annihilation
to SM final states, which yields the observed abundance for 
 \be\label{eq:annihilation}
\epsilon^2 \simeq  1.3 \times 10^{-8}  \left(\frac{m_{\apr}}{10 {\rm \ MeV }  }\right)^4       \left(\frac{ \rm MeV}{ m_\chi} \right)^2  \left(\frac{10^{-2}}{ \alpha_D }\right)~.
  \ee
The mass hierarchy $m_{\apr} > m_\chi$ and resulting dominant $\chi \bar \chi \to e^+ e^-$ annihilation channel allow this 
scenario to remain compatible with CMB constraints (see below)\footnote{If $m_{\apr} < m_\chi$, the dominant 
annihilation channel is $\bar \chi \chi \to \apr \apr$, which is not suppressed by $\epsilon$, is more readily constrained by late time CMB measurements, 
and easily leads to thermal underproduction in the early universe unless $\alpha_D \ll \alpha$. In this region of parameter space, $\apr$ decays visibly 
and doesn't contribute to the observables considered in this paper.}. 
Larger values of $\epsilon$ yield $\Omega_\chi < \Omega_{DM}$,
so $\chi$ can still be a subdominant fraction of the dark sector, but smaller values overclose the universe if $\chi$ was {\it ever} in thermal
equilibrium with the visible sector, so this places a generic constraint on the parameter space.  Indeed, even if the initial  $\chi$ population is matter-asymmetric,  
the annihilation rate must still exceed the thermal-relic value to erase the matter-symmetric $\chi \bar \chi$ population. 
The lowest black curve in Fig. \ref{fig:money} is the region for which which a 
thermal relic $\chi$ constitutes all of the dark matter for $m_{\apr} = 3 m_\chi$ and $\alpha_D = 1$. For lower $\alpha_D$ or a greater $m_{\apr}/m_\chi$ ratio,  
the relic density curve moves upward on the plot, so experimentally probing down to this diagonal suffices to cover the entire parameter 
space for which the DM-SM coupling is appreciable enough to keep the $\chi$ relic density below $\Omega_{DM}$.
The condition for $\chi$ to thermalize with the radiation in the early universe is,
 \be
 \epsilon^2 \sim \frac{ T^2 H(T) }{ \alpha \alpha_D n_e(T)} \biggr|_{T = 2m_\chi} 
 \!\! \!\!  \gsim  2.1 \times 10^{-17}  \left(\frac{   m_{\chi}       }{     10 \rm \ MeV     }\right) \!\!  \left(\frac{   0.1   }{  \alpha_D    }\right),~~ 
 \ee
assuming ${m_{\apr} \sim m_\chi}$. The parameter space along the relic density curve in Fig. \ref{fig:money}  (black, solid) 
trivially satisfies this requirement over the full MeV-GeV range, so  $\chi$ will have a thermal abundance in the early universe, and the only
viable parameter space is above the relic density curve. 
  
 \medskip
{\noindent \bf Beam-Dump Constraints \\ }  
The parameter space for an invisibly decaying $\apr$ in the MeV-GeV mass range is  constrained by various
electron and proton beam dump experiments. The strongest constraint over most of this range comes from 
the LSND measurement of the $e-\nu$ cross section \cite{deNiverville:2011it,Auerbach:2001wg}, which can be reinterpreted as a bound on the 
DM production via $\pi^0 \to \gamma\apr \to \gamma \bar \chi\chi$  followed by scattering off detector electrons 
$\chi e \to \chi e$, which has the same final state as the neutrino search. 
Similarly the E137 axion search is sensitive to light DM via radiative $\apr$ production followed by the 
decay to $\bar\chi\chi$ and scattering via   $\chi e \to \chi e$ to induce 
GeV-scale electron recoils in a downstream detector \cite{Batell:2014mga}. Finally, the E787 \cite{Adler:2004hp} and E949 \cite{Artamonov:2009sz} experiments, which  
measure the $K^+\to \pi^+\nu\bar\nu$ branching ratio are  sensitive 
to light DM via $K^+\to \pi^+\apr\to \pi^+ \bar \chi\chi$, where the DM carries away missing energy
in place of neutrinos.


 \begin{figure}[t!]
 \hspace{-0.5cm}
\includegraphics[width=8.6cm]{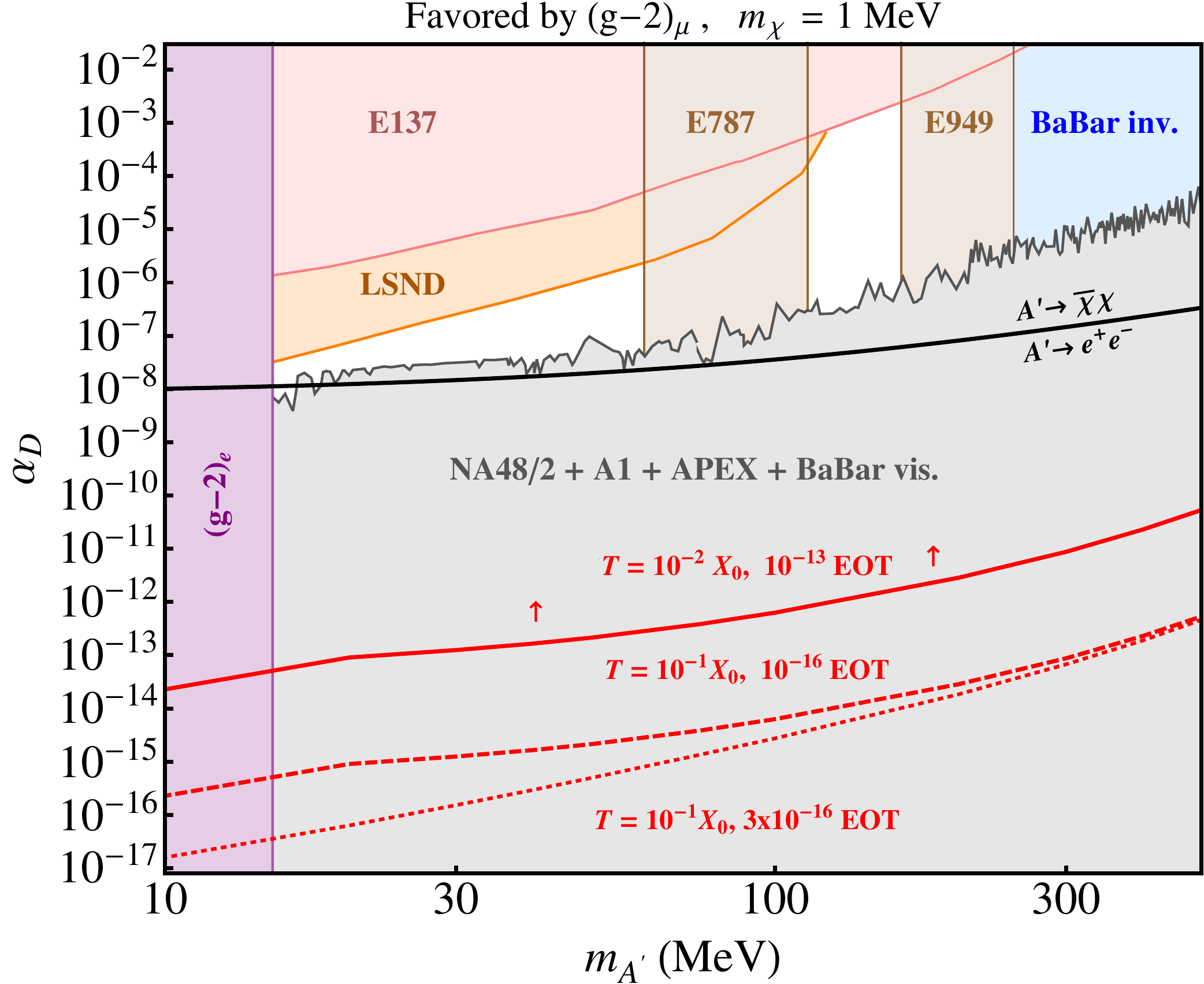}
\caption{Remaining parameter space for which $\apr$ can resolve the $(g-2)_\mu$ anomaly for 
both the visibly and invisibly decaying scenarios. Here $\epsilon$ is fixed to be the
 smallest value constant with experiment (e.g. the lower rim of the green bands in Fig. \ref{fig:money}). 
 The solid black line represents the $Br(\apr \to e^+e^-) = Br(\apr \to e^+e^-) = 0.5$ boundary.
 The missing momentum fixed-target approaches labeled I, II, and II described in this paper can cover
 the entire range of parameters that can resolve the $(g-2)_\mu$ anomaly.
It is, however, still possible to evade some of the constraints if $\apr$ decays in partially-visible cascades
\cite{Izaguirre:2014dua}.}\label{fig:setup}
\end{figure}

\medskip
{\noindent \bf Precision QED Constraints \\}
Since $\apr$ introduce corrections to leptonic vertices in QED diagrams,
both the visibly and invisibly decaying $\apr$ scenarios are  
constrained by measurements $(g-2)_{e,\mu}$ \cite{Giudice:2012ms,Pospelov:2008zw} for $\epsilon
 \gsim  10^{-2} -  10^{-3}$ and $m_\apr \lsim 100 $ MeV. These are the only model-independent
 constraints that arise purely from virtual quantum effects that renormalize leptonic couplings to photons.
 However, for most choices of parameter space for both the visibly and invisibly decaying $\apr$, beam dump constraints
 will be applicable unless the decays yield elaborate visible cascades that may be vetoed by direct searches \footnote{See \cite{Izaguirre:2014dua} for a
 case study in cascade decays whose experimental signatures do not satisfy conventional visible/invisible search criteria.}. 

\medskip
{\noindent \bf Cosmological Bounds \\}  
Model dependent bounds also arise from out-of-equilibrium annihilation 
$\chi \bar\chi \to \ell \ell$ near  $T\sim$ eV, which would partially re-ionize hydrogen and modify the power spectrum of the CMB \cite{Slatyer:2009yq,Galli:2009zc,Galli:2011rz,Hutsi:2011vx}. 
However, this bound is easily easily evaded if DM is matter-asymmetric, the annihilation cross section is $p$-wave suppressed, or if 
$\chi$ is a pseduo-Dirac fermion; see \cite{Izaguirre:2014dua} for a thorough discussion of this constraint.
 Other bounds from DM self interactions arise from 
the bullet cluster \cite{Markevitch:2003at}. These are evaded so long as the self interaction cross section satisfies $\sigma_{\chi\chi}/m_{\chi} \lsim {\rm cm}^2/ {\rm g}$, which implies
\be
\alpha_D \lsim  6 \times 10^{-2}\,\left(  \frac{m_{\apr}}{\rm 10 \>MeV} \right)^2 \left(\frac{\rm MeV}{m_\chi}\right)^{1/2}~.
\ee

Finally, so long as $\apr$ is in thermal equilibrium with the SM in the early universe, there are also constraints from $N_{\rm eff.}$  \cite{Ade:2013ktc,Brust:2013ova} as additional
radiation changes the expansion history during primordial nucleosynthesis. However for $m_{\chi, \apr} \gsim $ 100 keV, the bound is irrelevant.  

\medskip
{ \noindent \bf Supernovae \\ }  
In the absence of additional physics in the dark sector, an invisibly-decaying $\apr$ faces astrophysical constraints 
by potentially allowing supernovae (SN) to release energy more rapidly than is observed. 
We outline the constraint briefly here, but such constraints deserve more careful study in the future. 
For on-shell $A'$ production and decay, the luminosity of $A'$ production is too small to be constrained for 
$\epsilon^2 \lesssim 10^{-20}$ with $m_{\apr} \sim T_{SN} \sim 10$ MeV \cite{Izaguirre:2013uxa}. 
For larger values of $\epsilon^2$, dark photons decaying to $\chi$ can be produced in the SN core with an appreciable luminosity, 
and these can diffuse out of the core carrying energy away. 
The scattering cross section of $\chi$ with baryons at finite temperature $T\sim m_{\chi}$ is  $\sigma =4\pi\alpha_D\epsilon^2\frac{T^2}{m_{A'}^4}$.
Taking $m_{A'}\sim m_{\chi}\sim T$, we can use this result to calculate the mean free path $d$ for hard scattering of the $\chi$
off nucleons in the SN. 
For the range of $\epsilon$ and $\alpha_D$ we consider, the mean free path $d$ is much smaller than the distance $R$ over which 
$\chi$ must propagate to expel energy. 
However, $\chi$ can still diffuse out of the SN with a timescale $\tau_{\rm diff.}\sim R^2/dv$ (where $v$ is the velocity of a typical $\chi$),
and when this is less than a few seconds, the luminosity is accordingly constrained. 
In the other limit where $\tau_{\rm diff.}$ exceeds a few seconds, a small fraction of trapped $\chi$ being produced
can still escape with small probability $e^{-R/d}$ after each $\chi$-nucleon interaction, so the total probability of 
scattering out of the SN after a typical $N\sim (1 \ {\rm sec} \times v)/d$ scattering reactions is $Ne^{-R/d}$.
Using this escape probability and the $\chi$ luminosity of \cite{Izaguirre:2013uxa}, we find that SN requires 
$\alpha_D\epsilon^2 > O({\rm few})\times 10^{-15}$ in the $m_{\apr} \sim m_{\chi}\sim T_{SN} \sim 10$ MeV range. 
Thus, even for $\alpha_D\sim \alpha$, SN constraints do not provide an important constraint on the 
range of couplings required for an acceptable $\chi$ relic density.  
At very small $\alpha_D$, the SN constraints do become increasingly relevant. 

\section{Characteristics of Dark Matter Production}\label{sec:signal}

\begin{figure}[t!]
 \vspace{-0.2cm}
 \hspace{-0.3cm}
\includegraphics[width=8.4cm]{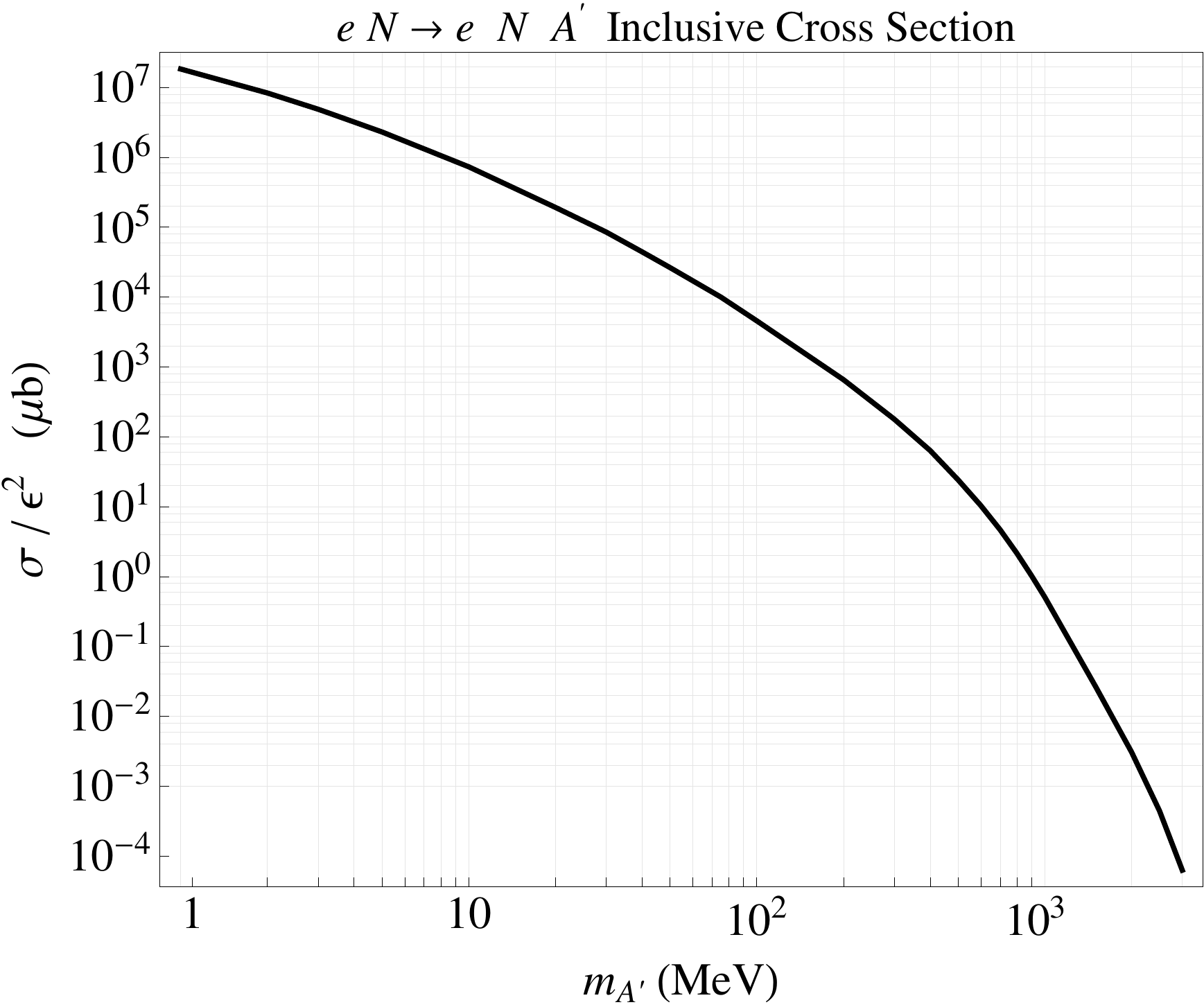}
\caption{  Inclusive cross section for radiative $\apr$ production in fixed-target electron-Tungsten collisions with a 10 GeV beam. 
  }\label{fig:TungCross}
\end{figure}

\begin{figure}[t!]
 \vspace{-0.4cm}
\includegraphics[width=8.4cm]{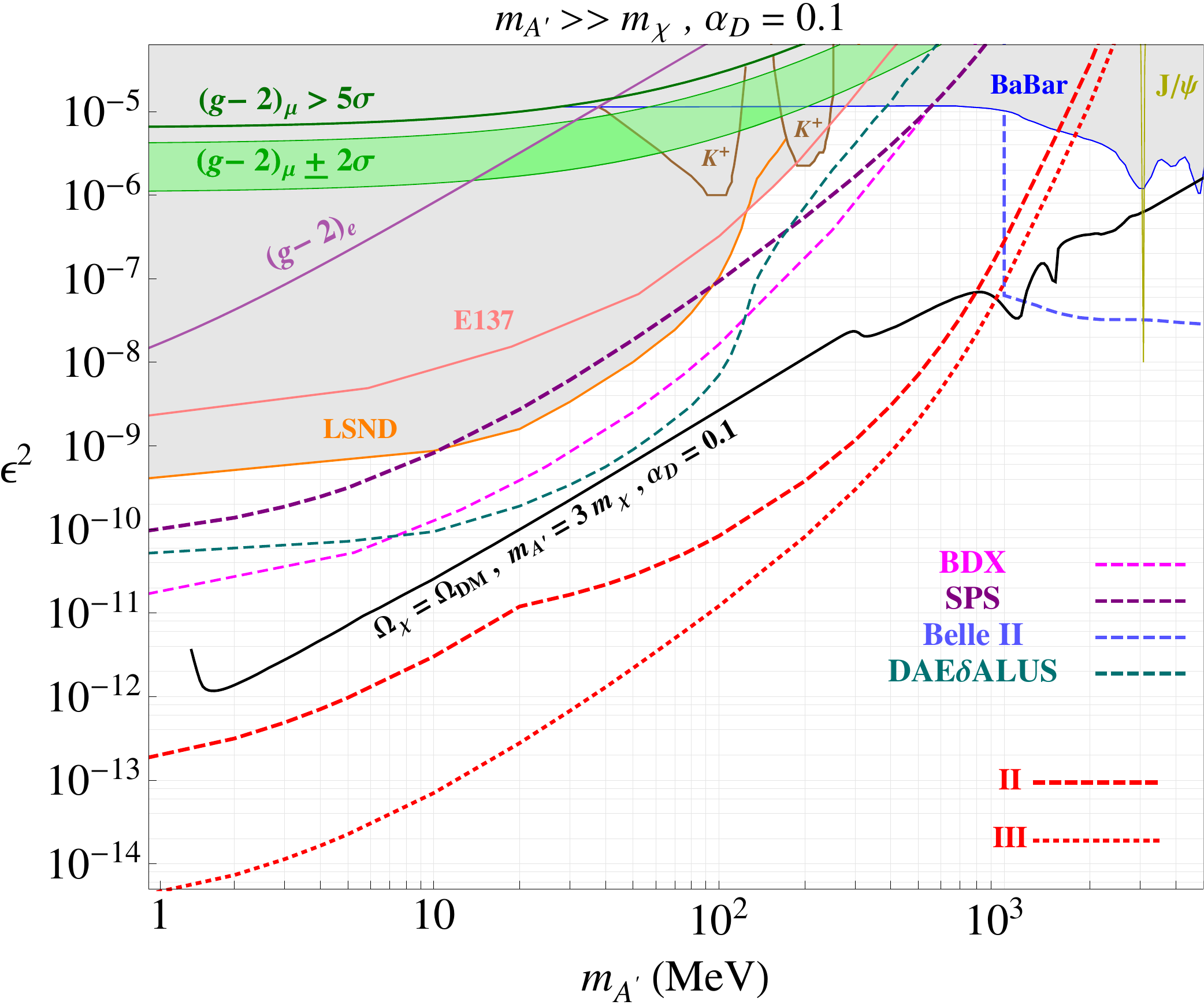}
\caption{ 
Kinetic mixing parameter space for a particular choice of  dark-sector parameters $m_\chi = 1$ MeV and $\alpha_D = 0.1$. 
The dashed and dotted red curves are described in the caption of Fig. \ref{fig:moneyA} and Section~\ref{sec:forward} 
and depict the reach of a JLab style missing energy-momentum experimental program with 
a Tungsten target.  In addition to the more robust constraints plotted in Fig. \ref{fig:moneyA}, here 
there are also bounds from from E137 \cite{Batell:2014mga}, LSND \cite{deNiverville:2011it,Auerbach:2001wg}, and the visible BaBar search \cite{Aubert:2008as,Izaguirre:2013uxa,Essig:2013vha}. Also shown are projected sensitivities for SPS  \cite{Andreas:2013lya},
  BDX \cite{Battaglieri:2014qoa}, 
  DAE$\delta$ALUS \cite{Kahn:2014sra}, and  
 Belle II \cite{Essig:2013vha} assuming the latter can implement  a mono photon trigger. 
}\label{fig:money}
\end{figure}

We use the simple model of Eq. \ref{eq:lagrangian} to characterize the sensitivity of the experimental approaches described in this paper. 
This model allows for direct $A'$ production in a fixed target setting. The $\apr$ can then decay invisibly to $\chi$ pairs when $2m_{\chi}< m_{A'}$, or 
propagate virtually to allow direct $\chi\bar{\chi}$ production. 
This later process is representative of direct dark matter production through a four-fermion operator with electrons. 
The former process covers the well-studied invisible decay of a dark photon \cite{Essig:2013lka}. 
We use a complete Monte Carlo model of $\chi$ production described in \cite{Bjorken:2009mm}, which uses a modified version of \texttt{Madgraph 4} \cite{Alwall:2007st}, to calculate the signal yields. The modification of \texttt{Madgraph 4} includes: (1) initial-state particle masses, (2) a new-physics model including a massive $\apr$ gauge boson coupled to electrons with coupling $e\epsilon$, and (3) introduction of a momentum-dependent form factor for photon-nucleus interactions.  In Figure \ref{fig:TungCross} we show
 the  $e^- N \to e^- N A^\prime$   production cross section for our mass range of interest. This section summarizes the important physics of production that will be exploited by the techniques described below.


We focus on an electron beam with energy $E_0$ exceeding $O(1) \ \GeV$ (suitable for JLab, for example) incident on a fixed target.
On-shell $A'$ production is described in detail in the literature \cite{Bjorken:2009mm,Andreas:2012mt}, so we will only summarize key features. 
In terms of the energy fraction $x=E_{A'}/E_0$ carried by the $A'$,
a simple approximate differential cross-section for production is
\be
\frac{d\sigma}{dx} = (4\alpha^3 \epsilon^2 )\bar\Phi(m_{A'},E_0) \frac{x^2 + 3 (1-x)}{3(1-x)m_{A'}^2} ~~ ,\label{dsigmadx}
\ee
dominated by $A'$ bremsstrahlung off the electron in scattering off target nuclei.
where $\bar\Phi(m_{A'},E_0)$ is defined in Eq.~A18 of \cite{Bjorken:2009mm} and describes the coherent nuclear form factor. 
Note that this expression differs by a factor of 2 relative to Eq.  A14 of  \cite{Bjorken:2009mm}, which contains a typographical 
error; the corrected version is given in Eq.~5 of \cite{Andreas:2012mt}.
The minimum momentum transfer to the nucleus is $q_{\rm min}\approx m_{A'}^2/2E_0$. When $q_{\rm min}$ is smaller than the 
inverse nuclear size $\approx 0.4 \, \GeV/A^{1/3}$, $\bar\Phi(m_{A'},E_0)\approx Z^2 \times ``Log''$ (see
Fig. 10 of \cite{Bjorken:2009mm}), a logarithmic factor which for an 11 GeV beam on Tungsten is $O(10)$ for $A'$ masses below a few hundred
MeV, dropping rapidly below 1 for $O(\GeV)$ masses.
This expression is dominated by the range $(1-x)\lesssim \delta$ where
 \be
 \delta \equiv \max\left(  \frac{m_{A'} }{E_0}, \frac{m_e^2}{m_{A'}^2}, \frac{ m_e}{E_0}\right)~.
 \ee
The total $A'$ production cross-section scales as 
\be
\sigma_{A'} \approx \frac{4}{3} \frac{\alpha^3 \epsilon^2}{m_{A'}^2} \bar\Phi(m_{A'},E_0)   \left[   \log(\delta^{-1})+O(1) \right]~~ . \label{crossSectionWW}
\ee

The $A'$ yield for a mono-energetic beam on a Tungsten (W) target of $T$ radiation lengths is given by 
\be \label{eq:Aprod}
N_{A'} &=& \sigma_{A'} \cdot \pf{T X_0 N_e N_0}{A}, \nonumber \\
&\approx&1.6\times10^{-2} N_e T \epsilon^2 \left(\frac{10 \ \MeV }{m_{A'}}   \right)^2
\ee
where $X_0$ is the radiation length of W in $\g/\cm^2$, $A$ the atomic mass in $\g/\rm{mole}$, $N_e$ is the number of electrons on target, 
and $N_0$ Avogadro's number (the latter factor is the ``luminosity''). For the second line, we quote the raw yield 
for a benchmark beam energy of $E_0=11 \ \GeV$, $m_{A'}=10 \ \MeV$ computed 
in full Monte Carlo.  

Four essential kinematic features of production are:
\begin{itemize}
\item The $A'$ \textbf{energy} is peaked at $x \approx 1$, with median value $\langle 1-x \rangle\sim O(\sqrt{\delta}\,)$. 
 From our full simulation, we find $0.02 < \langle 1-x\rangle  < 0.2$ for $A'$ in the  MeV--GeV range and an 11 GeV beam energy.  
\item The $A'$ \textbf{angle} relative to the beam-line is also peaked forward (roughly as $m_{A'}/E_0 \times \delta^{1/4}$) in a narrower region than the typical  opening angle for the $A'$, i.e. $m_{A'}/(E_0 x)$.   
\item The outgoing electron $\mathbf{p_T}$ has a median value well parametrized by $\langle p_T/\MeV \rangle\sim (m_{A'}/4\MeV)^{0.9}$.
\item Production is dominated by momentum transfers of order $\vec{\, q}_{\rm min}$, so the recoiling {\bf nucleus} has kinetic energy of order $|\vec{\, q}|^{\,2}_{\rm min}/2m_N \approx m_{\apr}^4/(8E_0^2 m_N) \sim$ few keV,  so it plays little role in identifying signal
production events.
\end{itemize}


The qualitative features of on-shell $A'$ production off an electron beam apply equally to the case of 
$\chi\bar\chi$ production mediated by an off-shell $A'$ (see \cite{Bjorken:2009mm} for a detailed discussion).   
Far above the $A'$ resonance, the cross-section differential in $\schi \equiv (p_\chi+p_{\bar\chi})^2$ can be written simply as
\be\label{ddschi}
\frac{d\sigma}{d\schi} =  \frac{4\alpha_D}{3 \pi}  \frac{(\alpha^3 \epsilon^2 )\Phi}{\schi^2} \log(\delta^{-1}) \sqrt{1-4 y} (1+2 y)~.
\ee 
where $y = (E_0-E_1)/E_0$ and $E_1$ is the energy of the scattered electron. The $1/s^2$ behavior implies that the production is dominated near threshold, at $\sqrt{\schi}\sim (2-4) m_{\chi}$.   The peaking of the angle--energy distribution at forward angles and high $\chi\bar\chi$ pair energy that were noted above continue to hold, with the role of $m_{A'}$ now played by (few)$\times m_{\chi}$.  A reasonable approximation to this scaling in the case of fermionic $\chi$ is 
\be
N_{\chi\bar\chi} \approx \pf{\alpha_D}{\pi} N_{A'}\bigg|_{m_{A'} = \sqrt{10} m_{\chi}},\label{offshellYield}
\ee
where the second factor denotes the result of \eqref{eq:Aprod} 
 at the fictitious $A'$ mass that dominates the $\schi$ integral.  For bosonic $\chi$ produced through an off-shell $A'$, the differential cross-section analogous to \eqref{ddschi} is $p$-wave suppressed near threshold, resulting in a further suppression of yield by roughly an order of magnitude.  

For concreteness, we focus throughout the following discussion on the case $m_\chi < m_{A'}/2$ where on-shell $A'$ production dominates, but the experimental scenarios are equally applicable to the off-shell case.


\section{The Limits of A Forward Calorimeter Experiment}\label{sec:forwardCal}

As a baseline setup, we consider the scenario depicted in Figure \ref{fig:SchematicFwd}, based on the SPS proposal to search for light dark matter \cite{Andreas:2013lya}.  
In addition to a full calculation of dark matter production rates, 
our goal is to calculate the irreducible physics limitations of this approach and identify those aspects of experimental performance 
that are likely to limit the sensitivity. This will directly motivate our proposal in section \ref{sec:forward}. 

We consider a very low-current beam of $\sim 10\GeV$ electrons (the SPS proposal uses $30-50 \ \GeV$) directed into a detector with excellent forward calorimetry with the goal of measuring missing energy in the forward direction.  Figure~\ref{fig:Schematic} illustrates the layout of a calorimeter approach as ``scenario A'' -- this is directly inspired by \cite{Andreas:2013lya}. 
In typical signal events, the (invisible) $A'$ carries most of the beam energy, so that visible products 
carry only a small fraction of the energy and forward momentum of the beam.
For the low-energy electron recoil to be an effective signal, it is essential to detect the passage of \emph{one electron at a time} through the detector and the \emph{absence} of other detected objects originating from the collision carrying momentum along the beam direction.  
For concreteness, we identify as ``signal-like'' events with a \emph{single} EM shower in the calorimeter, consistent with a recoil electron in the energy range $50\ \MeV< E_{e} < 0.1 E_{\rm beam} \approx 1 \GeV$, with a veto on any additional EM showers, hadronic showers, or minimum-ionizing particles traveling along the beamline in time with the incoming electron.  Each incoming beam electron is tagged as a high energy electron, either using a magnetic selector, near target tracking in a magnetic field, synchrotron radiation tagging techniques (see \cite{Andreas:2013lya}), or a combination.

In this scenario, the target is a calorimeter in the forward part of the detector, much thicker than one radiation length.  
A clean detection requires that the production reaction happen in the first few radiation lengths and before any hard radiation processes or appreciable straggling.  
The physics of straggling (electron energy loss) in the first radiation length of materials is discussed in detail in Tsai \cite{Tsai:1973py}. 
Using the results of section IV.A. of that paper, we find that 70\% of hard-scattering events in a thick target are preceded by photon emissions totalling $>0.1 E_b$.  The showers initiated by these photons would contribute to the measured energy of the recoil electron, pushing it outside the signal window.  Thus, only the remaining 30\% of signal events can pass the recoil energy selection.  
Additionally, the efficiency for the recoil electron to carry less than $10\%$ of the beam energy is about $50-75\%$ (depending on the $A'$ mass), with even more suppression for $A'$ masses below 5 MeV. 
These two effects lead to an overall signal efficiency of $\kappa\approx 0.15-0.23$ (depending on mass).
In all calculations throughout this paper, we use full Monte Carlo simulation of signal $eN \rightarrow eNA'$
production with a mono-energetic electron beam, including form factor suppression associated with the coherent nuclear scattering off of a nucleus $N$, following the 
treatment of \cite{Izaguirre:2013uxa}. We additionally include the inefficiency penalties for straggling and recoil-electron energy fraction  $E_e<0.1 E_{beam}$ as described above, with the latter evaluated in Monte Carlo for each $A'$ mass. 
For reference, the signal yield for the benchmark model of on-shell $\apr$ production is 
\be
N_{A'\rightarrow \mbox{invisible}} &\approx& 3.9 N_e T \epsilon^2 \left(\frac{m_e}{m_{A'} } \right)^2 \kappa
\ee
where $T\sim 1$ is the thickness in radiation lengths, $\kappa\approx 0.2$ (at $m_{A'}=100 \ \MeV$), and $N_e$
is the number of electrons on target. Additional form-factor suppression occurs for $m_{A'}\gtrsim 100\ \MeV$ in Tungsten.
For future reference, $N_e=10^{16}$ electrons impinging on one radiation length of Tungsten corresponds to an integrated electron-nucleon 
luminosity of $40 \ \fb^{-1}$ and electron-nucleus luminosity of $0.2 \ \fb^{-1}$. 

\subsection{Beam and Timing Characteristic Limitations}

To cleanly measure missing energy with the calorimeters, one must avoid pile-up reactions where the calorimeter response to different electrons' showers overlaps.  This constraint dictates the maximum beam current for the experiment.  If all electrons are hitting the same point on the calorimeter, they should therefore be spaced by at least the decay-time of the detector (preferably a few decay times), or at least $30-100$ ns apart \cite{Agashe:2014kda}, i.e. beam currents $\sim (1-3)\cdot 10^{7} e/\s$ or a few pA. By using a diffuse secondary beam or high-frequency beam rastering, the pile-up of electrons could be spread out over the calorimeter's area allowing rates as high as $10^9 e/\s = 160 \ {\rm pA}$  for a calorimeter 10's of cm in transverse size. This calorimeter size ensures that electromagnetic showers from ${\cal O}(10-100)$  near-simultaneous electrons can be spatially separated relative to the calorimeter's Moli\'ere radius, typically of order $\sim$1 cm in dense  materials such as Tungsten \cite{Agashe:2014kda}.  Over several months $\sim 10^7\,\s$, these two scenarios would deliver $10^{14}$ or $10^{16}$ electrons on target (EOT). This is not an irreducible limit, but a challenging technological one to overcome. 

To keep beam related backgrounds minimized, high purity beams are required. The only high purity ($10^{-4}-10^{-6}$ or better) 
beams capable of supporting this mode of operation are continuous wave (CW), such as those found at Jefferson Lab or Mainz.  
Multi-GeV electron beam energies place the missing energy signal from $\apr$ or $\chi\bar\chi$ production far from detector energy thresholds and allow higher-mass dark matter production, so we will focus on the $11$ GeV CW beam capabilities of JLab's Continuous Electron Beam Accelerator Facility (CEBAF).
CEBAF has achieved currents as low as $100$ pA, and could go at least a factor of 2 lower\footnote{We thank Jay Benesch and Arne Freyberger for useful conversations on this topic.}, but beam diagnostics at such low currents are challenging. From the detector timing point of view, a CW beam current of $\sim 2$ pA would allow for difficult but still well separated $10$ MHz timing, while an easier to achieve $\sim 20$ pA would only allow the more challenging $100$ MHz operational mode, in which case some level of rastering would be desirable.  
In practice, operation at few-pA (or lower) currents may be most readily realized using a secondary beam of elastically scattered electrons from a thin target, which are then magnetically deflected (and collimated) into the opening aperture of the detector. This has the advantage of spreading the incoming electrons from each other so that they react in different parts of the detector, but it will likely introduce beam impurities that must be specifically rejected. 

\subsection{Real Missing Energy Background}

The physics backgrounds for this scenario involve the production of high-energy neutrinos which carry away the beam energy.  
Relatively few Standard Model processes produce \emph{only} neutrinos, without accompanying charged particles which could be vetoed, 
and those that do are weak-interaction processes whose cross-sections are suppressed by $G_F$.  
The two leading (effectively) irreducible reactions are:

\textbf{Moller + Charged-Current Exchange:} 
Even with electron- and photon-initiated shower separation, a Moller scattered electron can initiate the recoil shower while the beam 
electron undergoes a charged-current quasi-elastic (CCQE) reaction $eN\rightarrow N'\nu$. 
The CCQE reaction $e^-\, p \rightarrow \nu\,n$ has a cross-section of $8\ \fb$ per nucleon at $10 \ \GeV$ incident beam energy (from the Particle Data Group NUANCE fit to measured $\bar \nu p \rightarrow e^+ n$ cross-sections \cite{Agashe:2014kda}).
Requiring the Moller electron to have energy exceeding $E_{\rm min} = 50 \MeV$, the Moller cross-section per target atom is approximately
\be
\sigma_{Moller} \approx \frac{2 \pi Z \alpha^2}{m_e E_{\rm min}} \approx 0.4 {\rm b}
\ee
 in $W$, bringing the rate for both processes to occur together in the first radiation length to $1-2$ events per $10^{16}$ 
electrons.  
These rate scales with the atomic mass and number of the calorimeter material as $A/(Z (Z+1)^2)$, and so can be significantly larger for low-$Z$ organic materials than the rates given above.  For example, the yields in elemental Carbon are $\sim (0.25)\times 3000$ CCQE events and $\sim 100$ Moller + CCQE events per $10^{16}$ EOT. 

\textbf{Neutrino Pair Production:} The production of neutrino pairs, $e\,N\rightarrow e\,\nu\,\bar\nu\,N$, through electroweak interactions diagrams analogous to electromagnetic ``trident'' processes produces a final state much like our signal, because the invisible $\nu \bar\nu$ pair frequently carries away most of the beam energy.  For a 10 GeV beam incident on Tungsten, the process has a 
cross-section $\sigma_{eN\rightarrow eN\nu\bar{\nu}}=0.03 \ \fb$ computed using MadGraph \cite{Alwall:2007st}, corresponding to a mere $\sim 6 \cdot 10^{-3}$ events per $10^{16}$ electrons on target (roughly independent of target material). 

Additionally, for typical calorimeter performance, there are two more backgrounds
that in practice dominate over the above:  

\textbf{Bremsstrahlung + Charged-Current Exchange:}
If efficient electron- and photon-initiated shower separation is not readily possible in the calorimeter using shower shape discrimination, 
then the largest background is a low energy electromagnetic shower initiated by soft bremsstrahlung
followed by a CCQE reaction.
The probability of the incident electron not initiating a shower with $5-10\%$ of the beam energy over the 
first radiation length is $0.25-0.3$ (see above). Using the above CCQE cross section we obtain a yield of 
$80$ events per $10^{16}$ incident electrons.  The inclusive charged-current rate is a factor of 4 larger, but the resulting hadronic showers will fail the 
requirement of low recoil shower energy. Of course, with efficient (i.e. $99\%$) electron- and photon-initiated shower separation, one can reject CCQE events by 
requiring electron initiated showers, but this is likely not feasible.  

\textbf{Charged-Current Exchange with Exclusive $\pi^0$ Final State:} 
A significant fraction of charged-current reactions result in a single-pion final state, e.g. $e^-\, p \rightarrow \nu\,n\, \pi^0$.  As before, this is technically reducible with efficient electron- and photon-initiated shower separation, but difficult. 
The cross-section at $10$ GeV incident energy can be estimated from related neutrino-induced processes \cite{Agashe:2014kda} to be of order $3-5$ fb per nucleon, 
for a yield of $\sim 25-50$ events per $10^{16}$ EOT in Tungsten (scaling as $A/(Z(Z+1))$.  We have again included the factor $0.25-0.3$ corresponding to 
the probability of the incident electron not initiating a shower with $5-10\%$ of the beam energy over the first radiation length. 

In summary, background-free sensitivity is limited by irreducible weak backgrounds for $10^{16}$ EOT, comparable to the (practical) maximum deliverable charge. 
Depending on the composition of the calorimeter (lower background for high-$Z$ materials) and the ability to discriminate $\pi^0$'s and photon-initiated showers from recoiling electrons, the more realistic limit may be $10^{14}$ EOT.
We emphasize that realizing the full $10^{16}$ EOT limit requires a technologically advanced tracker/calorimeter in the first couple 
of radiation lengths that can separate recoiling electrons vs. $\pi^0$ decays and soft brem (electron straggling) initiated showers.
$98-99\%$ rejection of these backgrounds is required to reach the background-free luminosity limit close to $10^{16}$ EOT,
but this hardly seems realistic -- the next section will describe how these can be reduced somewhat with a modified approach. 
With a more attainable background-free limit of $10^{14}$, the sensitivity of such an experiment would have $90\%$ exclusion sensitivity of  
\be
\epsilon_{irreducible}^2 \lesssim 1\times10^{-9} \left( \frac{  m_{A'} }{  100 \ \MeV  } \right)^2,
\ee
corresponding to $2.3$ signal events.  

\subsection{Reducible Physics Backgrounds}
The main class of reducible physics backgrounds, which we focus on here, is where most of the beam's forward momentum is carried by ordinary visible particles (hard photon(s), lepton(s), and hadron(s)), but these are not efficiently detected.  

Because hard bremsstrahlung occurs with O(1) probability in a thick target, it is a background that must be mitigated quite dramatically.  
With $10^{16\,(14)}$ EOT, we expect $\sim 3\cdot 10^{14\,(12)}$ incident electrons to deposit at least 90\% of their energy in a \emph{single} hard bremsstrhalung photon (events in which two or more photons share this energy are more numerous, but also much easier to detect).
The concept of a conventional ``photon veto'' is not very useful for this level of rejection.  
Rather, it is imperative that the detector be sensitive to rare photon-initiated processes (especially hadronic reactions and muon pair-production, which occur before an electromagnetic shower is initiated with probabilities $\sim 3\cdot 10^{-3} A/Z^2$ and $10^{-5}$, respectively).  We therefore begin by estimating the requirements for an experiment to achieve $\lesssim 1$ background event per $10^{16}$ EOT, starting with the rejection of real photons whether they induce standard electromagnetic showers or other hard reactions. 

\textbf{Electromagnetic Showers}
The desired sensitivity imposes two requirements on the electromagnetic calorimeter (ECAL) for the experiment: it must be thick enough that an incident photon will interact with a probability $>1-10^{-15}$, and must not have cracks or dead regions in which the interaction/detection probability is substantially reduced. 
The first requirement dictates the minimum depth of the calorimeter to be $\sim 35$ photon conversion lengths or 45 radiation lengths.  For a high-$Z$ material like Lead Tungstate, this can be achieved in a very reasonable 35 cm of active material.  

It remains to consider the relatively rare hard interactions of a $\approx 10$ GeV photon in such a detector. Since the depth requirements motivate a high-$Z$ calorimeter material, we focus for concreteness on interactions in elemental Tungsten, whose yields are within $O(1)$ of the expected yields in Lead Tungstate or a tungsten-based sampling calorimeter.  A useful point of reference is the photon-conversion cross-section in Tungsten, $\sigma_{\gamma N \rightarrow e^+e^- N} \approx 35$ barns.  

\textbf{Hard-Photon-Induced Hadronic Showers}
The inclusive photon-proton(neutron) cross-sections for $\approx 10 \ \GeV$ photon energies are $\approx 120$ (110) $\mub$ \cite{Caldwell:1973bu}, for a per-nucleus cross-section in Tungsten of about 21 mb and an inclusive probability $\sim 10^{-3}$ per photon.  The majority of these reactions produce multiple charged particles and/or $\pi^0$'s (see \cite{PhysRevD.8.1277} for a detailed breakdown in $\gamma p$ scattering at 9.3 GeV).  For example, the results of \cite{PhysRevD.8.1277} report at least three final-state charged hadrons in 93\% of these events; one would expect comparable two or more charged hadrons in a comparable fraction of $\gamma n$ events.  

Though \cite{PhysRevD.8.1277}, a bubble chamber experiment, did not directly detect $\pi^0$'s, one can infer from their data that about 70\% of these events contain multiple charged hadrons and at least one $\pi^0$ or photon; a further 15\% have either five or more charged hadrons in the final state.  In the present context, these events are relatively easy to reject compared to exclusive processes with relatively few-body final states (in particular, $\gamma N\rightarrow N\pi^+\pi^-$ which is large and dominated by a diffractive process, and $\gamma N\rightarrow n\pi^\pm$ because it contains only one charged final-state particle), which we consider more carefully below.

A large hadronic background arises from diffractive $\gamma N \rightarrow \rho N$ scattering, with $\rho\rightarrow \pi^+\pi^-$.  The measured cross sections for 10 GeV incident photons is $\sigma_{\gamma p \rightarrow \rho p}\approx 15 \ \mu b$ \cite{Butterworth:884219}, consistent with diffractive $\rho$ production dominating the inclusive $\pi^+\pi^- p$ final state of \cite{PhysRevD.8.1277}, for a probability of $7\times 10^{-5}$ per incident photon (accounting for a similar rate from neutrons) or $2\cdot 10^{10}$ events per $10^{16}$ EOT.  In a diffractive process, the final-state nucleon is not energetic.  If both pions are energetic and travel through the detector, then pion rejection with $10^{-5}$ inefficiency would be required to bring this background down to the single-event level.  

$K_SK_L$ production (via the $\phi$ or otherwise) is another potential source of background, with a cross-section per nucleon that should be comparable to 
$\sigma_{\gamma\,p\rightarrow p K^+K^-} \approx 0.6 \mub$ from \cite{PhysRevD.8.1277}, yielding $\sim 10^{9}$ expected $K_L K_S$ events per $10^{16}$ EOT.  
For $10 \ \GeV$ photon energies, the decay products of the $K_S$ typically carry a few $\GeV$. So the resulting prompt $K_S\rightarrow 2\pi^0$ 
decay yields an electromagnetic shower that is well above veto threshold, while $K_S\rightarrow \pi^+\pi^-$
yields a lower two-pion rate than the $\rho$ (with the added possibility of detecting the more energetic recoiling nucleon).

Though small in rate, processes $\gamma\, N\rightarrow \pi \, N$ can be challenging to detect.  The process $\gamma p\rightarrow \pi^+ n$ has an inclusive cross-section $\sigma_{\gamma p\rightarrow \pi^+ n}\approx 0.2 \mub$ per proton \cite{1968PhRvL..20..300B}, or $\sigma_{\gamma Z\rightarrow \pi^+ n + (Z-1)}\approx 15 \ \mub$ per nucleus for a relative probability of $4\times 10^{-7}$ per incident hard photon and $\sim 10^{8}$ such events per $10^{16}$ incident electron.
The process is dominated by a $t$-channel Regge trajectory, and so usually both the neutron and the pion will carry appreciable energy and forward momentum.  
In a suitably-built meter-scale hadronic calorimeter, the neutron energy can be be converted into a visible hadronic shower 
with an inefficiency of order $10^{-3}$. 
A charged pion inefficiency at the level of $10^{-5}$ would then be necessary to remove this background.

Another failure mode for $\gamma p\rightarrow \pi^+ n$ reaction is the region of phase near $u\rightarrow 0$, where the $\pi^+$ recoils backwards or transverse to the beam direction with sub-GeV momentum and the neutron carries most of the hard photon's energy.  The cross-section for this region of phase space is $\sim 1$ nb/nucleon for 10 GeV photons \cite{Anderson:1969bq,1997NuPhA.627..645G}, leading to $\sim 10^{6}$ such events per $10^{16}$ EOT (accounting also for a similar rate from $\gamma p\rightarrow \pi^+ n$).  Rejecting all such events would require an additional detector upstream of the main target specifically designed to reject back-scattered pions, and/or a significantly higher neutron detection efficiency $\sim 10^{-6}$ (for reference, the proposal \cite{Andreas:2013lya} finds a neutron detection inefficiency $\sim 10^{-9}$. By similar considerations, potential pure multi-neutron backgrounds reactions $\gamma n\rightarrow n\bar{n}n$ have
a cross section of $0.1 \ \mu b$ per nucleon yielding $10^{8}$ such events per $10^{16}$ EOT, but these too can be vetoed for inefficiency less than $10^{-3}$. 

\textbf{Muon Conversion}
Photon conversion into muon pairs presents another source of background, with a probability $\sim 2\cdot 10^{-5}$ per incident photon. 
The simplest background occurs from missing the muons altogether, but so long as the muon inefficiency is comparable 
to the pion inefficiency, this background is subdominant compared to the $\rho\rightarrow \pi^+\pi^-$ background.
In-flight decays that occur within the first few cm of the interaction point can yield an electron with lab-frame energy $\lesssim \GeV$ if its center-of-mass-frame momentum points backward.  Without near target track 
tagging, this electron will easily be lost in the recoil electron shower. 
The probability for a muon to decay in this configuration ranges from $10^{-6}$ to $10^{-5}$
for muons in the $0.5-5 \ \GeV$ range. 
This yields a probability for in-flight decay followed by missing the remaining muon at the $10^{-16}$ level
for $10^{-5}$ muon veto inefficiency.  

\textbf{Electron-Induced Backgrounds}
Further reducible backgrounds arise from either hadronic interactions of the incident beam electron or muon trident reactions, in the phase-space where the final-state electron carries relatively little energy.  These reactions proceed dominantly (indeed exclusively, for the hadronic interactions) through virtual-photon emission, and are therefore quite closely related to the interactions of real photons mentioned above.  They are, however, subdominant, because they are suppressed by a factor of $\sim \frac{\alpha}{2\pi} \log(\mu/m_e) \ll 1$ relative to the analogous real-photon-induced processes. 


\subsection{Summary of Sensitivity Estimate (and Limitations)}

\begin{table}[t] 
\begin{center}
\begin{tabular*}{0.5\textwidth}{@{\extracolsep{\fill}}lcl}
\hline\hline
\\[-7pt]
$\quad$  {\bf Real Missing Energy}  &  & {\bf Magnitude ($10^{16}$ EOT)} \\[2pt]
\hline
\\[-6pt]
Brem+CCQE  && $\sim 80$ (reduce with $e$-tag)  \\[2pt]
CCQE+$\pi^0$ && $\sim 25-50$ (reduce with $\pi^0$-tag) \\[2pt]
Moller+CCQE   &&  $\sim 1-2$ \\[2pt]
$eN\rightarrow eN\nu\bar{\nu}$ &&  $\sim10^{-2}$  \\[2pt]
             	\\[-6pt]
\hline \hline
\\[-6pt]
$\quad$  {\bf Reducible Backgrounds}  &  & {\bf Fake Rate/$10^{16}$ EOT} \\[2pt] 
\hline
\\[-6pt]
 $\gamma$ non-interaction  && $\sim3\times 10^{14}e^{-\tfrac{7}{9}(T/X_0=45)}< 1$   \\[2pt]
 $\gamma p\rightarrow \pi^+n$ && $\sim10^{8}\times \epsilon_{\pi}\epsilon_{n}$	\\
 $\gamma p\rightarrow \pi^+n$ (backscatter $\pi^+$) && $\sim10^{6}\times \epsilon_{n}$	 (see text)\\
 $\gamma N\rightarrow (\rho,\omega,\phi)N\rightarrow \pi^+\pi^-N$ & &  $\sim2\times10^{10}\epsilon_{\pi}^2$	\\
 $\gamma n\rightarrow n\bar{n}n$ && $\sim10^{8}\times \epsilon_n^3$	\\
 $\gamma N\rightarrow N\mu^+\mu^-$ && $\sim 6\times 10^9 \times \epsilon_{\mu}^2$	\\
 $\mu^+(\mu^-\rightarrow e\bar{\nu}\nu$ in-flight decay) & & $\sim6\times 10^4\times\epsilon_{\mu}$    \\[2pt] 
 \hline
 	\end{tabular*}
\caption{\label{tab:backgroundsA} Summary of ``real'' missing energy backgrounds and reducible ``fake''
missing energy backgrounds for the calorimetry concept illustrated in Figure~\ref{fig:Schematic} (A),
and described in more detail in the text. The magnitude of these 
dominant backgrounds are given in terms of the muon veto inefficiency $\epsilon_{\mu}$,
pion veto inefficiency $\epsilon_{\pi}$, and neutron veto inefficiency $\epsilon_{n}$ for $10^{16}$
electrons on target (EOT). Active efficient ($\sim 98\%$) tagging of $\pi^0$ and electron-initiated showers 
is required to bring real missing energy backgrounds down to the level of O(1) events per 
$10^{16}$ EOT, and this would additionally require $\epsilon_n\sim10^{-6}$ and $\epsilon_{\mu/\pi}\lesssim10^{-5}$
to control other backgrounds. More realistically, the ultimate background-free luminosity limit of 
this approach is $10^{13}$ EOT.}
\end{center}
\end{table}

The primary backgrounds and corresponding veto inefficiency are summarized in Table \ref{tab:backgroundsA}. 
Hadronic and muon reactions provide ways for photon energy to be missed in any reasonable 
forward calorimeter detector (even without cracks and other dead spots). 
Aggressive charged pion/muon veto inefficiency of $\lesssim 10^{-5}$
and neutron veto inefficiency of $\lesssim 10^{-6}$ is needed to control reducible backgrounds 
for $10^{16}$ EOT, but real missing energy will likely limit the background-free luminosity 
to $10^{14}$ EOT. 
Should charged pion inefficiency enter at the more mild $10^{-4}$ level and energetic neutron 
inefficiency at the $10^{-3}$ level, hard $\gamma p\rightarrow \pi^+n$ reactions 
could dominate the ``missing photon'' background (at the $10^{-13}$ level) 
while two-pronged diffractive hadronic reactions are comparable (at the $10^{-14}$ level). 
In this case, the background-free luminosity cannot realistically exceed 
$10^{13}$ EOT with a calorimetric approach alone,
and the limiting exclusion sensitivity of such an experiment would reach $\epsilon^2$ down to,
\be
\epsilon^2 \lesssim 6.9\times10^{-8} \left(\frac{ m_{A'} }{200 \ \MeV}\right)^2 ~,
\ee
corresponding to $2.3$ events. 
To achieve this level of sensitivity, an experiment running with a beam current close 
to $10^{6} s^{-1}$ ($0.16 pA$) integrated for $10^{7} s$ or potentially 
 $10^{7} s^{-1}$ integrated for $10^{6} s$ would be reasonable. 

This level of sensitivity is really exclusion sensitivity, not discovery potential. 
The basic difficulty is that using calorimetric techniques alone, one cannot {\it in-situ} 
measure the backgrounds using isolated control samples. 
So if excess events with large missing energy-momentum are seen, it would be difficult to justify that 
it is new physics and not one of the above reducible backgrounds entering at a larger than expected rate. 
This fact, and detailed knowledge of the reducible physics backgrounds described above motivate 
a new modified approach described below. 

\section{Near Target Tracking and Calorimeter Approach}\label{sec:forward}

The calorimetry approach is premised on the distinctive kinematics \cite{Bjorken:2009mm} of the \emph{recoiling electron} in dark-matter production, 
which typically carries only a small fraction of the beam energy (Figure~\ref{fig:ptVsY}) for dark matter (or A') mass larger than the electron mass. 
Realistically, however, missing energy due to missed pions from photo-produced $\rho$'s, pion and muon trident reactions, and hard few-pronged hadronic 
 reactions will be difficult to reject below the $10^{-10}$ level without very aggressive veto performance. 

In this section, we outline a scenario that exploits a dedicated thin target foil with near-target tracking of 
charged particles. This approach enhances the physics performance in the following crucial ways:
\begin{itemize} 
\item Reduce and/or reject all neutrino-related backgrounds down the level $10^{-16} N_{e} T$ for $N_e$ electrons on a target of thickness $T$
\item Reject hard brem backgrounds by an additional factor of $100-10^4$ using the recoil electron's transverse momentum and energy characteristics
\item Facilitate rejection of virtual photo-production backgrounds by using tracking and by spatially separating the final-state particles' energy deposition in the calorimeter
\item Using the recoil electron's $p_T$ and energy, one can define control regions for measuring backgrounds \emph{in situ} and discriminating between the kinematics of signals and backgrounds, allowing a compelling demonstration of a signal interpretation for any excess of missing energy events
\end{itemize}
The strategy, summarized by Figure \ref{fig:SchematicFwd}, draws on aspects of Scenario ``C'' in \cite{Bjorken:2009mm}
as well as the scenario described in the previous section.  Instead of directing the beam directly into an active
calorimeter target, the target is a thin ($T\sim 0.01-0.1$) isolated foil of high-Z material embedded  in a tracking
region designed to detect outgoing charged particles. This way, one can specifically identify events with only one
charged particle (identified as an electron downstream) emanating from the target interaction point.  Downstream of the
target/tracker, one could use the same detector configuration as used for the calorimeter approach. 
The tracking elements can be either silicon, scintillator, or drift chamber based (the main considerations are
minimizing material thickness and maintaining low dead-time in a flux of $10^{7}-10^{9} e^-/\s$).  The calorimeter remains the
primary tool for energy measurements, so the tracker could be placed in an analyzing
magnet for momentum measurements or used without a magnet simply to identify charged particles. 
In either case, as with the calorimetry alone approach, the incoming electron will need to be tagged as carrying high momentum.

\subsection{Reduction of Neutrino Backgrounds}

The largest neutrino-related backgrounds identified in Section~\ref{sec:forwardCal} were due to (1) electron straggling (energy loss) followed by CCQE neutrino reactions,
(2) CCQE with a $\pi^0$, and (3) Moller electrons produced before a CCQE reaction. 
In a thin-target configuration, the photon- and $\pi^0$-induced showers from processes (1)  and (2) can be distinguished from recoil electrons by the absence of a charged track and, in (2), a single EM shower.  In addition, (1) and (3) require a sequence of two interactions in the target and therefore are reduced by lower target thicknesses.
In the case of (1), a recoil electron can only be faked if a bremsstrahlung photon above the detection threshold of $50 \MeV$ is produced in a portion of the target preceding the CCQE event, then converts into an $e^+e^-$ pair --- a triple interaction suppressed by a factor of $\sim 1/T^2$ relative to CCQE alone in a target of $T$ radiation lengths (to be compared to the probability $\sim 0.2$ of straggling before CCQE in a thick target).
For (3), the suppression relative to a thick target is simply $1/T$.  
So for a target thickness of $T\lesssim10\%$, we expect these backgrounds to fall to a negligible level for EOT$_{\rm eff}\equiv$EOT$\times T=10^{16}$. 


\subsection{Reduction of Hard Bremsstrahlung and Muon Trident Backgrounds}

The ability to measure the recoil electron's energy and angle emanating from the target allows kinematic separation of the $\apr$ or DM signal from the background. 
The leading backgrounds for the calorimeter approach involve a hard bremsstrahlung photon which is not detected because it interacts with a nucleus, yielding a small number of neutrons, charged pions, and/or muons which can be difficult to detect.  Rejecting hard brem events using the recoil electron's kinematics relaxes the demanding veto inefficiency requirements for the experiment to reach a given sensitivity.  


\begin{figure}[tbp!]
\includegraphics[width=\columnwidth]{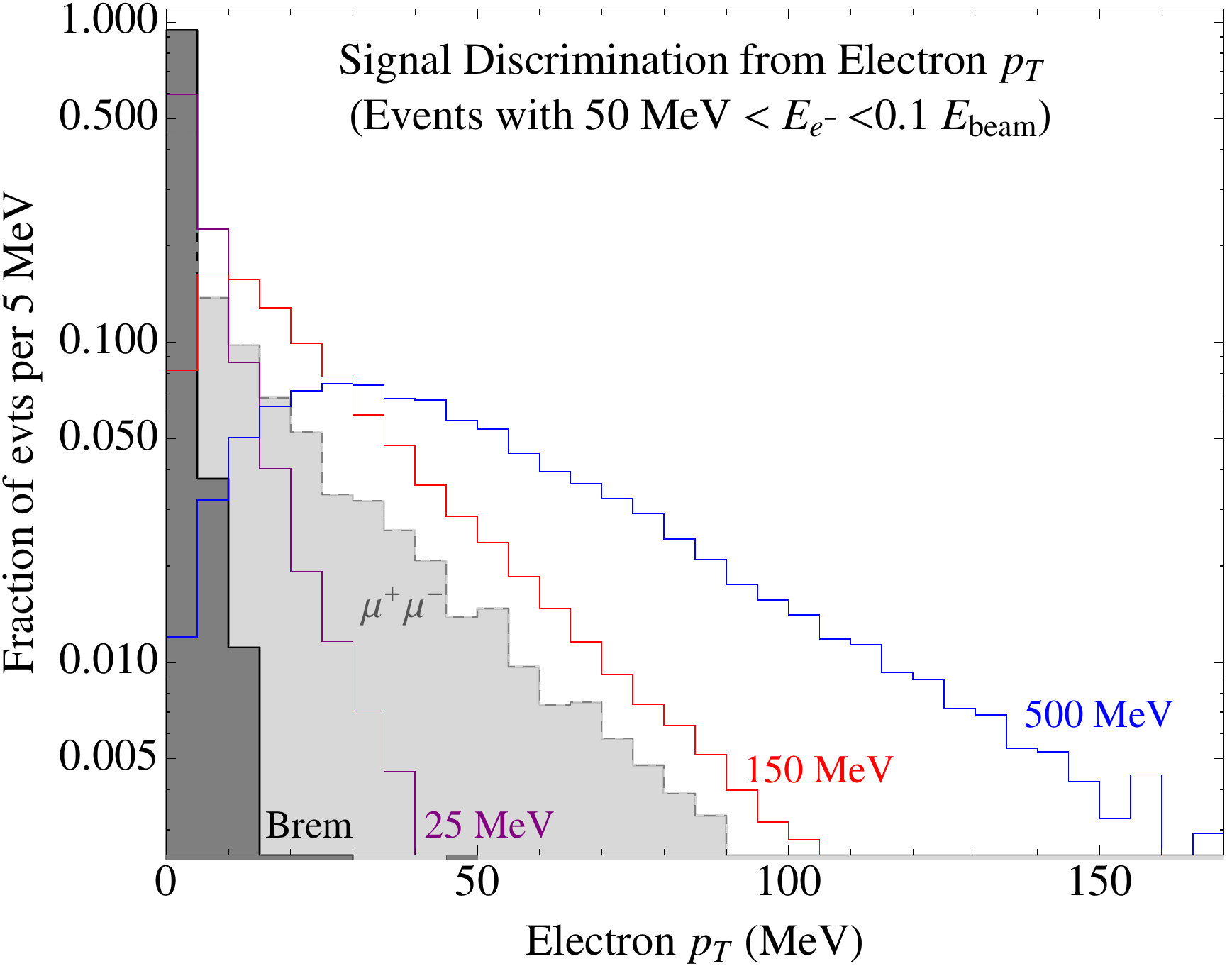}
\caption{\label{fig:ptDist}
Recoil $e^-$ $p_T$ distributions for two backgrounds (filled histograms from left to right: bremsstrahlung and $\mu^+\mu^-$ trident) and for the dark matter signal (unfilled histograms with dark photon mass of 25, 150, and 500 MeV from left to right).  Histograms are normalized to unit area and include only events in the region $50 \MeV < E_{e-} < 0.1 E_{beam}$ and are generated with a beam energy of 10 GeV and a Tungsten target.  This figure demonstrates that electron $p_T$ alone can be used to discriminate a dark matter signal from the dominant bremsstrahlung background; higher-order QED reactions like multiple photon emission and electron pair production have similarly peaked $p_T$ distributions.  Used alone, $p_T$ distributions cannot be used to reject photo-production of massive particle final states, such as $\mu^+\mu^-$ or hadronic final states, from \emph{internal} (virtual) photons.  However, these can still be differentiated from signal by using the recoil electron's energy spectrum (see Figure \ref{fig:ptVsY}).
}
\end{figure}

Two kinematic features can be used to distinguish $A'$ signals from bremsstrahlung: low electron recoil energy (used already in the previous section) and high recoil $p_T$.   Multiple scattering creates an intrinsic $p_T$ resolution $\sigma(p_T) \approx (19.2 \MeV) \times \sqrt{T}$ for the recoiling electron, in a target $T$ radiation lengths thick.   Thus, it is never practical to consider $p_T$ cuts below a few times this $\sigma$.  At $p_T$ larger than a few $\sigma$, the single Coulomb scattering approximation is reasonable (within a factor of 2 or better) and the $p_T$ falls off with the same power-law as the intrinsic angular distribution of bremsstrahlung.  Thus the fraction of bremsstrahlung recoil electrons that carry low energy \emph{and} have a sufficiently large $p_T$ is given approximately by
\be
P_{low-E,wide } \sim T (\Delta y) \pf{m_e}{\bar p_T^{min}}^2 \left[ 1 + \frac{\pi T}{2 \alpha \ln(184 Z^{-1/3})}\right].
\ee
The first term in brackets is due to the intrinsic bremsstrahlung angular distribution and the second due to wide-angle scattering of the recoiling electron, which dominates for targets thicker than about $2\% X_0$.

\begin{figure}[t]  
\includegraphics[width=\columnwidth]{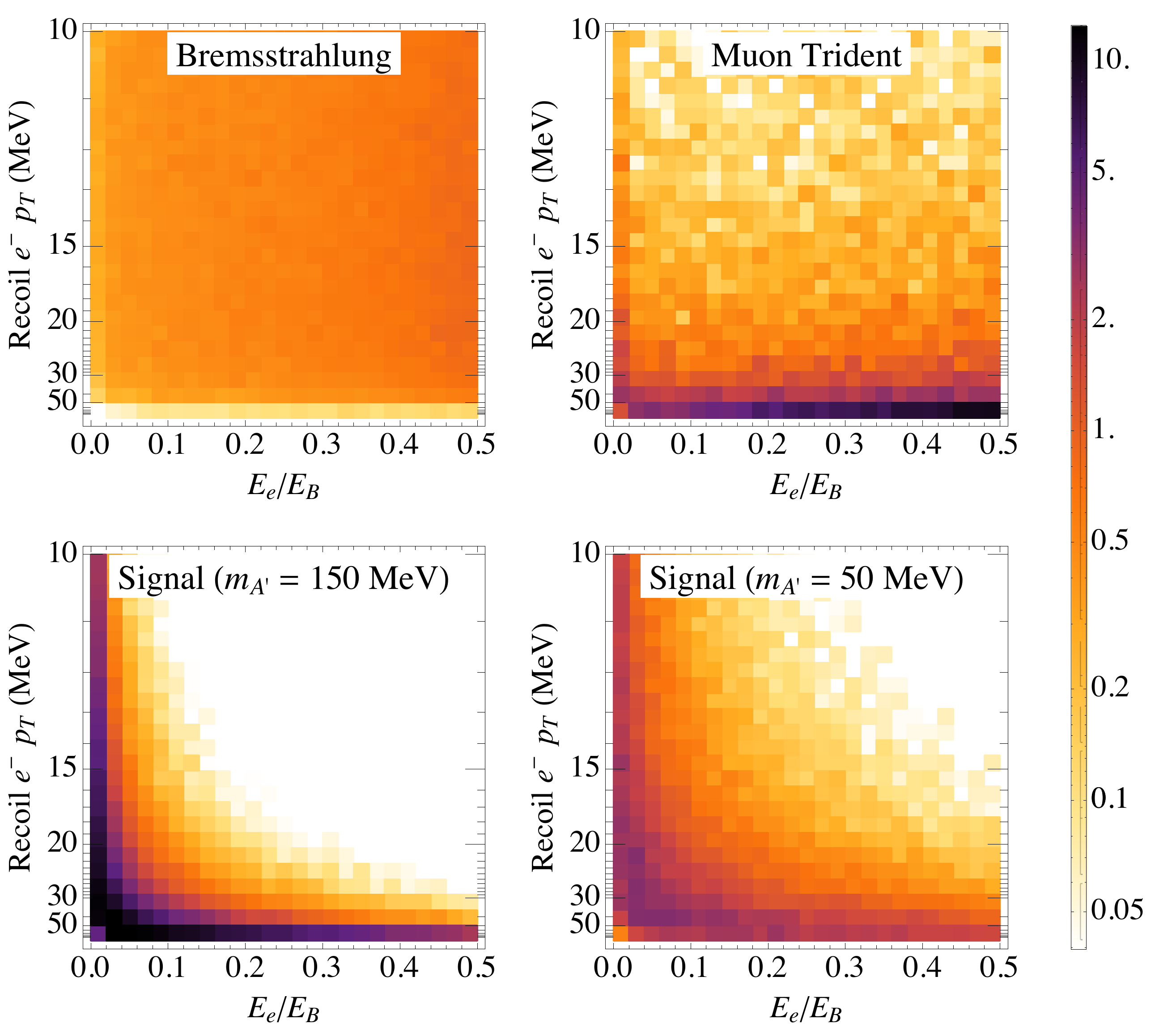}
\caption{\label{fig:ptVsY}
The four panels show the density of events in recoil electron $p_T$ and energy fraction $y_e \equiv E_{e^-}/E_{B}$, for bremsstrahlung and muon trident backgrounds (top) and dark matter signal with two representative $A'$ masses (bottom).  The $y$-axis is scaled as $1/p_T^2$, so that the bremsstrahlung distribution is very nearly uniform for $p_T \gg m_e$ and $y \ll 1$.  The color denotes the density of events in a given bin, relative to a uniform density (dark colors denote high event densities and light colors denote under-densities).  The dark matter signal events accumulate at relatively high $p_T$ and low recoil energy (with the scale of each dependent on the $A'$ mass), while bremsstrahlung is roughly uniform.  
Muon trident events are also clustered at large  $p_T$, but differ considerably from the signal in their recoil-energy distribution.  Other heavy-particle-production processes initiated by a virtual photon are expected  to share the qualitative features of the muon trident process.}
\end{figure}

By requiring a recoil $e^-$ with $p_T>20 \ (50) \MeV$, hard brem can be rejected by a factor of $10^{-3\,(-4)}$ for thin targets (see Figure \ref{fig:ptDist}).  In a $0.1 X_0$ target, multiple scattering dominates over the intrinsic bremsstrahlung $p_T$ spectrum reducing the rejection power by a factor of 5-10.  
These cuts have reasonable signal efficiency for $m_{A'} \gtrsim p_{T,min}$.  A more sophisticated cut, using correlations between electron $p_T$ and energy (see Figure \ref{fig:ptVsY}), should achieve comparable background rejection with much higher signal acceptance at slightly lower $A'$ masses but we have not attempted to optimize this.   

To a very good approximation, the rejection power of the electron $p_T$ cut and final-state veto are uncorrelated (events surviving the $p_T$ cuts considered above will have a typical photon angle relative to the beamline of only $p_{T,min}/E_0 \approx 2-5\cdot 10^{-3}$, well below the angular scale of the apparatus).  Thus, the factor of $10^2-10^4$ rejection obtained by kinematics is \emph{in addition to} that obtained from the veto of energetic final-state particles.

\subsection{Reduction of Virtual Photo-Production $\mu^+\mu^-$ and Hadronic Backgrounds}
In Section \ref{sec:forwardCal}, we argued that trident reactions $eN\rightarrow eN + \mu^+\mu^-$ and virtual-photon-induced reactions
$eN\rightarrow eN + X$ with missed pions, muons, and/or neutrons were suppressed relative to photoproduction by the effective photon flux
$\frac{\alpha}{2\pi} \log(\mu/m_e) \ll 1$.  However, with hard brem dramatically reduced by kinematic cuts these backgrounds can no longer be neglected.
To a good approximation, both types of reactions can be well approximated using the equivalent photon approximation: 
\be
\sigma(eN\! \rightarrow \! eX) \! = \!\! \int_{0}^{1} \!\! dz \! \int \! \frac{dp_T^2}{p_T^2} \! \frac{\alpha}{2\pi}\frac{1+(1-z)^2}{z}\sigma{(\gamma N\rightarrow X)},~~
\ee
where $z$ is the energy fraction of the photon, and $p_T$ is the transverse momentum given to the
photon. For a real electron, the photon virtuality is $q^2=p_T^2/(1-z)$ and the momentum integral
should be taken over the appropriate range for a given set of kinematics.  This appproximation breaks down for photon virtualities large enough to alter the effective $\gamma N$ cross-section --- $q^2 \approx m_\mu^2$ for muon tridents and slightly larger for most QCD processes (the scale we denoted by $\mu$ above).  So the distribution of electron recoil $p_T$'s is only logarithmically enhanced at low $p_T$.  In other words, with a $p_{T,min}$ cut the cross-section for a  given virtual-photon-induced background is suppressed by $\frac{\alpha}{2\pi} \log(\mu (1-z)^{1/2}/p_{T,min}) \sim \rm{few}\cdot 10^{-3}$, and can dominate over real-photon-induced reactions surviving the $p_T$ cut.  For the case of muon trident events, we can see the lack of low-$p_T$ enhancement in a full QED Monte Carlo in Figure~\ref{fig:ptDist}.

To be of interest to us, however, such reactions must occur inside the target.  
The near-target tracking provides an additional handle to further reduce these backgrounds. 
For example, trident reactions $eN\rightarrow eN + \mu^+\mu^-$ and diffractive $eN\rightarrow eN \rho \rightarrow \pi^+\pi^-$ will have three charged particle tracks emanating from the target, rather than the single charged track expected in the signal.  
Even with a tracking inefficiency of $10\%$, such three-pronged events can be rejected at the level of factor of $100$ level.
Thus, these virtual photo-production trident reactions should be down by $\sim 3\times 10^{-5}$ 
compared to their real-brem-induced counterparts (before kinematic rejection), and remain subdominant even after a hard $p_T$ cut.

Similar rejections apply to the $K^0_L K^0_S$ background (two extra tracks) and the $\pi^+ n$ background in its dominant kinematics, with a hard forward $\pi^0$ (one extra track).  The only virtual-photon-induced backgrounds we have identified for which this rejection is not relevant are the $u$-channel $n\pi^+$ process (where the pion goes backwards) and $2n\,\bar n$.  However, these two processes were already fairly low in rate, so a reasonable veto efficiency suffices to bring them to a negligible rate.  Accounting for the $3\cdot 10^{-3}$ lower inclusive rate for virtual-photon-induced reactions over real photons bring these backgrounds down to the level of $\sim 3\cdot 10^{3} \epsilon_n$ and $\sim 3\cdot 10^5 \epsilon_n^3$ events respectively per $10^{16}$ EOT$_{\rm eff}$, where $\epsilon_n$ is the neutron detection inefficiency.  

\subsection{In-Situ Measurement of Residual Backgrounds}
We have already highlighted the importance of recoil electron $p_T$ and energy as discriminating variables to
\emph{reject} background.  They can also be used to \emph{measure} background rates associated with instrumental
inefficiencies for detecting real- or virtual-photon-initiated reactions.  
Figure \ref{fig:ptVsY} recasts the energy and $p_T$ variables in terms of the electron's energy
fraction $y_e = E_e/E_B$ and $1/p_T^2$.  In these variables, the bremsstrahlung background distribution is nearly
uniform (for $y_e$ far from 1), while signals are peaked in the lower left corner (higher $p_T$ and lower $y_e$).  
The distribution for muon trident events is also shown, and is dominated at relatively high $p_T$ but high $y_e$
relative to the signal.  Virtual-photon-induced hadronic reactions are expected to have the same qualitative distribution as
muon tridents.

This kinematic separation of the signal from backgrounds allows the use of kinematically selected control regions in
which to measure or bound \emph{in situ} the rates for failing to detect these two types of background reactions: a low-$p_T$ and
moderate $y_e$ control region for bremsstrahlung, and a high-$p_T$ moderate $y_e$ control region for
virtual-photon-induced processes like muon tridents.  For the high $A'$ mass hypotheses, very clean background-enriched
samples can be produced.  

Of course, for  many backgrounds  we also envision many independent factors being used in the rejection (e.g. tracking and
calorimetry).  Inverting these cuts one at a time can also be used to determine the total rejection for a specific
exclusive process often with better statistical uncertainty than the kinematic control regions alone.  But the virtue of the
kinematic control regions is their inclusiveness -- their effectiveness relies only on the dominance of electromagnetic
interactions for electrons, the non-zero mass of the $A'$ (for separation of signal from real-photon backgrounds), and the
difference in masses between electrons and muon/hadrons (for separation of virtual-photon backgrounds)!  
Therefore, kinematic separation allows reliable estimates or bounds on the background \emph{even from final states whose importance
  has not been anticipated}. 

Put another way, if a reasonably large excess of $O(10)$ events or more is observed, these
kinematic handles can be used to credibly identify it as a new-physics signal or as probable background.  This is an
important handle for an experiment to have real discovery potential.  

\subsection{Performance and Sensitivity Summary}

\begin{table}[t] 
\begin{center}
\begin{tabular*}{0.5\textwidth}{@{\extracolsep{\fill}}lcl}
\hline\hline
\\[-7pt]
$\quad$  {\bf Real Missing Energy}  &  & {\bf Magnitude ($10^{16}$ EOT$_{eff}$)} \\[2pt]
\hline
\\[-6pt]
Brem+CCQE  && $<1$ ($T\lesssim 0.1$)  \\[2pt]
CCQE+$\pi^0$ && $<1$ ($T\lesssim 0.1$) \\[2pt]
Moller+CCQE   &&  $\ll 1$ ($T\lesssim 0.1$) \\[2pt]
$eN\rightarrow eN\nu\bar{\nu}$ &&  $\sim10^{-2}$  \\[2pt]
             	\\[-6pt]
\hline \hline
\\[-6pt]
$\quad$  {\bf Reducible Backgrounds}  &  & {\bf Fake Rate/$10^{14}$ EOT$_{eff}$} \\[2pt] 
\hline
\\[-6pt]
 $\gamma$ non-interaction  && $\sim3\times 10^{8}e^{-\tfrac{7}{9}(T/X_0=45)}\ll 1$   \\[2pt]
 $\gamma p\rightarrow \pi^+n$ && $\sim10^{2}\times \epsilon_{\pi}\epsilon_{n}$	\\
  $\gamma^* p\rightarrow \pi^+n$ (backscatter $\pi^+$) && $\sim3\times 10^{1}\times \epsilon_{n}$	 (see text)\\
 $\gamma N\rightarrow (\rho,\omega,\phi)N\rightarrow \pi^+\pi^-N$ & &  $\sim2\times10^{4}\epsilon_{\pi}^2$	\\
 $\gamma^* n\rightarrow n\bar{n}n$ && $\sim3\times 10^{3}\times \epsilon_n^3$	\\
  $e N\rightarrow e N(\mu^+\mu^-,\pi^+\pi^-)$ && $\sim 10^4 \times \epsilon_{\mu/\pi}^2$	\\
 $\gamma N\rightarrow N\mu^+\mu^-$ && $\sim 6\times 10^3 \times \epsilon_{\mu}^2$	\\[2pt]
 \hline
 	\end{tabular*}
\caption{\label{tab:backgroundsB} Summary of ``real'' missing energy backgrounds and reducible ``fake''
missing energy backgrounds for the near-target tracking and calorimetry concept illustrated in Figure~\ref{fig:Schematic} (B),
and described in more detail in the text. 
In practice, $T\approx 0.1$ (with a $p_{T}>20 \ \MeV$ selection) is sufficient to control CCQE backgrounds
for $10^{16}$ EOT$_{\rm eff}$.
For a thin $T\sim 0.01$ target with $p_{T}>50 \ \MeV$ selection, real photon backgrounds can be kinematically reduced 
by $10^4$, in which case readily attainable $\epsilon_n\sim10^{-2}$ and $\epsilon_{\mu/\pi}\sim10^{-3}$
are sufficient to control fake ``missing'' photon backgrounds for $10^{16}\times 0.01=10^{14}$ EOT$_{\rm eff}$.
Going to a thicker target $T=0.1$ reduces the effectiveness of the $p_T$ selection 
down to $\sim200$ rejection of real photon backgrounds, and requires 
a corresponding improvement for the veto inefficiencies.}
\end{center}
\end{table}

%
%

The near-target tracking layout offers several advantages over a target-calorimeter 
based approach,  which together improve its overall sensitivity reach as a function 
of veto performance, as well as enhancing the potential for a credible discovery.

Figure \ref{fig:moneyA} summarizes the sensitivity reach for several benchmark cases. 
The red curves in Fig. \ref{fig:moneyA} depict expected $90\%$ exclusion regions for various realizations of the near target tracking scenario (Scenario B). 
The solid  curve labeled I assumes $10^{13}$ EOT$_{\rm eff}$ and target thickness of $T = 0.01 X_{0}$, while
the  dashed red curve  labeled II  assumes $10^{15}$ EOT$_{\rm eff}$ and $T = 0.1 X_{0}$. Both solid and dashed lines compute signal yield 
requiring either ($P_T(e) > 20 \ \MeV $ and $50\ \MeV < E_e < 0.1 E_{beam}$) with 2.3 event sensitivity for a $90\%$ exclusion  or 
requiring just ($50 \ \MeV < E_e < 0.1 E_{beam}$) with 35 event sensitivity for a $90\%$ exclusion; whichever yields a smaller $\epsilon$ for a given value of $m_\apr$.  This corresponds to a scenario with a  total of $\sim 300$ background events, dominated by real-photon conversions.  For high $A'$ masses an effective search strategy is to cut away from these events using recoil electron $p_T$; for lower $A'$ masses, it is more efficient to measure the backgrounds in a control region and statistically subtract them. 
The dotted red line labeled III represents the ultimate limit of this experimental program and assumes $3 \times10^{15}$ EOT$_{eff}$ incident on a $T = 0.1 X_0$ target, assuming zero backgroun in the range ($50 \ \MeV < E_e < 0.1 E_{beam}$) for a 90\% sensitivity limit of 2.3 signal events. We also show our estimated $90\%$ exclusion sensitivity for an SPS configuration ($30 \ \GeV$ beam energy on Tungsten) with $10^9$ and $10^{12}$ EOT$_{\rm eff}$ \footnote{Our signal yield estimate for the SPS set-up at $90\%$ C.L. exclusion is $\approx 20-30$ times lower than what is inferred from Fig. 19 of  Ref.~\cite{Andreas:2013lya}. The difference is due to including full Monte Carlo simulation of the form factor suppression as a function of $A'$ mass, including $\mathcal{O}(50-70\%)$ efficiency for a signal event to have electron recoil energy below $0.1E_{\rm beam}$ (even smaller than $50\%$ for $m_{A'}\lesssim 10 \ \MeV$), and a $30\%$ efficiency that we apply to the signal to account for straggling effects --- there's a large probability that multiple (forward) Bremsstrahlung will reduce the incident electron energy to $< 0.9 E_{\rm beam}$ before the hard scatter that produces DM.  In that case, one would see a shower above $0.1E_{\rm beam}$ even if the $e^-$ recoil energy is low, and veto the event. These factors combined account for the factor of $20-30$ discrepancy in yields and corresponding sensitivity to $\epsilon$.}.

To see how background yields are reduced by the factors discussed above, 
we consider a benchmark neutron veto inefficiency of $\epsilon_n\sim10^{-3}$ and 
muon/pion inefficiency of $\epsilon_{\mu/\pi}\lesssim10^{-3}$. In this case, 
referring to Table~\ref{tab:backgroundsA}, the target-calorimetry approach 
would be background limited at the level of $N_e\times T=10^{12}$ EOT$_{\rm eff}$,
while the near-target tracking could reach $10^{14}$ EOT$_{\rm eff}$.
Signal production is reduced by the thinner $T$ and $p_{T}$ selection, but that is partially compensated 
for by the complete reduction of straggling losses yielding an overall $\sim 70$ reduction 
in signal yields when compared to $T=1$ (at $m_{A'}\gsim 50 \ \MeV$ for example), so that signal over background can be improved by 
$\gsim 100$ in this case. This leads to a $\sim 100$ improvement in background limited $\epsilon^2$ sensitivity.

In practice, a larger $T\approx 0.1$ is sufficient to control CCQE backgrounds, and so with better 
veto inefficiency of $\epsilon_n\sim10^{-5}$ and $\epsilon_{\mu/\pi}\lesssim 10^{-4}$
one could reach $3\times 10^{15}$ EOT$_{\rm eff}$ background limited luminosity. 
In this case, signal yields are only reduced by a factor of 
$\sim 3$ (only mild straggling reduction for $T=0.1$) relative to $T=1$ for $m_{A'}\gsim 50 \ \MeV$, 
and the limiting exclusion sensitivity of such an experiment would reach $\epsilon^2$ down to,
\be
\epsilon^2 \lesssim 4.2\times10^{-10}\left( \frac{m_{A'} }{   200 \ \MeV} \right)^2
\ee
corresponding to $2.3$ signal events. 

In the above, we have not considered the effects of finite angular acceptance. These merit a more careful study but should be relatively small for $\sim 30^\circ$ forward acceptance.  For most of the incident electron's energy to be carried by wide-angle final states, there must be \emph{two} high-$p_T$ objects --- either a high-mass pair or a substantially recoiling nucleus.  In the first case, to escape a detector with angular size $\theta$ a pair must have invariant mass exceeding $E_0\sin\theta \sim 5\ \GeV$ for beam energy $E_0 \sim 10$ GeV and $\theta\sim 30^{\circ}$.  This is above the kinematic threshold for incoherent scattering off a nucleon, so can only occur through the exponentially suppressed tail of the coherent scattering form factor (for a 2-parameter Fermi model of Tungsten, for example, the form-factor suppression alone is $10^{-12}$).  Wide-angle coherent scattering of a single high-energy particle requires even higher momentum transfer ($q \sim E \sin\theta$ rather than $E \sin^2\theta/2$), and so is even more exponentially suppressed, while inelastic wide-angle scattering will typically produce a substantial forward hadronic shower.

To reach the full potential of a fixed-target missing momentum approach, 
it appears essential to use some form of near-target charged particle tracking and basic kinematic handles of the recoil electron, as we've described. 
In this case it may be possible to robustly cover essentially all of the theoretically interesting range 
of $\alpha_D\epsilon^2$ (over the MeV--GeV mass range) in Eq.~\eqref{eq:goal}.

\section{Discussion and Conclusion}\label{sec:conclusion}

In this paper, we evaluated the irreducible physics limitations for fixed-target DM searches using missing energy-momentum.  
For the pure calorimetric concept as outlined in Section~\ref{sec:forwardCal} (Scenario A in Figure~\ref{fig:Schematic}), 
based on the SPS proposal in \cite{Andreas:2013lya}, the realistic limitation will likely be detection inefficiencies that allow 
rare photo-production reactions to mimic the missing energy-momentum signature.
This assumes that beam impurity backgrounds can be kept small (perhaps more realistic for JLab CW beam than SPS secondary beam). 
Nonetheless, with aggressive veto inefficiency performance, the ultimate reach of this approach may only be limited 
by irreducible neutrino (quasi-elastic charged current) backgrounds and neutrino trident production. 

To enhance the sensitivity of this approach, we advocate the use of a separated (high-Z) thin target 
with near-target tracking and calorimetry (Scenario B in Figure~\ref{fig:Schematic}) to measure the kinematics of the recoiling electron. 
The thin target reduces straggling and charged-current neutrino reaction backgrounds. Spatially
 separating the target from the calorimeter and adding tracking allows clean identification of exclusive single charged particle 
 final states and corresponding reduction of virtual photon backgrounds. 
 Additionally, the energy-momentum measurement of the recoil electron can be used for signal discrimination, 
 reduction of backgrounds associated with hard bremsstrahlung followed by rare real photon reactions,
 and to measure residual backgrounds \emph{in situ} with well-defined data-driven control regions.   
In this case, the eventual sensitivity is sufficient to cover the entire range of vector portal dark matter couplings $\alpha_D$
and $\epsilon$ required for an acceptable relic density in the regime $m_{A'}>m_{\chi}$ (see Sec.~\ref{sec:model}). 
With this level of sensitivity, the discovery potential is excellent, and a null result would be decisive. 

One powerful advantage of the approach introduced in Sec.~\ref{sec:forward} is that it is amenable to data-driven estimates of the backgrounds by defining suitable low electron recoil $p_T$ control regions.  In the case of an excess of events in the signal region, 
this approach then lends itself to a more robust discovery and characterization of a new physics signal then is available through the use of 
a pure calorimetry (Sec.~\ref{sec:forwardCal}) approach. 

The proposal to use the SPS secondary electron beam \cite{Andreas:2013lya} with a missing energy calorimetry approach (scenario A) is 
an excellent first step in a physics program to cover the GeV-scale vector portal dark matter parameter space. 
In practice, the detector layout described in that proposal could be augmented in a second stage with the kind of near-target tracking setup advocated here. 
Eventually, the higher intensity beams available at JLab would need to be exploited to achieve the full 
science potential of this approach -- it is both interesting and valuable to investigate this possibility in more detail.

\section*{Acknowledgements}
We thank Marco Battaglieri, Jay Benesch, Arne Freyberger, Sergei Gninenko, Yonatan Kahn,  Maxim Pospelov, Stepan Stepanayan, and Jesse Thaler for helpful conversations. We also thank 
the University of Victoria workshop on Dark Forces, during which this work was initiated. 
This research was supported in part by Perimeter Institute for Theoretical Physics. Research at Perimeter Institute is supported by the Government of Canada through Industry Canada and by the Province of Ontario through the Ministry of Research and Innovation. NT and EI are partially supported by the Ministry of Research and Innovation - ERA (Early Research Awards) program.

\bibliographystyle{apsrevM}
\bibliography{epsilon2}

\ifx\mcitethebibliography\mciteundefinedmacro
\PackageError{apsrevM.bst}{mciteplus.sty has not been loaded}
{This bibstyle requires the use of the mciteplus package.}\fi
\begin{mcitethebibliography}{115}
\expandafter\ifx\csname natexlab\endcsname\relax\def\natexlab#1{#1}\fi
\expandafter\ifx\csname bibnamefont\endcsname\relax
  \def\bibnamefont#1{#1}\fi
\expandafter\ifx\csname bibfnamefont\endcsname\relax
  \def\bibfnamefont#1{#1}\fi
\expandafter\ifx\csname citenamefont\endcsname\relax
  \def\citenamefont#1{#1}\fi
\expandafter\ifx\csname url\endcsname\relax
  \def\url#1{\texttt{#1}}\fi
\expandafter\ifx\csname urlprefix\endcsname\relax\def\urlprefix{URL }\fi
\providecommand{\bibinfo}[2]{#2}
\providecommand{\eprint}[2][]{\url{#2}}

\bibitem[{\citenamefont{Akerib et~al.}(2014)}]{Akerib:2013tjd}
\bibinfo{author}{\bibfnamefont{D.}~\bibnamefont{Akerib}} \bibnamefont{et~al.}
  (\bibinfo{collaboration}{LUX Collaboration}),
  \bibinfo{journal}{Phys.Rev.Lett.} \textbf{\bibinfo{volume}{112}},
  \bibinfo{pages}{091303} (\bibinfo{year}{2014}), \eprint{1310.8214}\relax
\mciteBstWouldAddEndPuncttrue
\mciteSetBstMidEndSepPunct{\mcitedefaultmidpunct}
{\mcitedefaultendpunct}{\mcitedefaultseppunct}\relax
\EndOfBibitem
\bibitem[{\citenamefont{Aalseth et~al.}(2013)}]{Aalseth:2012if}
\bibinfo{author}{\bibfnamefont{C.}~\bibnamefont{Aalseth}} \bibnamefont{et~al.}
  (\bibinfo{collaboration}{CoGeNT Collaboration}), \bibinfo{journal}{Physical
  Review D 88,} \textbf{\bibinfo{volume}{012002}} (\bibinfo{year}{2013}),
  \eprint{1208.5737}\relax
\mciteBstWouldAddEndPuncttrue
\mciteSetBstMidEndSepPunct{\mcitedefaultmidpunct}
{\mcitedefaultendpunct}{\mcitedefaultseppunct}\relax
\EndOfBibitem
\bibitem[{\citenamefont{Aprile et~al.}(2013)}]{Aprile:2013doa}
\bibinfo{author}{\bibfnamefont{E.}~\bibnamefont{Aprile}} \bibnamefont{et~al.}
  (\bibinfo{collaboration}{XENON100 Collaboration}) (\bibinfo{year}{2013}),
  \eprint{1301.6620}\relax
\mciteBstWouldAddEndPuncttrue
\mciteSetBstMidEndSepPunct{\mcitedefaultmidpunct}
{\mcitedefaultendpunct}{\mcitedefaultseppunct}\relax
\EndOfBibitem
\bibitem[{Fermi-LAT Collaboration()}]{Fermi-LAT:2013uma}
Fermi-LAT Collaboration (\bibinfo{year}{2013}), \eprint{1305.5597}\relax
\mciteBstWouldAddEndPuncttrue
\mciteSetBstMidEndSepPunct{\mcitedefaultmidpunct}
{\mcitedefaultendpunct}{\mcitedefaultseppunct}\relax
\EndOfBibitem
\bibitem[{\citenamefont{Bernabei et~al.}(2010)}]{DAMA}
\bibinfo{author}{\bibfnamefont{R.}~\bibnamefont{Bernabei}} \bibnamefont{et~al.}
  (\bibinfo{collaboration}{DAMA Collaboration, LIBRA Collaboration}),
  \bibinfo{journal}{Eur.Phys.J.} \textbf{\bibinfo{volume}{C67}},
  \bibinfo{pages}{39} (\bibinfo{year}{2010}), \eprint{1002.1028}\relax
\mciteBstWouldAddEndPuncttrue
\mciteSetBstMidEndSepPunct{\mcitedefaultmidpunct}
{\mcitedefaultendpunct}{\mcitedefaultseppunct}\relax
\EndOfBibitem
\bibitem[{\citenamefont{Chatrchyan et~al.}(2012)}]{Chatrchyan:2012me}
\bibinfo{author}{\bibfnamefont{S.}~\bibnamefont{Chatrchyan}}
  \bibnamefont{et~al.} (\bibinfo{collaboration}{CMS Collaboration}),
  \bibinfo{journal}{JHEP} \textbf{\bibinfo{volume}{1209}}, \bibinfo{pages}{094}
  (\bibinfo{year}{2012}), \eprint{1206.5663}\relax
\mciteBstWouldAddEndPuncttrue
\mciteSetBstMidEndSepPunct{\mcitedefaultmidpunct}
{\mcitedefaultendpunct}{\mcitedefaultseppunct}\relax
\EndOfBibitem
\bibitem[{ATLAS-CONF-2012-147()}]{ATLAS:2012zim}
ATLAS-CONF-2012-147 (\bibinfo{year}{2012})\relax
\mciteBstWouldAddEndPuncttrue
\mciteSetBstMidEndSepPunct{\mcitedefaultmidpunct}
{\mcitedefaultendpunct}{\mcitedefaultseppunct}\relax
\EndOfBibitem
\bibitem[{\citenamefont{Asztalos et~al.}(2001)}]{Asztalos:2001tf}
\bibinfo{author}{\bibfnamefont{S.~J.} \bibnamefont{Asztalos}}
  \bibnamefont{et~al.} (\bibinfo{collaboration}{ADMX Collaboration}),
  \bibinfo{journal}{Phys.Rev.} \textbf{\bibinfo{volume}{D64}},
  \bibinfo{pages}{092003} (\bibinfo{year}{2001})\relax
\mciteBstWouldAddEndPuncttrue
\mciteSetBstMidEndSepPunct{\mcitedefaultmidpunct}
{\mcitedefaultendpunct}{\mcitedefaultseppunct}\relax
\EndOfBibitem
\bibitem[{\citenamefont{Bähre et~al.}(2013)\citenamefont{Bähre, Döbrich,
  Dreyling-Eschweiler, Ghazaryan, Hodajerdi et~al.}}]{Bahre:2013ywa}
\bibinfo{author}{\bibfnamefont{R.}~\bibnamefont{Bähre}},
  \bibinfo{author}{\bibfnamefont{B.}~\bibnamefont{Döbrich}},
  \bibinfo{author}{\bibfnamefont{J.}~\bibnamefont{Dreyling-Eschweiler}},
  \bibinfo{author}{\bibfnamefont{S.}~\bibnamefont{Ghazaryan}},
  \bibinfo{author}{\bibfnamefont{R.}~\bibnamefont{Hodajerdi}},
  \bibnamefont{et~al.}, \bibinfo{journal}{JINST} \textbf{\bibinfo{volume}{8}},
  \bibinfo{pages}{T09001} (\bibinfo{year}{2013}), \eprint{1302.5647}\relax
\mciteBstWouldAddEndPuncttrue
\mciteSetBstMidEndSepPunct{\mcitedefaultmidpunct}
{\mcitedefaultendpunct}{\mcitedefaultseppunct}\relax
\EndOfBibitem
\bibitem[{\citenamefont{Irastorza et~al.}(2011)\citenamefont{Irastorza,
  Avignone, Caspi, Carmona, Dafni et~al.}}]{Irastorza:2011gs}
\bibinfo{author}{\bibfnamefont{I.}~\bibnamefont{Irastorza}},
  \bibinfo{author}{\bibfnamefont{F.}~\bibnamefont{Avignone}},
  \bibinfo{author}{\bibfnamefont{S.}~\bibnamefont{Caspi}},
  \bibinfo{author}{\bibfnamefont{J.}~\bibnamefont{Carmona}},
  \bibinfo{author}{\bibfnamefont{T.}~\bibnamefont{Dafni}},
  \bibnamefont{et~al.}, \bibinfo{journal}{JCAP}
  \textbf{\bibinfo{volume}{1106}}, \bibinfo{pages}{013} (\bibinfo{year}{2011}),
  \eprint{1103.5334}\relax
\mciteBstWouldAddEndPuncttrue
\mciteSetBstMidEndSepPunct{\mcitedefaultmidpunct}
{\mcitedefaultendpunct}{\mcitedefaultseppunct}\relax
\EndOfBibitem
\bibitem[{\citenamefont{Vogel et~al.}(2013)\citenamefont{Vogel, Avignone,
  Cantatore, Carmona, Caspi et~al.}}]{Vogel:2013bta}
\bibinfo{author}{\bibfnamefont{J.}~\bibnamefont{Vogel}},
  \bibinfo{author}{\bibfnamefont{F.}~\bibnamefont{Avignone}},
  \bibinfo{author}{\bibfnamefont{G.}~\bibnamefont{Cantatore}},
  \bibinfo{author}{\bibfnamefont{J.}~\bibnamefont{Carmona}},
  \bibinfo{author}{\bibfnamefont{S.}~\bibnamefont{Caspi}}, \bibnamefont{et~al.}
  (\bibinfo{year}{2013}), \eprint{1302.3273}\relax
\mciteBstWouldAddEndPuncttrue
\mciteSetBstMidEndSepPunct{\mcitedefaultmidpunct}
{\mcitedefaultendpunct}{\mcitedefaultseppunct}\relax
\EndOfBibitem
\bibitem[{\citenamefont{Irastorza et~al.}(2012)}]{Irastorza:2012qf}
\bibinfo{author}{\bibfnamefont{I.}~\bibnamefont{Irastorza}}
  \bibnamefont{et~al.} (\bibinfo{collaboration}{IAXO Collaboration})
  (\bibinfo{year}{2012}), \eprint{1201.3849}\relax
\mciteBstWouldAddEndPuncttrue
\mciteSetBstMidEndSepPunct{\mcitedefaultmidpunct}
{\mcitedefaultendpunct}{\mcitedefaultseppunct}\relax
\EndOfBibitem
\bibitem[{\citenamefont{Horns et~al.}(2013)\citenamefont{Horns, Jaeckel,
  Lindner, Lobanov, Redondo et~al.}}]{Horns:2012jf}
\bibinfo{author}{\bibfnamefont{D.}~\bibnamefont{Horns}},
  \bibinfo{author}{\bibfnamefont{J.}~\bibnamefont{Jaeckel}},
  \bibinfo{author}{\bibfnamefont{A.}~\bibnamefont{Lindner}},
  \bibinfo{author}{\bibfnamefont{A.}~\bibnamefont{Lobanov}},
  \bibinfo{author}{\bibfnamefont{J.}~\bibnamefont{Redondo}},
  \bibnamefont{et~al.}, \bibinfo{journal}{JCAP}
  \textbf{\bibinfo{volume}{1304}}, \bibinfo{pages}{016} (\bibinfo{year}{2013}),
  \eprint{1212.2970}\relax
\mciteBstWouldAddEndPuncttrue
\mciteSetBstMidEndSepPunct{\mcitedefaultmidpunct}
{\mcitedefaultendpunct}{\mcitedefaultseppunct}\relax
\EndOfBibitem
\bibitem[{\citenamefont{Stadnik and
  Flambaum}(2014{\natexlab{a}})}]{Stadnik:2014ala}
\bibinfo{author}{\bibfnamefont{Y.~V.} \bibnamefont{Stadnik}} \bibnamefont{and}
  \bibinfo{author}{\bibfnamefont{V.~V.} \bibnamefont{Flambaum}}
  (\bibinfo{year}{2014}{\natexlab{a}}), \eprint{1409.2986}\relax
\mciteBstWouldAddEndPuncttrue
\mciteSetBstMidEndSepPunct{\mcitedefaultmidpunct}
{\mcitedefaultendpunct}{\mcitedefaultseppunct}\relax
\EndOfBibitem
\bibitem[{\citenamefont{Stadnik and
  Flambaum}(2014{\natexlab{b}})}]{Stadnik:2013raa}
\bibinfo{author}{\bibfnamefont{Y.}~\bibnamefont{Stadnik}} \bibnamefont{and}
  \bibinfo{author}{\bibfnamefont{V.}~\bibnamefont{Flambaum}},
  \bibinfo{journal}{Phys.Rev.} \textbf{\bibinfo{volume}{D89}},
  \bibinfo{pages}{043522} (\bibinfo{year}{2014}{\natexlab{b}}),
  \eprint{1312.6667}\relax
\mciteBstWouldAddEndPuncttrue
\mciteSetBstMidEndSepPunct{\mcitedefaultmidpunct}
{\mcitedefaultendpunct}{\mcitedefaultseppunct}\relax
\EndOfBibitem
\bibitem[{\citenamefont{Stadnik and
  Flambaum}(2014{\natexlab{c}})}]{Stadnik:2014xja}
\bibinfo{author}{\bibfnamefont{Y.}~\bibnamefont{Stadnik}} \bibnamefont{and}
  \bibinfo{author}{\bibfnamefont{V.}~\bibnamefont{Flambaum}}
  (\bibinfo{year}{2014}{\natexlab{c}}), \eprint{1408.2184}\relax
\mciteBstWouldAddEndPuncttrue
\mciteSetBstMidEndSepPunct{\mcitedefaultmidpunct}
{\mcitedefaultendpunct}{\mcitedefaultseppunct}\relax
\EndOfBibitem
\bibitem[{\citenamefont{Kaplan}(1992)}]{Kaplan:1991ah}
\bibinfo{author}{\bibfnamefont{D.~B.} \bibnamefont{Kaplan}},
  \bibinfo{journal}{Phys.Rev.Lett.} \textbf{\bibinfo{volume}{68}},
  \bibinfo{pages}{741} (\bibinfo{year}{1992})\relax
\mciteBstWouldAddEndPuncttrue
\mciteSetBstMidEndSepPunct{\mcitedefaultmidpunct}
{\mcitedefaultendpunct}{\mcitedefaultseppunct}\relax
\EndOfBibitem
\bibitem[{\citenamefont{Kaplan et~al.}(2009)\citenamefont{Kaplan, Luty, and
  Zurek}}]{Kaplan:2009ag}
\bibinfo{author}{\bibfnamefont{D.~E.} \bibnamefont{Kaplan}},
  \bibinfo{author}{\bibfnamefont{M.~A.} \bibnamefont{Luty}}, \bibnamefont{and}
  \bibinfo{author}{\bibfnamefont{K.~M.} \bibnamefont{Zurek}},
  \bibinfo{journal}{Phys.Rev.} \textbf{\bibinfo{volume}{D79}},
  \bibinfo{pages}{115016} (\bibinfo{year}{2009}), \eprint{0901.4117}\relax
\mciteBstWouldAddEndPuncttrue
\mciteSetBstMidEndSepPunct{\mcitedefaultmidpunct}
{\mcitedefaultendpunct}{\mcitedefaultseppunct}\relax
\EndOfBibitem
\bibitem[{\citenamefont{Essig et~al.}(2013{\natexlab{a}})\citenamefont{Essig,
  Jaros, Wester, Adrian, Andreas et~al.}}]{Essig:2013lka}
\bibinfo{author}{\bibfnamefont{R.}~\bibnamefont{Essig}},
  \bibinfo{author}{\bibfnamefont{J.~A.} \bibnamefont{Jaros}},
  \bibinfo{author}{\bibfnamefont{W.}~\bibnamefont{Wester}},
  \bibinfo{author}{\bibfnamefont{P.~H.} \bibnamefont{Adrian}},
  \bibinfo{author}{\bibfnamefont{S.}~\bibnamefont{Andreas}},
  \bibnamefont{et~al.} (\bibinfo{year}{2013}{\natexlab{a}}),
  \eprint{1311.0029}\relax
\mciteBstWouldAddEndPuncttrue
\mciteSetBstMidEndSepPunct{\mcitedefaultmidpunct}
{\mcitedefaultendpunct}{\mcitedefaultseppunct}\relax
\EndOfBibitem
\bibitem[{\citenamefont{Andreas
  et~al.}(2013{\natexlab{a}})\citenamefont{Andreas, Donskov, Crivelli,
  Gardikiotis, Gninenko et~al.}}]{Andreas:2013lya}
\bibinfo{author}{\bibfnamefont{S.}~\bibnamefont{Andreas}},
  \bibinfo{author}{\bibfnamefont{S.}~\bibnamefont{Donskov}},
  \bibinfo{author}{\bibfnamefont{P.}~\bibnamefont{Crivelli}},
  \bibinfo{author}{\bibfnamefont{A.}~\bibnamefont{Gardikiotis}},
  \bibinfo{author}{\bibfnamefont{S.}~\bibnamefont{Gninenko}},
  \bibnamefont{et~al.} (\bibinfo{year}{2013}{\natexlab{a}}),
  \eprint{1312.3309}\relax
\mciteBstWouldAddEndPuncttrue
\mciteSetBstMidEndSepPunct{\mcitedefaultmidpunct}
{\mcitedefaultendpunct}{\mcitedefaultseppunct}\relax
\EndOfBibitem
\bibitem[{\citenamefont{Olive et~al.}(2014)}]{Agashe:2014kda}
\bibinfo{author}{\bibfnamefont{K.}~\bibnamefont{Olive}} \bibnamefont{et~al.}
  (\bibinfo{collaboration}{Particle Data Group}), \bibinfo{journal}{Chin.Phys.}
  \textbf{\bibinfo{volume}{C38}}, \bibinfo{pages}{090001}
  (\bibinfo{year}{2014})\relax
\mciteBstWouldAddEndPuncttrue
\mciteSetBstMidEndSepPunct{\mcitedefaultmidpunct}
{\mcitedefaultendpunct}{\mcitedefaultseppunct}\relax
\EndOfBibitem
\bibitem[{\citenamefont{Essig et~al.}(2013{\natexlab{b}})\citenamefont{Essig,
  Mardon, Papucci, Volansky, and Zhong}}]{Essig:2013vha}
\bibinfo{author}{\bibfnamefont{R.}~\bibnamefont{Essig}},
  \bibinfo{author}{\bibfnamefont{J.}~\bibnamefont{Mardon}},
  \bibinfo{author}{\bibfnamefont{M.}~\bibnamefont{Papucci}},
  \bibinfo{author}{\bibfnamefont{T.}~\bibnamefont{Volansky}}, \bibnamefont{and}
  \bibinfo{author}{\bibfnamefont{Y.-M.} \bibnamefont{Zhong}},
  \bibinfo{journal}{JHEP} \textbf{\bibinfo{volume}{1311}}, \bibinfo{pages}{167}
  (\bibinfo{year}{2013}{\natexlab{b}}), \eprint{1309.5084}\relax
\mciteBstWouldAddEndPuncttrue
\mciteSetBstMidEndSepPunct{\mcitedefaultmidpunct}
{\mcitedefaultendpunct}{\mcitedefaultseppunct}\relax
\EndOfBibitem
\bibitem[{\citenamefont{Giudice et~al.}(2012)\citenamefont{Giudice, Paradisi,
  and Passera}}]{Giudice:2012ms}
\bibinfo{author}{\bibfnamefont{G.}~\bibnamefont{Giudice}},
  \bibinfo{author}{\bibfnamefont{P.}~\bibnamefont{Paradisi}}, \bibnamefont{and}
  \bibinfo{author}{\bibfnamefont{M.}~\bibnamefont{Passera}},
  \bibinfo{journal}{JHEP} \textbf{\bibinfo{volume}{1211}}, \bibinfo{pages}{113}
  (\bibinfo{year}{2012}), \eprint{1208.6583}\relax
\mciteBstWouldAddEndPuncttrue
\mciteSetBstMidEndSepPunct{\mcitedefaultmidpunct}
{\mcitedefaultendpunct}{\mcitedefaultseppunct}\relax
\EndOfBibitem
\bibitem[{\citenamefont{Izaguirre et~al.}(2013)\citenamefont{Izaguirre,
  Krnjaic, Schuster, and Toro}}]{Izaguirre:2013uxa}
\bibinfo{author}{\bibfnamefont{E.}~\bibnamefont{Izaguirre}},
  \bibinfo{author}{\bibfnamefont{G.}~\bibnamefont{Krnjaic}},
  \bibinfo{author}{\bibfnamefont{P.}~\bibnamefont{Schuster}}, \bibnamefont{and}
  \bibinfo{author}{\bibfnamefont{N.}~\bibnamefont{Toro}},
  \bibinfo{journal}{Phys.Rev.} \textbf{\bibinfo{volume}{D88}},
  \bibinfo{pages}{114015} (\bibinfo{year}{2013}), \eprint{1307.6554}\relax
\mciteBstWouldAddEndPuncttrue
\mciteSetBstMidEndSepPunct{\mcitedefaultmidpunct}
{\mcitedefaultendpunct}{\mcitedefaultseppunct}\relax
\EndOfBibitem
\bibitem[{\citenamefont{Pospelov}(2009)}]{Pospelov:2008zw}
\bibinfo{author}{\bibfnamefont{M.}~\bibnamefont{Pospelov}},
  \bibinfo{journal}{Phys.Rev.} \textbf{\bibinfo{volume}{D80}},
  \bibinfo{pages}{095002} (\bibinfo{year}{2009}), \eprint{0811.1030}\relax
\mciteBstWouldAddEndPuncttrue
\mciteSetBstMidEndSepPunct{\mcitedefaultmidpunct}
{\mcitedefaultendpunct}{\mcitedefaultseppunct}\relax
\EndOfBibitem
\bibitem[{\citenamefont{Ablikim et~al.}(2008)}]{Ablikim:2007ek}
\bibinfo{author}{\bibfnamefont{M.}~\bibnamefont{Ablikim}} \bibnamefont{et~al.}
  (\bibinfo{collaboration}{BES Collaboration}),
  \bibinfo{journal}{Phys.Rev.Lett.} \textbf{\bibinfo{volume}{100}},
  \bibinfo{pages}{192001} (\bibinfo{year}{2008}), \eprint{0710.0039}\relax
\mciteBstWouldAddEndPuncttrue
\mciteSetBstMidEndSepPunct{\mcitedefaultmidpunct}
{\mcitedefaultendpunct}{\mcitedefaultseppunct}\relax
\EndOfBibitem
\bibitem[{\citenamefont{Adler et~al.}(2004)}]{Adler:2004hp}
\bibinfo{author}{\bibfnamefont{S.}~\bibnamefont{Adler}} \bibnamefont{et~al.}
  (\bibinfo{collaboration}{E787 Collaboration}), \bibinfo{journal}{Phys.Rev.}
  \textbf{\bibinfo{volume}{D70}}, \bibinfo{pages}{037102}
  (\bibinfo{year}{2004}), \eprint{hep-ex/0403034}\relax
\mciteBstWouldAddEndPuncttrue
\mciteSetBstMidEndSepPunct{\mcitedefaultmidpunct}
{\mcitedefaultendpunct}{\mcitedefaultseppunct}\relax
\EndOfBibitem
\bibitem[{\citenamefont{Artamonov et~al.}(2009)}]{Artamonov:2009sz}
\bibinfo{author}{\bibfnamefont{A.}~\bibnamefont{Artamonov}}
  \bibnamefont{et~al.} (\bibinfo{collaboration}{BNL-E949 Collaboration}),
  \bibinfo{journal}{Phys.Rev.} \textbf{\bibinfo{volume}{D79}},
  \bibinfo{pages}{092004} (\bibinfo{year}{2009}), \eprint{0903.0030}\relax
\mciteBstWouldAddEndPuncttrue
\mciteSetBstMidEndSepPunct{\mcitedefaultmidpunct}
{\mcitedefaultendpunct}{\mcitedefaultseppunct}\relax
\EndOfBibitem
\bibitem[{\citenamefont{Okun}(1982)}]{Okun:1982xi}
\bibinfo{author}{\bibfnamefont{L.}~\bibnamefont{Okun}},
  \bibinfo{journal}{Sov.Phys.JETP} \textbf{\bibinfo{volume}{56}},
  \bibinfo{pages}{502} (\bibinfo{year}{1982})\relax
\mciteBstWouldAddEndPuncttrue
\mciteSetBstMidEndSepPunct{\mcitedefaultmidpunct}
{\mcitedefaultendpunct}{\mcitedefaultseppunct}\relax
\EndOfBibitem
\bibitem[{\citenamefont{Holdom}(1986)}]{Holdom:1985ag}
\bibinfo{author}{\bibfnamefont{B.}~\bibnamefont{Holdom}},
  \bibinfo{journal}{Phys.Lett.} \textbf{\bibinfo{volume}{B166}},
  \bibinfo{pages}{196} (\bibinfo{year}{1986})\relax
\mciteBstWouldAddEndPuncttrue
\mciteSetBstMidEndSepPunct{\mcitedefaultmidpunct}
{\mcitedefaultendpunct}{\mcitedefaultseppunct}\relax
\EndOfBibitem
\bibitem[{\citenamefont{deNiverville et~al.}(2012)\citenamefont{deNiverville,
  McKeen, and Ritz}}]{deNiverville:2012ij}
\bibinfo{author}{\bibfnamefont{P.}~\bibnamefont{deNiverville}},
  \bibinfo{author}{\bibfnamefont{D.}~\bibnamefont{McKeen}}, \bibnamefont{and}
  \bibinfo{author}{\bibfnamefont{A.}~\bibnamefont{Ritz}},
  \bibinfo{journal}{Phys.Rev.} \textbf{\bibinfo{volume}{D86}},
  \bibinfo{pages}{035022} (\bibinfo{year}{2012}), \eprint{1205.3499}\relax
\mciteBstWouldAddEndPuncttrue
\mciteSetBstMidEndSepPunct{\mcitedefaultmidpunct}
{\mcitedefaultendpunct}{\mcitedefaultseppunct}\relax
\EndOfBibitem
\bibitem[{\citenamefont{Dharmapalan et~al.}(2012)}]{Dharmapalan:2012xp}
\bibinfo{author}{\bibfnamefont{R.}~\bibnamefont{Dharmapalan}}
  \bibnamefont{et~al.} (\bibinfo{collaboration}{MiniBooNE Collaboration})
  (\bibinfo{year}{2012}), \eprint{1211.2258}\relax
\mciteBstWouldAddEndPuncttrue
\mciteSetBstMidEndSepPunct{\mcitedefaultmidpunct}
{\mcitedefaultendpunct}{\mcitedefaultseppunct}\relax
\EndOfBibitem
\bibitem[{\citenamefont{deNiverville et~al.}(2011)\citenamefont{deNiverville,
  Pospelov, and Ritz}}]{deNiverville:2011it}
\bibinfo{author}{\bibfnamefont{P.}~\bibnamefont{deNiverville}},
  \bibinfo{author}{\bibfnamefont{M.}~\bibnamefont{Pospelov}}, \bibnamefont{and}
  \bibinfo{author}{\bibfnamefont{A.}~\bibnamefont{Ritz}},
  \bibinfo{journal}{Phys.Rev.} \textbf{\bibinfo{volume}{D84}},
  \bibinfo{pages}{075020} (\bibinfo{year}{2011}), \eprint{1107.4580}\relax
\mciteBstWouldAddEndPuncttrue
\mciteSetBstMidEndSepPunct{\mcitedefaultmidpunct}
{\mcitedefaultendpunct}{\mcitedefaultseppunct}\relax
\EndOfBibitem
\bibitem[{\citenamefont{Batell et~al.}(2009{\natexlab{a}})\citenamefont{Batell,
  Pospelov, and Ritz}}]{Batell:2009di}
\bibinfo{author}{\bibfnamefont{B.}~\bibnamefont{Batell}},
  \bibinfo{author}{\bibfnamefont{M.}~\bibnamefont{Pospelov}}, \bibnamefont{and}
  \bibinfo{author}{\bibfnamefont{A.}~\bibnamefont{Ritz}},
  \bibinfo{journal}{Phys.Rev.} \textbf{\bibinfo{volume}{D80}},
  \bibinfo{pages}{095024} (\bibinfo{year}{2009}{\natexlab{a}}),
  \eprint{0906.5614}\relax
\mciteBstWouldAddEndPuncttrue
\mciteSetBstMidEndSepPunct{\mcitedefaultmidpunct}
{\mcitedefaultendpunct}{\mcitedefaultseppunct}\relax
\EndOfBibitem
\bibitem[{\citenamefont{Batell et~al.}(2014{\natexlab{a}})\citenamefont{Batell,
  Essig, and Surujon}}]{Batell:2014mga}
\bibinfo{author}{\bibfnamefont{B.}~\bibnamefont{Batell}},
  \bibinfo{author}{\bibfnamefont{R.}~\bibnamefont{Essig}}, \bibnamefont{and}
  \bibinfo{author}{\bibfnamefont{Z.}~\bibnamefont{Surujon}},
  \bibinfo{journal}{Phys.Rev.Lett.} \textbf{\bibinfo{volume}{113}},
  \bibinfo{pages}{171802} (\bibinfo{year}{2014}{\natexlab{a}}),
  \eprint{1406.2698}\relax
\mciteBstWouldAddEndPuncttrue
\mciteSetBstMidEndSepPunct{\mcitedefaultmidpunct}
{\mcitedefaultendpunct}{\mcitedefaultseppunct}\relax
\EndOfBibitem
\bibitem[{\citenamefont{Battaglieri et~al.}(2014)}]{Battaglieri:2014qoa}
\bibinfo{author}{\bibfnamefont{M.}~\bibnamefont{Battaglieri}}
  \bibnamefont{et~al.} (\bibinfo{collaboration}{BDX Collaboration})
  (\bibinfo{year}{2014}), \eprint{1406.3028}\relax
\mciteBstWouldAddEndPuncttrue
\mciteSetBstMidEndSepPunct{\mcitedefaultmidpunct}
{\mcitedefaultendpunct}{\mcitedefaultseppunct}\relax
\EndOfBibitem
\bibitem[{\citenamefont{Batell et~al.}(2014{\natexlab{b}})\citenamefont{Batell,
  deNiverville, McKeen, Pospelov, and Ritz}}]{Batell:2014yra}
\bibinfo{author}{\bibfnamefont{B.}~\bibnamefont{Batell}},
  \bibinfo{author}{\bibfnamefont{P.}~\bibnamefont{deNiverville}},
  \bibinfo{author}{\bibfnamefont{D.}~\bibnamefont{McKeen}},
  \bibinfo{author}{\bibfnamefont{M.}~\bibnamefont{Pospelov}}, \bibnamefont{and}
  \bibinfo{author}{\bibfnamefont{A.}~\bibnamefont{Ritz}}
  (\bibinfo{year}{2014}{\natexlab{b}}), \eprint{1405.7049}\relax
\mciteBstWouldAddEndPuncttrue
\mciteSetBstMidEndSepPunct{\mcitedefaultmidpunct}
{\mcitedefaultendpunct}{\mcitedefaultseppunct}\relax
\EndOfBibitem
\bibitem[{\citenamefont{Izaguirre
  et~al.}(2014{\natexlab{a}})\citenamefont{Izaguirre, Krnjaic, Schuster, and
  Toro}}]{Izaguirre:2014dua}
\bibinfo{author}{\bibfnamefont{E.}~\bibnamefont{Izaguirre}},
  \bibinfo{author}{\bibfnamefont{G.}~\bibnamefont{Krnjaic}},
  \bibinfo{author}{\bibfnamefont{P.}~\bibnamefont{Schuster}}, \bibnamefont{and}
  \bibinfo{author}{\bibfnamefont{N.}~\bibnamefont{Toro}}
  (\bibinfo{year}{2014}{\natexlab{a}}), \eprint{1403.6826}\relax
\mciteBstWouldAddEndPuncttrue
\mciteSetBstMidEndSepPunct{\mcitedefaultmidpunct}
{\mcitedefaultendpunct}{\mcitedefaultseppunct}\relax
\EndOfBibitem
\bibitem[{\citenamefont{Diamond and Schuster}(2013)}]{Diamond:2013oda}
\bibinfo{author}{\bibfnamefont{M.~D.} \bibnamefont{Diamond}} \bibnamefont{and}
  \bibinfo{author}{\bibfnamefont{P.}~\bibnamefont{Schuster}},
  \bibinfo{journal}{Phys.Rev.Lett.} \textbf{\bibinfo{volume}{111}},
  \bibinfo{pages}{221803} (\bibinfo{year}{2013}), \eprint{1307.6861}\relax
\mciteBstWouldAddEndPuncttrue
\mciteSetBstMidEndSepPunct{\mcitedefaultmidpunct}
{\mcitedefaultendpunct}{\mcitedefaultseppunct}\relax
\EndOfBibitem
\bibitem[{\citenamefont{Bjorken et~al.}(2009)\citenamefont{Bjorken, Essig,
  Schuster, and Toro}}]{Bjorken:2009mm}
\bibinfo{author}{\bibfnamefont{J.~D.} \bibnamefont{Bjorken}},
  \bibinfo{author}{\bibfnamefont{R.}~\bibnamefont{Essig}},
  \bibinfo{author}{\bibfnamefont{P.}~\bibnamefont{Schuster}}, \bibnamefont{and}
  \bibinfo{author}{\bibfnamefont{N.}~\bibnamefont{Toro}},
  \bibinfo{journal}{Phys.Rev.} \textbf{\bibinfo{volume}{D80}},
  \bibinfo{pages}{075018} (\bibinfo{year}{2009}), \eprint{0906.0580}\relax
\mciteBstWouldAddEndPuncttrue
\mciteSetBstMidEndSepPunct{\mcitedefaultmidpunct}
{\mcitedefaultendpunct}{\mcitedefaultseppunct}\relax
\EndOfBibitem
\bibitem[{\citenamefont{Bjorken et~al.}(1988)\citenamefont{Bjorken, Ecklund,
  Nelson, Abashian, Church et~al.}}]{Bjorken:1988as}
\bibinfo{author}{\bibfnamefont{J.}~\bibnamefont{Bjorken}},
  \bibinfo{author}{\bibfnamefont{S.}~\bibnamefont{Ecklund}},
  \bibinfo{author}{\bibfnamefont{W.}~\bibnamefont{Nelson}},
  \bibinfo{author}{\bibfnamefont{A.}~\bibnamefont{Abashian}},
  \bibinfo{author}{\bibfnamefont{C.}~\bibnamefont{Church}},
  \bibnamefont{et~al.}, \bibinfo{journal}{Phys.Rev.}
  \textbf{\bibinfo{volume}{D38}}, \bibinfo{pages}{3375}
  (\bibinfo{year}{1988})\relax
\mciteBstWouldAddEndPuncttrue
\mciteSetBstMidEndSepPunct{\mcitedefaultmidpunct}
{\mcitedefaultendpunct}{\mcitedefaultseppunct}\relax
\EndOfBibitem
\bibitem[{\citenamefont{Riordan et~al.}(1987)\citenamefont{Riordan, Krasny,
  Lang, De~Barbaro, Bodek et~al.}}]{Riordan:1987aw}
\bibinfo{author}{\bibfnamefont{E.}~\bibnamefont{Riordan}},
  \bibinfo{author}{\bibfnamefont{M.}~\bibnamefont{Krasny}},
  \bibinfo{author}{\bibfnamefont{K.}~\bibnamefont{Lang}},
  \bibinfo{author}{\bibfnamefont{P.}~\bibnamefont{De~Barbaro}},
  \bibinfo{author}{\bibfnamefont{A.}~\bibnamefont{Bodek}},
  \bibnamefont{et~al.}, \bibinfo{journal}{Phys.Rev.Lett.}
  \textbf{\bibinfo{volume}{59}}, \bibinfo{pages}{755}
  (\bibinfo{year}{1987})\relax
\mciteBstWouldAddEndPuncttrue
\mciteSetBstMidEndSepPunct{\mcitedefaultmidpunct}
{\mcitedefaultendpunct}{\mcitedefaultseppunct}\relax
\EndOfBibitem
\bibitem[{\citenamefont{Bross et~al.}(1991)\citenamefont{Bross, Crisler,
  Pordes, Volk, Errede et~al.}}]{Bross:1989mp}
\bibinfo{author}{\bibfnamefont{A.}~\bibnamefont{Bross}},
  \bibinfo{author}{\bibfnamefont{M.}~\bibnamefont{Crisler}},
  \bibinfo{author}{\bibfnamefont{S.~H.} \bibnamefont{Pordes}},
  \bibinfo{author}{\bibfnamefont{J.}~\bibnamefont{Volk}},
  \bibinfo{author}{\bibfnamefont{S.}~\bibnamefont{Errede}},
  \bibnamefont{et~al.}, \bibinfo{journal}{Phys.Rev.Lett.}
  \textbf{\bibinfo{volume}{67}}, \bibinfo{pages}{2942}
  (\bibinfo{year}{1991})\relax
\mciteBstWouldAddEndPuncttrue
\mciteSetBstMidEndSepPunct{\mcitedefaultmidpunct}
{\mcitedefaultendpunct}{\mcitedefaultseppunct}\relax
\EndOfBibitem
\bibitem[{\citenamefont{Essig et~al.}(2009)\citenamefont{Essig, Schuster, and
  Toro}}]{Essig:2009nc}
\bibinfo{author}{\bibfnamefont{R.}~\bibnamefont{Essig}},
  \bibinfo{author}{\bibfnamefont{P.}~\bibnamefont{Schuster}}, \bibnamefont{and}
  \bibinfo{author}{\bibfnamefont{N.}~\bibnamefont{Toro}},
  \bibinfo{journal}{Phys.Rev.} \textbf{\bibinfo{volume}{D80}},
  \bibinfo{pages}{015003} (\bibinfo{year}{2009}), \eprint{0903.3941}\relax
\mciteBstWouldAddEndPuncttrue
\mciteSetBstMidEndSepPunct{\mcitedefaultmidpunct}
{\mcitedefaultendpunct}{\mcitedefaultseppunct}\relax
\EndOfBibitem
\bibitem[{\citenamefont{Essig et~al.}(2011)\citenamefont{Essig, Schuster, Toro,
  and Wojtsekhowski}}]{Essig:2010xa}
\bibinfo{author}{\bibfnamefont{R.}~\bibnamefont{Essig}},
  \bibinfo{author}{\bibfnamefont{P.}~\bibnamefont{Schuster}},
  \bibinfo{author}{\bibfnamefont{N.}~\bibnamefont{Toro}}, \bibnamefont{and}
  \bibinfo{author}{\bibfnamefont{B.}~\bibnamefont{Wojtsekhowski}},
  \bibinfo{journal}{JHEP} \textbf{\bibinfo{volume}{1102}}, \bibinfo{pages}{009}
  (\bibinfo{year}{2011}), \eprint{1001.2557}\relax
\mciteBstWouldAddEndPuncttrue
\mciteSetBstMidEndSepPunct{\mcitedefaultmidpunct}
{\mcitedefaultendpunct}{\mcitedefaultseppunct}\relax
\EndOfBibitem
\bibitem[{\citenamefont{Fayet}(2007)}]{Fayet:2007ua}
\bibinfo{author}{\bibfnamefont{P.}~\bibnamefont{Fayet}},
  \bibinfo{journal}{Phys.Rev.} \textbf{\bibinfo{volume}{D75}},
  \bibinfo{pages}{115017} (\bibinfo{year}{2007}), \eprint{hep-ph/0702176}\relax
\mciteBstWouldAddEndPuncttrue
\mciteSetBstMidEndSepPunct{\mcitedefaultmidpunct}
{\mcitedefaultendpunct}{\mcitedefaultseppunct}\relax
\EndOfBibitem
\bibitem[{\citenamefont{Freytsis et~al.}(2010)\citenamefont{Freytsis,
  Ovanesyan, and Thaler}}]{Freytsis:2009bh}
\bibinfo{author}{\bibfnamefont{M.}~\bibnamefont{Freytsis}},
  \bibinfo{author}{\bibfnamefont{G.}~\bibnamefont{Ovanesyan}},
  \bibnamefont{and} \bibinfo{author}{\bibfnamefont{J.}~\bibnamefont{Thaler}},
  \bibinfo{journal}{JHEP} \textbf{\bibinfo{volume}{1001}}, \bibinfo{pages}{111}
  (\bibinfo{year}{2010}), \eprint{0909.2862}\relax
\mciteBstWouldAddEndPuncttrue
\mciteSetBstMidEndSepPunct{\mcitedefaultmidpunct}
{\mcitedefaultendpunct}{\mcitedefaultseppunct}\relax
\EndOfBibitem
\bibitem[{\citenamefont{Essig et~al.}(2010)\citenamefont{Essig, Harnik, Kaplan,
  and Toro}}]{Essig:2010gu}
\bibinfo{author}{\bibfnamefont{R.}~\bibnamefont{Essig}},
  \bibinfo{author}{\bibfnamefont{R.}~\bibnamefont{Harnik}},
  \bibinfo{author}{\bibfnamefont{J.}~\bibnamefont{Kaplan}}, \bibnamefont{and}
  \bibinfo{author}{\bibfnamefont{N.}~\bibnamefont{Toro}},
  \bibinfo{journal}{Phys.Rev.} \textbf{\bibinfo{volume}{D82}},
  \bibinfo{pages}{113008} (\bibinfo{year}{2010}), \eprint{1008.0636}\relax
\mciteBstWouldAddEndPuncttrue
\mciteSetBstMidEndSepPunct{\mcitedefaultmidpunct}
{\mcitedefaultendpunct}{\mcitedefaultseppunct}\relax
\EndOfBibitem
\bibitem[{\citenamefont{Reece and Wang}(2009)}]{Reece:2009un}
\bibinfo{author}{\bibfnamefont{M.}~\bibnamefont{Reece}} \bibnamefont{and}
  \bibinfo{author}{\bibfnamefont{L.-T.} \bibnamefont{Wang}},
  \bibinfo{journal}{JHEP} \textbf{\bibinfo{volume}{0907}}, \bibinfo{pages}{051}
  (\bibinfo{year}{2009}), \eprint{0904.1743}\relax
\mciteBstWouldAddEndPuncttrue
\mciteSetBstMidEndSepPunct{\mcitedefaultmidpunct}
{\mcitedefaultendpunct}{\mcitedefaultseppunct}\relax
\EndOfBibitem
\bibitem[{\citenamefont{Wojtsekhowski}(2009)}]{Wojtsekhowski:2009vz}
\bibinfo{author}{\bibfnamefont{B.}~\bibnamefont{Wojtsekhowski}},
  \bibinfo{journal}{AIP Conf.Proc.} \textbf{\bibinfo{volume}{1160}},
  \bibinfo{pages}{149} (\bibinfo{year}{2009}), \eprint{0906.5265}\relax
\mciteBstWouldAddEndPuncttrue
\mciteSetBstMidEndSepPunct{\mcitedefaultmidpunct}
{\mcitedefaultendpunct}{\mcitedefaultseppunct}\relax
\EndOfBibitem
\bibitem[{\citenamefont{Amelino-Camelia
  et~al.}(2010)\citenamefont{Amelino-Camelia, Archilli, Babusci, Badoni,
  Bencivenni et~al.}}]{AmelinoCamelia:2010me}
\bibinfo{author}{\bibfnamefont{G.}~\bibnamefont{Amelino-Camelia}},
  \bibinfo{author}{\bibfnamefont{F.}~\bibnamefont{Archilli}},
  \bibinfo{author}{\bibfnamefont{D.}~\bibnamefont{Babusci}},
  \bibinfo{author}{\bibfnamefont{D.}~\bibnamefont{Badoni}},
  \bibinfo{author}{\bibfnamefont{G.}~\bibnamefont{Bencivenni}},
  \bibnamefont{et~al.}, \bibinfo{journal}{Eur.Phys.J.}
  \textbf{\bibinfo{volume}{C68}}, \bibinfo{pages}{619} (\bibinfo{year}{2010}),
  \eprint{1003.3868}\relax
\mciteBstWouldAddEndPuncttrue
\mciteSetBstMidEndSepPunct{\mcitedefaultmidpunct}
{\mcitedefaultendpunct}{\mcitedefaultseppunct}\relax
\EndOfBibitem
\bibitem[{\citenamefont{Batell et~al.}(2009{\natexlab{b}})\citenamefont{Batell,
  Pospelov, and Ritz}}]{Batell:2009yf}
\bibinfo{author}{\bibfnamefont{B.}~\bibnamefont{Batell}},
  \bibinfo{author}{\bibfnamefont{M.}~\bibnamefont{Pospelov}}, \bibnamefont{and}
  \bibinfo{author}{\bibfnamefont{A.}~\bibnamefont{Ritz}},
  \bibinfo{journal}{Phys.Rev.} \textbf{\bibinfo{volume}{D79}},
  \bibinfo{pages}{115008} (\bibinfo{year}{2009}{\natexlab{b}}),
  \eprint{0903.0363}\relax
\mciteBstWouldAddEndPuncttrue
\mciteSetBstMidEndSepPunct{\mcitedefaultmidpunct}
{\mcitedefaultendpunct}{\mcitedefaultseppunct}\relax
\EndOfBibitem
\bibitem[{\citenamefont{Baumgart et~al.}(2009)\citenamefont{Baumgart, Cheung,
  Ruderman, Wang, and Yavin}}]{Baumgart:2009tn}
\bibinfo{author}{\bibfnamefont{M.}~\bibnamefont{Baumgart}},
  \bibinfo{author}{\bibfnamefont{C.}~\bibnamefont{Cheung}},
  \bibinfo{author}{\bibfnamefont{J.~T.} \bibnamefont{Ruderman}},
  \bibinfo{author}{\bibfnamefont{L.-T.} \bibnamefont{Wang}}, \bibnamefont{and}
  \bibinfo{author}{\bibfnamefont{I.}~\bibnamefont{Yavin}},
  \bibinfo{journal}{JHEP} \textbf{\bibinfo{volume}{0904}}, \bibinfo{pages}{014}
  (\bibinfo{year}{2009}), \eprint{0901.0283}\relax
\mciteBstWouldAddEndPuncttrue
\mciteSetBstMidEndSepPunct{\mcitedefaultmidpunct}
{\mcitedefaultendpunct}{\mcitedefaultseppunct}\relax
\EndOfBibitem
\bibitem[{\citenamefont{Merkel et~al.}(2011)}]{Merkel:2011ze}
\bibinfo{author}{\bibfnamefont{H.}~\bibnamefont{Merkel}} \bibnamefont{et~al.}
  (\bibinfo{collaboration}{A1 Collaboration}),
  \bibinfo{journal}{Phys.Rev.Lett.} \textbf{\bibinfo{volume}{106}},
  \bibinfo{pages}{251802} (\bibinfo{year}{2011}), \eprint{1101.4091}\relax
\mciteBstWouldAddEndPuncttrue
\mciteSetBstMidEndSepPunct{\mcitedefaultmidpunct}
{\mcitedefaultendpunct}{\mcitedefaultseppunct}\relax
\EndOfBibitem
\bibitem[{\citenamefont{Abrahamyan et~al.}(2011)}]{Abrahamyan:2011gv}
\bibinfo{author}{\bibfnamefont{S.}~\bibnamefont{Abrahamyan}}
  \bibnamefont{et~al.} (\bibinfo{collaboration}{APEX Collaboration}),
  \bibinfo{journal}{Phys.Rev.Lett.} \textbf{\bibinfo{volume}{107}},
  \bibinfo{pages}{191804} (\bibinfo{year}{2011}), \eprint{1108.2750}\relax
\mciteBstWouldAddEndPuncttrue
\mciteSetBstMidEndSepPunct{\mcitedefaultmidpunct}
{\mcitedefaultendpunct}{\mcitedefaultseppunct}\relax
\EndOfBibitem
\bibitem[{\citenamefont{Aubert et~al.}(2009)}]{Aubert:2009cp}
\bibinfo{author}{\bibfnamefont{B.}~\bibnamefont{Aubert}} \bibnamefont{et~al.}
  (\bibinfo{collaboration}{BaBar Collaboration}),
  \bibinfo{journal}{Phys.Rev.Lett.} \textbf{\bibinfo{volume}{103}},
  \bibinfo{pages}{081803} (\bibinfo{year}{2009}), \eprint{0905.4539}\relax
\mciteBstWouldAddEndPuncttrue
\mciteSetBstMidEndSepPunct{\mcitedefaultmidpunct}
{\mcitedefaultendpunct}{\mcitedefaultseppunct}\relax
\EndOfBibitem
\bibitem[{\citenamefont{Babusci et~al.}(2013)}]{Babusci:2012cr}
\bibinfo{author}{\bibfnamefont{D.}~\bibnamefont{Babusci}} \bibnamefont{et~al.}
  (\bibinfo{collaboration}{KLOE-2 Collaboration}),
  \bibinfo{journal}{Phys.Lett.} \textbf{\bibinfo{volume}{B720}},
  \bibinfo{pages}{111} (\bibinfo{year}{2013}), \eprint{1210.3927}\relax
\mciteBstWouldAddEndPuncttrue
\mciteSetBstMidEndSepPunct{\mcitedefaultmidpunct}
{\mcitedefaultendpunct}{\mcitedefaultseppunct}\relax
\EndOfBibitem
\bibitem[{\citenamefont{Echenard}(2012)}]{Echenard:2012hq}
\bibinfo{author}{\bibfnamefont{B.}~\bibnamefont{Echenard}},
  \bibinfo{journal}{Adv.High Energy Phys.} \textbf{\bibinfo{volume}{2012}},
  \bibinfo{pages}{514014} (\bibinfo{year}{2012}), \eprint{1209.1143}\relax
\mciteBstWouldAddEndPuncttrue
\mciteSetBstMidEndSepPunct{\mcitedefaultmidpunct}
{\mcitedefaultendpunct}{\mcitedefaultseppunct}\relax
\EndOfBibitem
\bibitem[{\citenamefont{Adlarson et~al.}(2013)}]{Adlarson:2013eza}
\bibinfo{author}{\bibfnamefont{P.}~\bibnamefont{Adlarson}} \bibnamefont{et~al.}
  (\bibinfo{collaboration}{WASA-at-COSY Collaboration}) (\bibinfo{year}{2013}),
  \eprint{1304.0671}\relax
\mciteBstWouldAddEndPuncttrue
\mciteSetBstMidEndSepPunct{\mcitedefaultmidpunct}
{\mcitedefaultendpunct}{\mcitedefaultseppunct}\relax
\EndOfBibitem
\bibitem[{\citenamefont{Hook et~al.}(2011)\citenamefont{Hook, Izaguirre, and
  Wacker}}]{Hook:2010tw}
\bibinfo{author}{\bibfnamefont{A.}~\bibnamefont{Hook}},
  \bibinfo{author}{\bibfnamefont{E.}~\bibnamefont{Izaguirre}},
  \bibnamefont{and} \bibinfo{author}{\bibfnamefont{J.~G.}
  \bibnamefont{Wacker}}, \bibinfo{journal}{Adv.High Energy Phys.}
  \textbf{\bibinfo{volume}{2011}}, \bibinfo{pages}{859762}
  (\bibinfo{year}{2011}), \eprint{1006.0973}\relax
\mciteBstWouldAddEndPuncttrue
\mciteSetBstMidEndSepPunct{\mcitedefaultmidpunct}
{\mcitedefaultendpunct}{\mcitedefaultseppunct}\relax
\EndOfBibitem
\bibitem[{\citenamefont{Goudzovski}(2014)}]{NA48}
\bibinfo{author}{\bibfnamefont{E.}~\bibnamefont{Goudzovski}}
  (\bibinfo{collaboration}{NA48/2 Collaboration}) (\bibinfo{year}{2014})\relax
\mciteBstWouldAddEndPuncttrue
\mciteSetBstMidEndSepPunct{\mcitedefaultmidpunct}
{\mcitedefaultendpunct}{\mcitedefaultseppunct}\relax
\EndOfBibitem
\bibitem[{\citenamefont{Boehm et~al.}(2004)\citenamefont{Boehm, Hooper, Silk,
  Casse, and Paul}}]{Boehm:2003bt}
\bibinfo{author}{\bibfnamefont{C.}~\bibnamefont{Boehm}},
  \bibinfo{author}{\bibfnamefont{D.}~\bibnamefont{Hooper}},
  \bibinfo{author}{\bibfnamefont{J.}~\bibnamefont{Silk}},
  \bibinfo{author}{\bibfnamefont{M.}~\bibnamefont{Casse}}, \bibnamefont{and}
  \bibinfo{author}{\bibfnamefont{J.}~\bibnamefont{Paul}},
  \bibinfo{journal}{Phys.Rev.Lett.} \textbf{\bibinfo{volume}{92}},
  \bibinfo{pages}{101301} (\bibinfo{year}{2004}),
  \eprint{astro-ph/0309686}\relax
\mciteBstWouldAddEndPuncttrue
\mciteSetBstMidEndSepPunct{\mcitedefaultmidpunct}
{\mcitedefaultendpunct}{\mcitedefaultseppunct}\relax
\EndOfBibitem
\bibitem[{\citenamefont{Huh et~al.}(2008)\citenamefont{Huh, Kim, Park, and
  Park}}]{Huh:2007zw}
\bibinfo{author}{\bibfnamefont{J.-H.} \bibnamefont{Huh}},
  \bibinfo{author}{\bibfnamefont{J.~E.} \bibnamefont{Kim}},
  \bibinfo{author}{\bibfnamefont{J.-C.} \bibnamefont{Park}}, \bibnamefont{and}
  \bibinfo{author}{\bibfnamefont{S.~C.} \bibnamefont{Park}},
  \bibinfo{journal}{Phys.Rev.} \textbf{\bibinfo{volume}{D77}},
  \bibinfo{pages}{123503} (\bibinfo{year}{2008}), \eprint{0711.3528}\relax
\mciteBstWouldAddEndPuncttrue
\mciteSetBstMidEndSepPunct{\mcitedefaultmidpunct}
{\mcitedefaultendpunct}{\mcitedefaultseppunct}\relax
\EndOfBibitem
\bibitem[{\citenamefont{Arkani-Hamed et~al.}(2009)\citenamefont{Arkani-Hamed,
  Finkbeiner, Slatyer, and Weiner}}]{ArkaniHamed:2008qn}
\bibinfo{author}{\bibfnamefont{N.}~\bibnamefont{Arkani-Hamed}},
  \bibinfo{author}{\bibfnamefont{D.~P.} \bibnamefont{Finkbeiner}},
  \bibinfo{author}{\bibfnamefont{T.~R.} \bibnamefont{Slatyer}},
  \bibnamefont{and} \bibinfo{author}{\bibfnamefont{N.}~\bibnamefont{Weiner}},
  \bibinfo{journal}{Phys.Rev.} \textbf{\bibinfo{volume}{D79}},
  \bibinfo{pages}{015014} (\bibinfo{year}{2009}), \eprint{0810.0713}\relax
\mciteBstWouldAddEndPuncttrue
\mciteSetBstMidEndSepPunct{\mcitedefaultmidpunct}
{\mcitedefaultendpunct}{\mcitedefaultseppunct}\relax
\EndOfBibitem
\bibitem[{\citenamefont{Hooper and Zurek}(2008)}]{Hooper:2008im}
\bibinfo{author}{\bibfnamefont{D.}~\bibnamefont{Hooper}} \bibnamefont{and}
  \bibinfo{author}{\bibfnamefont{K.~M.} \bibnamefont{Zurek}},
  \bibinfo{journal}{Phys.Rev.} \textbf{\bibinfo{volume}{D77}},
  \bibinfo{pages}{087302} (\bibinfo{year}{2008}), \eprint{0801.3686}\relax
\mciteBstWouldAddEndPuncttrue
\mciteSetBstMidEndSepPunct{\mcitedefaultmidpunct}
{\mcitedefaultendpunct}{\mcitedefaultseppunct}\relax
\EndOfBibitem
\bibitem[{\citenamefont{Cirelli and Strumia}(2009)}]{Cirelli:2009uv}
\bibinfo{author}{\bibfnamefont{M.}~\bibnamefont{Cirelli}} \bibnamefont{and}
  \bibinfo{author}{\bibfnamefont{A.}~\bibnamefont{Strumia}},
  \bibinfo{journal}{New J.Phys.} \textbf{\bibinfo{volume}{11}},
  \bibinfo{pages}{105005} (\bibinfo{year}{2009}), \eprint{0903.3381}\relax
\mciteBstWouldAddEndPuncttrue
\mciteSetBstMidEndSepPunct{\mcitedefaultmidpunct}
{\mcitedefaultendpunct}{\mcitedefaultseppunct}\relax
\EndOfBibitem
\bibitem[{\citenamefont{Cyr-Racine and Sigurdson}(2013)}]{CyrRacine:2012fz}
\bibinfo{author}{\bibfnamefont{F.-Y.} \bibnamefont{Cyr-Racine}}
  \bibnamefont{and}
  \bibinfo{author}{\bibfnamefont{K.}~\bibnamefont{Sigurdson}},
  \bibinfo{journal}{Phys.Rev.} \textbf{\bibinfo{volume}{D87}},
  \bibinfo{pages}{103515} (\bibinfo{year}{2013}), \eprint{1209.5752}\relax
\mciteBstWouldAddEndPuncttrue
\mciteSetBstMidEndSepPunct{\mcitedefaultmidpunct}
{\mcitedefaultendpunct}{\mcitedefaultseppunct}\relax
\EndOfBibitem
\bibitem[{\citenamefont{Kaplinghat et~al.}(2014)\citenamefont{Kaplinghat,
  Keeley, Linden, and Yu}}]{Kaplinghat:2013xca}
\bibinfo{author}{\bibfnamefont{M.}~\bibnamefont{Kaplinghat}},
  \bibinfo{author}{\bibfnamefont{R.~E.} \bibnamefont{Keeley}},
  \bibinfo{author}{\bibfnamefont{T.}~\bibnamefont{Linden}}, \bibnamefont{and}
  \bibinfo{author}{\bibfnamefont{H.-B.} \bibnamefont{Yu}},
  \bibinfo{journal}{Phys.Rev.Lett.} \textbf{\bibinfo{volume}{113}},
  \bibinfo{pages}{021302} (\bibinfo{year}{2014}), \eprint{1311.6524}\relax
\mciteBstWouldAddEndPuncttrue
\mciteSetBstMidEndSepPunct{\mcitedefaultmidpunct}
{\mcitedefaultendpunct}{\mcitedefaultseppunct}\relax
\EndOfBibitem
\bibitem[{\citenamefont{Kaplan et~al.}(2011)\citenamefont{Kaplan, Krnjaic,
  Rehermann, and Wells}}]{Kaplan:2011yj}
\bibinfo{author}{\bibfnamefont{D.~E.} \bibnamefont{Kaplan}},
  \bibinfo{author}{\bibfnamefont{G.~Z.} \bibnamefont{Krnjaic}},
  \bibinfo{author}{\bibfnamefont{K.~R.} \bibnamefont{Rehermann}},
  \bibnamefont{and} \bibinfo{author}{\bibfnamefont{C.~M.} \bibnamefont{Wells}},
  \bibinfo{journal}{JCAP} \textbf{\bibinfo{volume}{1110}}, \bibinfo{pages}{011}
  (\bibinfo{year}{2011}), \eprint{1105.2073}\relax
\mciteBstWouldAddEndPuncttrue
\mciteSetBstMidEndSepPunct{\mcitedefaultmidpunct}
{\mcitedefaultendpunct}{\mcitedefaultseppunct}\relax
\EndOfBibitem
\bibitem[{\citenamefont{Brust et~al.}(2013)\citenamefont{Brust, Kaplan, and
  Walters}}]{Brust:2013ova}
\bibinfo{author}{\bibfnamefont{C.}~\bibnamefont{Brust}},
  \bibinfo{author}{\bibfnamefont{D.~E.} \bibnamefont{Kaplan}},
  \bibnamefont{and} \bibinfo{author}{\bibfnamefont{M.~T.}
  \bibnamefont{Walters}}, \bibinfo{journal}{JHEP}
  \textbf{\bibinfo{volume}{1312}}, \bibinfo{pages}{058} (\bibinfo{year}{2013}),
  \eprint{1303.5379}\relax
\mciteBstWouldAddEndPuncttrue
\mciteSetBstMidEndSepPunct{\mcitedefaultmidpunct}
{\mcitedefaultendpunct}{\mcitedefaultseppunct}\relax
\EndOfBibitem
\bibitem[{\citenamefont{Barger et~al.}(2011)\citenamefont{Barger, Chiang,
  Keung, and Marfatia}}]{Barger:2010aj}
\bibinfo{author}{\bibfnamefont{V.}~\bibnamefont{Barger}},
  \bibinfo{author}{\bibfnamefont{C.-W.} \bibnamefont{Chiang}},
  \bibinfo{author}{\bibfnamefont{W.-Y.} \bibnamefont{Keung}}, \bibnamefont{and}
  \bibinfo{author}{\bibfnamefont{D.}~\bibnamefont{Marfatia}},
  \bibinfo{journal}{Phys.Rev.Lett.} \textbf{\bibinfo{volume}{106}},
  \bibinfo{pages}{153001} (\bibinfo{year}{2011}), \eprint{1011.3519}\relax
\mciteBstWouldAddEndPuncttrue
\mciteSetBstMidEndSepPunct{\mcitedefaultmidpunct}
{\mcitedefaultendpunct}{\mcitedefaultseppunct}\relax
\EndOfBibitem
\bibitem[{\citenamefont{Izaguirre
  et~al.}(2014{\natexlab{b}})\citenamefont{Izaguirre, Krnjaic, and
  Pospelov}}]{Izaguirre:2014cza}
\bibinfo{author}{\bibfnamefont{E.}~\bibnamefont{Izaguirre}},
  \bibinfo{author}{\bibfnamefont{G.}~\bibnamefont{Krnjaic}}, \bibnamefont{and}
  \bibinfo{author}{\bibfnamefont{M.}~\bibnamefont{Pospelov}}
  (\bibinfo{year}{2014}{\natexlab{b}}), \eprint{1405.4864}\relax
\mciteBstWouldAddEndPuncttrue
\mciteSetBstMidEndSepPunct{\mcitedefaultmidpunct}
{\mcitedefaultendpunct}{\mcitedefaultseppunct}\relax
\EndOfBibitem
\bibitem[{\citenamefont{Batell et~al.}(2011)\citenamefont{Batell, McKeen, and
  Pospelov}}]{Batell:2011qq}
\bibinfo{author}{\bibfnamefont{B.}~\bibnamefont{Batell}},
  \bibinfo{author}{\bibfnamefont{D.}~\bibnamefont{McKeen}}, \bibnamefont{and}
  \bibinfo{author}{\bibfnamefont{M.}~\bibnamefont{Pospelov}},
  \bibinfo{journal}{Phys.Rev.Lett.} \textbf{\bibinfo{volume}{107}},
  \bibinfo{pages}{011803} (\bibinfo{year}{2011}), \eprint{1103.0721}\relax
\mciteBstWouldAddEndPuncttrue
\mciteSetBstMidEndSepPunct{\mcitedefaultmidpunct}
{\mcitedefaultendpunct}{\mcitedefaultseppunct}\relax
\EndOfBibitem
\bibitem[{\citenamefont{Tucker-Smith and Yavin}(2011)}]{TuckerSmith:2010ra}
\bibinfo{author}{\bibfnamefont{D.}~\bibnamefont{Tucker-Smith}}
  \bibnamefont{and} \bibinfo{author}{\bibfnamefont{I.}~\bibnamefont{Yavin}},
  \bibinfo{journal}{Phys.Rev.} \textbf{\bibinfo{volume}{D83}},
  \bibinfo{pages}{101702} (\bibinfo{year}{2011}), \eprint{1011.4922}\relax
\mciteBstWouldAddEndPuncttrue
\mciteSetBstMidEndSepPunct{\mcitedefaultmidpunct}
{\mcitedefaultendpunct}{\mcitedefaultseppunct}\relax
\EndOfBibitem
\bibitem[{\citenamefont{Kahn et~al.}(2008)\citenamefont{Kahn, Schmitt, and
  Tait}}]{Kahn:2007ru}
\bibinfo{author}{\bibfnamefont{Y.}~\bibnamefont{Kahn}},
  \bibinfo{author}{\bibfnamefont{M.}~\bibnamefont{Schmitt}}, \bibnamefont{and}
  \bibinfo{author}{\bibfnamefont{T.~M.} \bibnamefont{Tait}},
  \bibinfo{journal}{Phys.Rev.} \textbf{\bibinfo{volume}{D78}},
  \bibinfo{pages}{115002} (\bibinfo{year}{2008}), \eprint{0712.0007}\relax
\mciteBstWouldAddEndPuncttrue
\mciteSetBstMidEndSepPunct{\mcitedefaultmidpunct}
{\mcitedefaultendpunct}{\mcitedefaultseppunct}\relax
\EndOfBibitem
\bibitem[{\citenamefont{Galison and Manohar}(1984)}]{Galison:1983pa}
\bibinfo{author}{\bibfnamefont{P.}~\bibnamefont{Galison}} \bibnamefont{and}
  \bibinfo{author}{\bibfnamefont{A.}~\bibnamefont{Manohar}},
  \bibinfo{journal}{Phys.Lett.} \textbf{\bibinfo{volume}{B136}},
  \bibinfo{pages}{279} (\bibinfo{year}{1984})\relax
\mciteBstWouldAddEndPuncttrue
\mciteSetBstMidEndSepPunct{\mcitedefaultmidpunct}
{\mcitedefaultendpunct}{\mcitedefaultseppunct}\relax
\EndOfBibitem
\bibitem[{\citenamefont{Boehm and Fayet}(2004)}]{Boehm:2003hm}
\bibinfo{author}{\bibfnamefont{C.}~\bibnamefont{Boehm}} \bibnamefont{and}
  \bibinfo{author}{\bibfnamefont{P.}~\bibnamefont{Fayet}},
  \bibinfo{journal}{Nucl.Phys.} \textbf{\bibinfo{volume}{B683}},
  \bibinfo{pages}{219} (\bibinfo{year}{2004}), \eprint{hep-ph/0305261}\relax
\mciteBstWouldAddEndPuncttrue
\mciteSetBstMidEndSepPunct{\mcitedefaultmidpunct}
{\mcitedefaultendpunct}{\mcitedefaultseppunct}\relax
\EndOfBibitem
\bibitem[{\citenamefont{Pospelov et~al.}(2008)\citenamefont{Pospelov, Ritz, and
  Voloshin}}]{Pospelov:2007mp}
\bibinfo{author}{\bibfnamefont{M.}~\bibnamefont{Pospelov}},
  \bibinfo{author}{\bibfnamefont{A.}~\bibnamefont{Ritz}}, \bibnamefont{and}
  \bibinfo{author}{\bibfnamefont{M.~B.} \bibnamefont{Voloshin}},
  \bibinfo{journal}{Phys.Lett.} \textbf{\bibinfo{volume}{B662}},
  \bibinfo{pages}{53} (\bibinfo{year}{2008}), \eprint{0711.4866}\relax
\mciteBstWouldAddEndPuncttrue
\mciteSetBstMidEndSepPunct{\mcitedefaultmidpunct}
{\mcitedefaultendpunct}{\mcitedefaultseppunct}\relax
\EndOfBibitem
\bibitem[{\citenamefont{Finkbeiner and Weiner}(2007)}]{Finkbeiner:2007kk}
\bibinfo{author}{\bibfnamefont{D.~P.} \bibnamefont{Finkbeiner}}
  \bibnamefont{and} \bibinfo{author}{\bibfnamefont{N.}~\bibnamefont{Weiner}},
  \bibinfo{journal}{Phys.Rev.} \textbf{\bibinfo{volume}{D76}},
  \bibinfo{pages}{083519} (\bibinfo{year}{2007}),
  \eprint{astro-ph/0702587}\relax
\mciteBstWouldAddEndPuncttrue
\mciteSetBstMidEndSepPunct{\mcitedefaultmidpunct}
{\mcitedefaultendpunct}{\mcitedefaultseppunct}\relax
\EndOfBibitem
\bibitem[{\citenamefont{Alves et~al.}(2010)\citenamefont{Alves, Behbahani,
  Schuster, and Wacker}}]{Alves:2009nf}
\bibinfo{author}{\bibfnamefont{D.~S.} \bibnamefont{Alves}},
  \bibinfo{author}{\bibfnamefont{S.~R.} \bibnamefont{Behbahani}},
  \bibinfo{author}{\bibfnamefont{P.}~\bibnamefont{Schuster}}, \bibnamefont{and}
  \bibinfo{author}{\bibfnamefont{J.~G.} \bibnamefont{Wacker}},
  \bibinfo{journal}{Phys.Lett.} \textbf{\bibinfo{volume}{B692}},
  \bibinfo{pages}{323} (\bibinfo{year}{2010}), \eprint{0903.3945}\relax
\mciteBstWouldAddEndPuncttrue
\mciteSetBstMidEndSepPunct{\mcitedefaultmidpunct}
{\mcitedefaultendpunct}{\mcitedefaultseppunct}\relax
\EndOfBibitem
\bibitem[{\citenamefont{Feng et~al.}(2008)\citenamefont{Feng, Tu, and
  Yu}}]{Feng:2008mu}
\bibinfo{author}{\bibfnamefont{J.~L.} \bibnamefont{Feng}},
  \bibinfo{author}{\bibfnamefont{H.}~\bibnamefont{Tu}}, \bibnamefont{and}
  \bibinfo{author}{\bibfnamefont{H.-B.} \bibnamefont{Yu}},
  \bibinfo{journal}{JCAP} \textbf{\bibinfo{volume}{0810}}, \bibinfo{pages}{043}
  (\bibinfo{year}{2008}), \eprint{0808.2318}\relax
\mciteBstWouldAddEndPuncttrue
\mciteSetBstMidEndSepPunct{\mcitedefaultmidpunct}
{\mcitedefaultendpunct}{\mcitedefaultseppunct}\relax
\EndOfBibitem
\bibitem[{\citenamefont{Feng and Kumar}(2008)}]{Feng:2008ya}
\bibinfo{author}{\bibfnamefont{J.~L.} \bibnamefont{Feng}} \bibnamefont{and}
  \bibinfo{author}{\bibfnamefont{J.}~\bibnamefont{Kumar}},
  \bibinfo{journal}{Phys.Rev.Lett.} \textbf{\bibinfo{volume}{101}},
  \bibinfo{pages}{231301} (\bibinfo{year}{2008}), \eprint{0803.4196}\relax
\mciteBstWouldAddEndPuncttrue
\mciteSetBstMidEndSepPunct{\mcitedefaultmidpunct}
{\mcitedefaultendpunct}{\mcitedefaultseppunct}\relax
\EndOfBibitem
\bibitem[{\citenamefont{Morrissey et~al.}(2009)\citenamefont{Morrissey, Poland,
  and Zurek}}]{Morrissey:2009ur}
\bibinfo{author}{\bibfnamefont{D.~E.} \bibnamefont{Morrissey}},
  \bibinfo{author}{\bibfnamefont{D.}~\bibnamefont{Poland}}, \bibnamefont{and}
  \bibinfo{author}{\bibfnamefont{K.~M.} \bibnamefont{Zurek}},
  \bibinfo{journal}{JHEP} \textbf{\bibinfo{volume}{0907}}, \bibinfo{pages}{050}
  (\bibinfo{year}{2009}), \eprint{0904.2567}\relax
\mciteBstWouldAddEndPuncttrue
\mciteSetBstMidEndSepPunct{\mcitedefaultmidpunct}
{\mcitedefaultendpunct}{\mcitedefaultseppunct}\relax
\EndOfBibitem
\bibitem[{\citenamefont{Chang et~al.}(2010)\citenamefont{Chang, Liu, Pierce,
  Weiner, and Yavin}}]{Chang:2010yk}
\bibinfo{author}{\bibfnamefont{S.}~\bibnamefont{Chang}},
  \bibinfo{author}{\bibfnamefont{J.}~\bibnamefont{Liu}},
  \bibinfo{author}{\bibfnamefont{A.}~\bibnamefont{Pierce}},
  \bibinfo{author}{\bibfnamefont{N.}~\bibnamefont{Weiner}}, \bibnamefont{and}
  \bibinfo{author}{\bibfnamefont{I.}~\bibnamefont{Yavin}},
  \bibinfo{journal}{JCAP} \textbf{\bibinfo{volume}{1008}}, \bibinfo{pages}{018}
  (\bibinfo{year}{2010}), \eprint{1004.0697}\relax
\mciteBstWouldAddEndPuncttrue
\mciteSetBstMidEndSepPunct{\mcitedefaultmidpunct}
{\mcitedefaultendpunct}{\mcitedefaultseppunct}\relax
\EndOfBibitem
\bibitem[{\citenamefont{Morris and Weiner}(2011)}]{Morris:2011dj}
\bibinfo{author}{\bibfnamefont{R.}~\bibnamefont{Morris}} \bibnamefont{and}
  \bibinfo{author}{\bibfnamefont{N.}~\bibnamefont{Weiner}}
  (\bibinfo{year}{2011}), \eprint{1109.3747}\relax
\mciteBstWouldAddEndPuncttrue
\mciteSetBstMidEndSepPunct{\mcitedefaultmidpunct}
{\mcitedefaultendpunct}{\mcitedefaultseppunct}\relax
\EndOfBibitem
\bibitem[{\citenamefont{Falkowski et~al.}(2011)\citenamefont{Falkowski,
  Ruderman, and Volansky}}]{Falkowski:2011xh}
\bibinfo{author}{\bibfnamefont{A.}~\bibnamefont{Falkowski}},
  \bibinfo{author}{\bibfnamefont{J.~T.} \bibnamefont{Ruderman}},
  \bibnamefont{and} \bibinfo{author}{\bibfnamefont{T.}~\bibnamefont{Volansky}},
  \bibinfo{journal}{JHEP} \textbf{\bibinfo{volume}{1105}}, \bibinfo{pages}{106}
  (\bibinfo{year}{2011}), \eprint{1101.4936}\relax
\mciteBstWouldAddEndPuncttrue
\mciteSetBstMidEndSepPunct{\mcitedefaultmidpunct}
{\mcitedefaultendpunct}{\mcitedefaultseppunct}\relax
\EndOfBibitem
\bibitem[{\citenamefont{Essig et~al.}(2012)\citenamefont{Essig, Mardon, and
  Volansky}}]{Essig:2011nj}
\bibinfo{author}{\bibfnamefont{R.}~\bibnamefont{Essig}},
  \bibinfo{author}{\bibfnamefont{J.}~\bibnamefont{Mardon}}, \bibnamefont{and}
  \bibinfo{author}{\bibfnamefont{T.}~\bibnamefont{Volansky}},
  \bibinfo{journal}{Phys.Rev.} \textbf{\bibinfo{volume}{D85}},
  \bibinfo{pages}{076007} (\bibinfo{year}{2012}), \eprint{1108.5383}\relax
\mciteBstWouldAddEndPuncttrue
\mciteSetBstMidEndSepPunct{\mcitedefaultmidpunct}
{\mcitedefaultendpunct}{\mcitedefaultseppunct}\relax
\EndOfBibitem
\bibitem[{\citenamefont{Andreas
  et~al.}(2013{\natexlab{b}})\citenamefont{Andreas, Goodsell, and
  Ringwald}}]{Andreas:2011in}
\bibinfo{author}{\bibfnamefont{S.}~\bibnamefont{Andreas}},
  \bibinfo{author}{\bibfnamefont{M.}~\bibnamefont{Goodsell}}, \bibnamefont{and}
  \bibinfo{author}{\bibfnamefont{A.}~\bibnamefont{Ringwald}},
  \bibinfo{journal}{Phys.Rev.} \textbf{\bibinfo{volume}{D87}},
  \bibinfo{pages}{025007} (\bibinfo{year}{2013}{\natexlab{b}}),
  \eprint{1109.2869}\relax
\mciteBstWouldAddEndPuncttrue
\mciteSetBstMidEndSepPunct{\mcitedefaultmidpunct}
{\mcitedefaultendpunct}{\mcitedefaultseppunct}\relax
\EndOfBibitem
\bibitem[{\citenamefont{An et~al.}(2012)\citenamefont{An, Ji, and
  Wang}}]{An:2012va}
\bibinfo{author}{\bibfnamefont{H.}~\bibnamefont{An}},
  \bibinfo{author}{\bibfnamefont{X.}~\bibnamefont{Ji}}, \bibnamefont{and}
  \bibinfo{author}{\bibfnamefont{L.-T.} \bibnamefont{Wang}},
  \bibinfo{journal}{JHEP} \textbf{\bibinfo{volume}{1207}}, \bibinfo{pages}{182}
  (\bibinfo{year}{2012}), \eprint{1202.2894}\relax
\mciteBstWouldAddEndPuncttrue
\mciteSetBstMidEndSepPunct{\mcitedefaultmidpunct}
{\mcitedefaultendpunct}{\mcitedefaultseppunct}\relax
\EndOfBibitem
\bibitem[{\citenamefont{Graham et~al.}(2012)\citenamefont{Graham, Kaplan,
  Rajendran, and Walters}}]{Graham:2012su}
\bibinfo{author}{\bibfnamefont{P.~W.} \bibnamefont{Graham}},
  \bibinfo{author}{\bibfnamefont{D.~E.} \bibnamefont{Kaplan}},
  \bibinfo{author}{\bibfnamefont{S.}~\bibnamefont{Rajendran}},
  \bibnamefont{and} \bibinfo{author}{\bibfnamefont{M.~T.}
  \bibnamefont{Walters}}, \bibinfo{journal}{Phys.Dark Univ.}
  \textbf{\bibinfo{volume}{1}}, \bibinfo{pages}{32} (\bibinfo{year}{2012}),
  \eprint{1203.2531}\relax
\mciteBstWouldAddEndPuncttrue
\mciteSetBstMidEndSepPunct{\mcitedefaultmidpunct}
{\mcitedefaultendpunct}{\mcitedefaultseppunct}\relax
\EndOfBibitem
\bibitem[{\citenamefont{Hooper et~al.}(2012)\citenamefont{Hooper, Weiner, and
  Xue}}]{Hooper:2012cw}
\bibinfo{author}{\bibfnamefont{D.}~\bibnamefont{Hooper}},
  \bibinfo{author}{\bibfnamefont{N.}~\bibnamefont{Weiner}}, \bibnamefont{and}
  \bibinfo{author}{\bibfnamefont{W.}~\bibnamefont{Xue}},
  \bibinfo{journal}{Phys.Rev.} \textbf{\bibinfo{volume}{D86}},
  \bibinfo{pages}{056009} (\bibinfo{year}{2012}), \eprint{1206.2929}\relax
\mciteBstWouldAddEndPuncttrue
\mciteSetBstMidEndSepPunct{\mcitedefaultmidpunct}
{\mcitedefaultendpunct}{\mcitedefaultseppunct}\relax
\EndOfBibitem
\bibitem[{\citenamefont{Cline et~al.}(2012)\citenamefont{Cline, Liu, and
  Xue}}]{Cline:2012is}
\bibinfo{author}{\bibfnamefont{J.~M.} \bibnamefont{Cline}},
  \bibinfo{author}{\bibfnamefont{Z.}~\bibnamefont{Liu}}, \bibnamefont{and}
  \bibinfo{author}{\bibfnamefont{W.}~\bibnamefont{Xue}},
  \bibinfo{journal}{Phys.Rev.} \textbf{\bibinfo{volume}{D85}},
  \bibinfo{pages}{101302} (\bibinfo{year}{2012}), \eprint{1201.4858}\relax
\mciteBstWouldAddEndPuncttrue
\mciteSetBstMidEndSepPunct{\mcitedefaultmidpunct}
{\mcitedefaultendpunct}{\mcitedefaultseppunct}\relax
\EndOfBibitem
\bibitem[{\citenamefont{Foot}(2014)}]{Foot:2014mia}
\bibinfo{author}{\bibfnamefont{R.}~\bibnamefont{Foot}},
  \bibinfo{journal}{Int.J.Mod.Phys.} \textbf{\bibinfo{volume}{A29}},
  \bibinfo{pages}{1430013} (\bibinfo{year}{2014}), \eprint{1401.3965}\relax
\mciteBstWouldAddEndPuncttrue
\mciteSetBstMidEndSepPunct{\mcitedefaultmidpunct}
{\mcitedefaultendpunct}{\mcitedefaultseppunct}\relax
\EndOfBibitem
\bibitem[{\citenamefont{Hochberg et~al.}(2014)\citenamefont{Hochberg, Kuflik,
  Volansky, and Wacker}}]{Hochberg:2014dra}
\bibinfo{author}{\bibfnamefont{Y.}~\bibnamefont{Hochberg}},
  \bibinfo{author}{\bibfnamefont{E.}~\bibnamefont{Kuflik}},
  \bibinfo{author}{\bibfnamefont{T.}~\bibnamefont{Volansky}}, \bibnamefont{and}
  \bibinfo{author}{\bibfnamefont{J.~G.} \bibnamefont{Wacker}},
  \bibinfo{journal}{Phys.Rev.Lett.} \textbf{\bibinfo{volume}{113}},
  \bibinfo{pages}{171301} (\bibinfo{year}{2014}), \eprint{1402.5143}\relax
\mciteBstWouldAddEndPuncttrue
\mciteSetBstMidEndSepPunct{\mcitedefaultmidpunct}
{\mcitedefaultendpunct}{\mcitedefaultseppunct}\relax
\EndOfBibitem
\bibitem[{\citenamefont{Shuve and Yavin}(2014)}]{Shuve:2014doa}
\bibinfo{author}{\bibfnamefont{B.}~\bibnamefont{Shuve}} \bibnamefont{and}
  \bibinfo{author}{\bibfnamefont{I.}~\bibnamefont{Yavin}},
  \bibinfo{journal}{Phys.Rev.} \textbf{\bibinfo{volume}{D89}},
  \bibinfo{pages}{113004} (\bibinfo{year}{2014}), \eprint{1403.2727}\relax
\mciteBstWouldAddEndPuncttrue
\mciteSetBstMidEndSepPunct{\mcitedefaultmidpunct}
{\mcitedefaultendpunct}{\mcitedefaultseppunct}\relax
\EndOfBibitem
\bibitem[{\citenamefont{Krnjaic and Sigurdson}(2014)}]{Krnjaic:2014xza}
\bibinfo{author}{\bibfnamefont{G.}~\bibnamefont{Krnjaic}} \bibnamefont{and}
  \bibinfo{author}{\bibfnamefont{K.}~\bibnamefont{Sigurdson}}
  (\bibinfo{year}{2014}), \eprint{1406.1171}\relax
\mciteBstWouldAddEndPuncttrue
\mciteSetBstMidEndSepPunct{\mcitedefaultmidpunct}
{\mcitedefaultendpunct}{\mcitedefaultseppunct}\relax
\EndOfBibitem
\bibitem[{\citenamefont{Detmold et~al.}(2014)\citenamefont{Detmold, McCullough,
  and Pochinsky}}]{Detmold:2014qqa}
\bibinfo{author}{\bibfnamefont{W.}~\bibnamefont{Detmold}},
  \bibinfo{author}{\bibfnamefont{M.}~\bibnamefont{McCullough}},
  \bibnamefont{and} \bibinfo{author}{\bibfnamefont{A.}~\bibnamefont{Pochinsky}}
  (\bibinfo{year}{2014}), \eprint{1406.2276}\relax
\mciteBstWouldAddEndPuncttrue
\mciteSetBstMidEndSepPunct{\mcitedefaultmidpunct}
{\mcitedefaultendpunct}{\mcitedefaultseppunct}\relax
\EndOfBibitem
\bibitem[{\citenamefont{Auerbach et~al.}(2001)}]{Auerbach:2001wg}
\bibinfo{author}{\bibfnamefont{L.}~\bibnamefont{Auerbach}} \bibnamefont{et~al.}
  (\bibinfo{collaboration}{LSND Collaboration}), \bibinfo{journal}{Phys.Rev.}
  \textbf{\bibinfo{volume}{D63}}, \bibinfo{pages}{112001}
  (\bibinfo{year}{2001}), \eprint{hep-ex/0101039}\relax
\mciteBstWouldAddEndPuncttrue
\mciteSetBstMidEndSepPunct{\mcitedefaultmidpunct}
{\mcitedefaultendpunct}{\mcitedefaultseppunct}\relax
\EndOfBibitem
\bibitem[{\citenamefont{Slatyer et~al.}(2009)\citenamefont{Slatyer,
  Padmanabhan, and Finkbeiner}}]{Slatyer:2009yq}
\bibinfo{author}{\bibfnamefont{T.~R.} \bibnamefont{Slatyer}},
  \bibinfo{author}{\bibfnamefont{N.}~\bibnamefont{Padmanabhan}},
  \bibnamefont{and} \bibinfo{author}{\bibfnamefont{D.~P.}
  \bibnamefont{Finkbeiner}}, \bibinfo{journal}{Phys.Rev.}
  \textbf{\bibinfo{volume}{D80}}, \bibinfo{pages}{043526}
  (\bibinfo{year}{2009}), \eprint{0906.1197}\relax
\mciteBstWouldAddEndPuncttrue
\mciteSetBstMidEndSepPunct{\mcitedefaultmidpunct}
{\mcitedefaultendpunct}{\mcitedefaultseppunct}\relax
\EndOfBibitem
\bibitem[{\citenamefont{Galli et~al.}(2009)\citenamefont{Galli, Iocco, Bertone,
  and Melchiorri}}]{Galli:2009zc}
\bibinfo{author}{\bibfnamefont{S.}~\bibnamefont{Galli}},
  \bibinfo{author}{\bibfnamefont{F.}~\bibnamefont{Iocco}},
  \bibinfo{author}{\bibfnamefont{G.}~\bibnamefont{Bertone}}, \bibnamefont{and}
  \bibinfo{author}{\bibfnamefont{A.}~\bibnamefont{Melchiorri}},
  \bibinfo{journal}{Phys.Rev.} \textbf{\bibinfo{volume}{D80}},
  \bibinfo{pages}{023505} (\bibinfo{year}{2009}), \eprint{0905.0003}\relax
\mciteBstWouldAddEndPuncttrue
\mciteSetBstMidEndSepPunct{\mcitedefaultmidpunct}
{\mcitedefaultendpunct}{\mcitedefaultseppunct}\relax
\EndOfBibitem
\bibitem[{\citenamefont{Galli et~al.}(2011)\citenamefont{Galli, Iocco, Bertone,
  and Melchiorri}}]{Galli:2011rz}
\bibinfo{author}{\bibfnamefont{S.}~\bibnamefont{Galli}},
  \bibinfo{author}{\bibfnamefont{F.}~\bibnamefont{Iocco}},
  \bibinfo{author}{\bibfnamefont{G.}~\bibnamefont{Bertone}}, \bibnamefont{and}
  \bibinfo{author}{\bibfnamefont{A.}~\bibnamefont{Melchiorri}},
  \bibinfo{journal}{Phys.Rev.} \textbf{\bibinfo{volume}{D84}},
  \bibinfo{pages}{027302} (\bibinfo{year}{2011}), \eprint{1106.1528}\relax
\mciteBstWouldAddEndPuncttrue
\mciteSetBstMidEndSepPunct{\mcitedefaultmidpunct}
{\mcitedefaultendpunct}{\mcitedefaultseppunct}\relax
\EndOfBibitem
\bibitem[{\citenamefont{Hutsi et~al.}(2011)\citenamefont{Hutsi, Chluba, Hektor,
  and Raidal}}]{Hutsi:2011vx}
\bibinfo{author}{\bibfnamefont{G.}~\bibnamefont{Hutsi}},
  \bibinfo{author}{\bibfnamefont{J.}~\bibnamefont{Chluba}},
  \bibinfo{author}{\bibfnamefont{A.}~\bibnamefont{Hektor}}, \bibnamefont{and}
  \bibinfo{author}{\bibfnamefont{M.}~\bibnamefont{Raidal}},
  \bibinfo{journal}{Astron.Astrophys.} \textbf{\bibinfo{volume}{535}},
  \bibinfo{pages}{A26} (\bibinfo{year}{2011}), \eprint{1103.2766}\relax
\mciteBstWouldAddEndPuncttrue
\mciteSetBstMidEndSepPunct{\mcitedefaultmidpunct}
{\mcitedefaultendpunct}{\mcitedefaultseppunct}\relax
\EndOfBibitem
\bibitem[{\citenamefont{Markevitch et~al.}(2004)\citenamefont{Markevitch,
  Gonzalez, Clowe, Vikhlinin, David et~al.}}]{Markevitch:2003at}
\bibinfo{author}{\bibfnamefont{M.}~\bibnamefont{Markevitch}},
  \bibinfo{author}{\bibfnamefont{A.}~\bibnamefont{Gonzalez}},
  \bibinfo{author}{\bibfnamefont{D.}~\bibnamefont{Clowe}},
  \bibinfo{author}{\bibfnamefont{A.}~\bibnamefont{Vikhlinin}},
  \bibinfo{author}{\bibfnamefont{L.}~\bibnamefont{David}},
  \bibnamefont{et~al.}, \bibinfo{journal}{Astrophys.J.}
  \textbf{\bibinfo{volume}{606}}, \bibinfo{pages}{819} (\bibinfo{year}{2004}),
  \eprint{astro-ph/0309303}\relax
\mciteBstWouldAddEndPuncttrue
\mciteSetBstMidEndSepPunct{\mcitedefaultmidpunct}
{\mcitedefaultendpunct}{\mcitedefaultseppunct}\relax
\EndOfBibitem
\bibitem[{\citenamefont{Ade et~al.}(2013)}]{Ade:2013ktc}
\bibinfo{author}{\bibfnamefont{P.}~\bibnamefont{Ade}} \bibnamefont{et~al.}
  (\bibinfo{collaboration}{Planck Collaboration}) (\bibinfo{year}{2013}),
  \eprint{1303.5062}\relax
\mciteBstWouldAddEndPuncttrue
\mciteSetBstMidEndSepPunct{\mcitedefaultmidpunct}
{\mcitedefaultendpunct}{\mcitedefaultseppunct}\relax
\EndOfBibitem
\bibitem[{\citenamefont{Alwall et~al.}(2007)\citenamefont{Alwall, Demin,
  de~Visscher, Frederix, Herquet et~al.}}]{Alwall:2007st}
\bibinfo{author}{\bibfnamefont{J.}~\bibnamefont{Alwall}},
  \bibinfo{author}{\bibfnamefont{P.}~\bibnamefont{Demin}},
  \bibinfo{author}{\bibfnamefont{S.}~\bibnamefont{de~Visscher}},
  \bibinfo{author}{\bibfnamefont{R.}~\bibnamefont{Frederix}},
  \bibinfo{author}{\bibfnamefont{M.}~\bibnamefont{Herquet}},
  \bibnamefont{et~al.}, \bibinfo{journal}{JHEP}
  \textbf{\bibinfo{volume}{0709}}, \bibinfo{pages}{028} (\bibinfo{year}{2007}),
  \eprint{0706.2334}\relax
\mciteBstWouldAddEndPuncttrue
\mciteSetBstMidEndSepPunct{\mcitedefaultmidpunct}
{\mcitedefaultendpunct}{\mcitedefaultseppunct}\relax
\EndOfBibitem
\bibitem[{\citenamefont{Andreas et~al.}(2012)\citenamefont{Andreas, Niebuhr,
  and Ringwald}}]{Andreas:2012mt}
\bibinfo{author}{\bibfnamefont{S.}~\bibnamefont{Andreas}},
  \bibinfo{author}{\bibfnamefont{C.}~\bibnamefont{Niebuhr}}, \bibnamefont{and}
  \bibinfo{author}{\bibfnamefont{A.}~\bibnamefont{Ringwald}},
  \bibinfo{journal}{Phys.Rev.} \textbf{\bibinfo{volume}{D86}},
  \bibinfo{pages}{095019} (\bibinfo{year}{2012}), \eprint{1209.6083}\relax
\mciteBstWouldAddEndPuncttrue
\mciteSetBstMidEndSepPunct{\mcitedefaultmidpunct}
{\mcitedefaultendpunct}{\mcitedefaultseppunct}\relax
\EndOfBibitem
\bibitem[{\citenamefont{Aubert et~al.}(2008)}]{Aubert:2008as}
\bibinfo{author}{\bibfnamefont{B.}~\bibnamefont{Aubert}} \bibnamefont{et~al.}
  (\bibinfo{collaboration}{BaBar Collaboration}) (\bibinfo{year}{2008}),
  \eprint{0808.0017}\relax
\mciteBstWouldAddEndPuncttrue
\mciteSetBstMidEndSepPunct{\mcitedefaultmidpunct}
{\mcitedefaultendpunct}{\mcitedefaultseppunct}\relax
\EndOfBibitem
\bibitem[{\citenamefont{Kahn et~al.}(2014)\citenamefont{Kahn, Krnjaic, Thaler,
  and Toups}}]{Kahn:2014sra}
\bibinfo{author}{\bibfnamefont{Y.}~\bibnamefont{Kahn}},
  \bibinfo{author}{\bibfnamefont{G.}~\bibnamefont{Krnjaic}},
  \bibinfo{author}{\bibfnamefont{J.}~\bibnamefont{Thaler}}, \bibnamefont{and}
  \bibinfo{author}{\bibfnamefont{M.}~\bibnamefont{Toups}}
  (\bibinfo{year}{2014}), \eprint{1411.1055}\relax
\mciteBstWouldAddEndPuncttrue
\mciteSetBstMidEndSepPunct{\mcitedefaultmidpunct}
{\mcitedefaultendpunct}{\mcitedefaultseppunct}\relax
\EndOfBibitem
\bibitem[{\citenamefont{Tsai}(1974)}]{Tsai:1973py}
\bibinfo{author}{\bibfnamefont{Y.-S.} \bibnamefont{Tsai}},
  \bibinfo{journal}{Rev.Mod.Phys.} \textbf{\bibinfo{volume}{46}},
  \bibinfo{pages}{815} (\bibinfo{year}{1974})\relax
\mciteBstWouldAddEndPuncttrue
\mciteSetBstMidEndSepPunct{\mcitedefaultmidpunct}
{\mcitedefaultendpunct}{\mcitedefaultseppunct}\relax
\EndOfBibitem
\bibitem[{\citenamefont{Caldwell et~al.}(1973)\citenamefont{Caldwell, Elings,
  Hesse, Morrison, Murphy, and {others}}}]{Caldwell:1973bu}
\bibinfo{author}{\bibfnamefont{D.~O.} \bibnamefont{Caldwell}},
  \bibinfo{author}{\bibfnamefont{V.~B.} \bibnamefont{Elings}},
  \bibinfo{author}{\bibfnamefont{W.~P.} \bibnamefont{Hesse}},
  \bibinfo{author}{\bibfnamefont{R.~J.} \bibnamefont{Morrison}},
  \bibinfo{author}{\bibfnamefont{F.~V.} \bibnamefont{Murphy}},
  \bibnamefont{and} \bibinfo{author}{\bibnamefont{{others}}},
  \bibinfo{journal}{Phys.Rev.} \textbf{\bibinfo{volume}{D7}},
  \bibinfo{pages}{1362} (\bibinfo{year}{1973})\relax
\mciteBstWouldAddEndPuncttrue
\mciteSetBstMidEndSepPunct{\mcitedefaultmidpunct}
{\mcitedefaultendpunct}{\mcitedefaultseppunct}\relax
\EndOfBibitem
\bibitem[{\citenamefont{Bingham et~al.}(1973)\citenamefont{Bingham, Fretter,
  Podolsky, Rabin, Rosenfeld, Smadja, Ballam, Chadwick, Eisenberg, Gearhart
  et~al.}}]{PhysRevD.8.1277}
\bibinfo{author}{\bibfnamefont{H.~H.} \bibnamefont{Bingham}},
  \bibinfo{author}{\bibfnamefont{W.~B.} \bibnamefont{Fretter}},
  \bibinfo{author}{\bibfnamefont{W.~J.} \bibnamefont{Podolsky}},
  \bibinfo{author}{\bibfnamefont{M.~S.} \bibnamefont{Rabin}},
  \bibinfo{author}{\bibfnamefont{A.~H.} \bibnamefont{Rosenfeld}},
  \bibinfo{author}{\bibfnamefont{G.}~\bibnamefont{Smadja}},
  \bibinfo{author}{\bibfnamefont{J.}~\bibnamefont{Ballam}},
  \bibinfo{author}{\bibfnamefont{G.~B.} \bibnamefont{Chadwick}},
  \bibinfo{author}{\bibfnamefont{Y.}~\bibnamefont{Eisenberg}},
  \bibinfo{author}{\bibfnamefont{R.}~\bibnamefont{Gearhart}},
  \bibnamefont{et~al.}, \bibinfo{journal}{Phys. Rev. D}
  \textbf{\bibinfo{volume}{8}}, \bibinfo{pages}{1277} (\bibinfo{year}{1973}),
  \urlprefix\url{http://link.aps.org/doi/10.1103/PhysRevD.8.1277}\relax
\mciteBstWouldAddEndPuncttrue
\mciteSetBstMidEndSepPunct{\mcitedefaultmidpunct}
{\mcitedefaultendpunct}{\mcitedefaultseppunct}\relax
\EndOfBibitem
\bibitem[{\citenamefont{Butterworth and Wing}(2005)}]{Butterworth:884219}
\bibinfo{author}{\bibfnamefont{J.~M.} \bibnamefont{Butterworth}}
  \bibnamefont{and} \bibinfo{author}{\bibfnamefont{M.}~\bibnamefont{Wing}},
  \bibinfo{journal}{Rep. Prog. Phys.} \textbf{\bibinfo{volume}{68}},
  \bibinfo{pages}{2773} (\bibinfo{year}{2005})\relax
\mciteBstWouldAddEndPuncttrue
\mciteSetBstMidEndSepPunct{\mcitedefaultmidpunct}
{\mcitedefaultendpunct}{\mcitedefaultseppunct}\relax
\EndOfBibitem
\bibitem[{\citenamefont{Boyarski et~al.}(1968)\citenamefont{Boyarski, Bulos,
  Busza, Diebold, Ecklund, Fischer, Rees, and Richter}}]{1968PhRvL..20..300B}
\bibinfo{author}{\bibfnamefont{A.~M.} \bibnamefont{Boyarski}},
  \bibinfo{author}{\bibfnamefont{F.}~\bibnamefont{Bulos}},
  \bibinfo{author}{\bibfnamefont{W.}~\bibnamefont{Busza}},
  \bibinfo{author}{\bibfnamefont{R.}~\bibnamefont{Diebold}},
  \bibinfo{author}{\bibfnamefont{S.~D.} \bibnamefont{Ecklund}},
  \bibinfo{author}{\bibfnamefont{G.~E.} \bibnamefont{Fischer}},
  \bibinfo{author}{\bibfnamefont{J.~R.} \bibnamefont{Rees}}, \bibnamefont{and}
  \bibinfo{author}{\bibfnamefont{B.}~\bibnamefont{Richter}},
  \bibinfo{journal}{Physical Review Letters} \textbf{\bibinfo{volume}{20}},
  \bibinfo{pages}{300} (\bibinfo{year}{1968})\relax
\mciteBstWouldAddEndPuncttrue
\mciteSetBstMidEndSepPunct{\mcitedefaultmidpunct}
{\mcitedefaultendpunct}{\mcitedefaultseppunct}\relax
\EndOfBibitem
\bibitem[{\citenamefont{Anderson et~al.}(1969)\citenamefont{Anderson,
  Gustavson, Johnson, Overman, Ritson, and Wiik}}]{Anderson:1969bq}
\bibinfo{author}{\bibfnamefont{R.}~\bibnamefont{Anderson}},
  \bibinfo{author}{\bibfnamefont{D.}~\bibnamefont{Gustavson}},
  \bibinfo{author}{\bibfnamefont{J.}~\bibnamefont{Johnson}},
  \bibinfo{author}{\bibfnamefont{I.}~\bibnamefont{Overman}},
  \bibinfo{author}{\bibfnamefont{D.}~\bibnamefont{Ritson}}, \bibnamefont{and}
  \bibinfo{author}{\bibfnamefont{B.}~\bibnamefont{Wiik}},
  \bibinfo{journal}{Phys.Rev.Lett.} \textbf{\bibinfo{volume}{23}},
  \bibinfo{pages}{721} (\bibinfo{year}{1969})\relax
\mciteBstWouldAddEndPuncttrue
\mciteSetBstMidEndSepPunct{\mcitedefaultmidpunct}
{\mcitedefaultendpunct}{\mcitedefaultseppunct}\relax
\EndOfBibitem
\bibitem[{\citenamefont{Guidal et~al.}(1997)\citenamefont{Guidal, Laget, and
  Vanderhaeghen}}]{1997NuPhA.627..645G}
\bibinfo{author}{\bibfnamefont{M.}~\bibnamefont{Guidal}},
  \bibinfo{author}{\bibfnamefont{J.~M.} \bibnamefont{Laget}}, \bibnamefont{and}
  \bibinfo{author}{\bibfnamefont{M.}~\bibnamefont{Vanderhaeghen}},
  \bibinfo{journal}{Nuclear Physics} \textbf{\bibinfo{volume}{627}},
  \bibinfo{pages}{645} (\bibinfo{year}{1997})\relax
\mciteBstWouldAddEndPuncttrue
\mciteSetBstMidEndSepPunct{\mcitedefaultmidpunct}
{\mcitedefaultendpunct}{\mcitedefaultseppunct}\relax
\EndOfBibitem
\end{mcitethebibliography}
\end{document}